\documentclass[aps,pra,showpacs,twoside,twocolumn,10pt]{revtex4-2}
\usepackage[colorlinks=true, citecolor=red, urlcolor=blue ]{hyperref}
\usepackage{epsfig,newlfont,amssymb,amsfonts,amsmath,bm,subfigure,palatino,mathtools,amsthm,braket,soul,enumitem,color,graphics,graphicx,times,physics,bbold}
\usepackage[normalem]{ulem}
\usepackage{xcolor}
\usepackage{physics}
\usepackage{dsfont}
\usepackage{mathrsfs}
\usepackage{verbatim}
\usepackage{amsthm,amssymb}
\usepackage{comment}
\usepackage{bigints}

\begin{document}

\title{Availing non-Markovian dynamics in effective negative temperature-based transient \\ quantum Otto engines}

\author{Arghya Maity and Ahana Ghoshal}
\affiliation{Harish-Chandra Research Institute, A CI of Homi Bhabha National
Institute, 
Chhatnag Road, Jhunsi, Allahabad - 211019, India}

\begin{abstract} 
We demonstrate that the efficiency of effective negative temperature-based quantum Otto engines, already known to outperform their traditional counterparts operating with positive-temperature thermal reservoirs, can be further improved by terminating the isochoric strokes before the working substance reaches perfect equilibrium with its environment. Our investigation encompasses both Markovian and non-Markovian dynamics during these finite-time isochoric processes while considering a weak coupling between the working substance and the reservoirs. We assess the performance of these engines as they undergo a transition from the Markovian to the non-Markovian regime using two figures of merit: maximum achievable efficiency at a certain finite time during the isochoric heating stroke, and overall performance of the engine over an extended period during the transient phase of this stroke. We show that the maximum efficiency increases with the increase of non-Markovianity. However, the overall engine performance decreases as non-Markovianity increases. Additionally, we discover the existence of effective negative temperature-based necessarily transient quantum Otto engines. These engines operate within an extended operational domain, reaching into temperature ranges where conventional effective negative temperature-based quantum Otto engines, which rely on perfect thermalization during the isochoric strokes, are unable to function. Furthermore, this extended operational domain of an effective negative temperature-based necessarily transient quantum Otto engine increases as non-Markovianity becomes more pronounced.
\end{abstract}

\maketitle

\section{Introduction}
\label{sec:intro}

In the quest for efficient and sustainable energy conversion processes, researchers have been exploring various avenues to push the boundaries of conventional thermodynamics. Among these emerging frontiers, quantum Otto engines stand out as a promising paradigm that capitalizes on the fundamental principles of quantum mechanics to redefine the limits of thermodynamic efficiency in the quantum regime~\cite{Scovil_PRL_1959, Kosloff_JOCP_1984, Scully_Science_2003, Gemmer_QT_book, Sebastian_book, Francisco_Entropy_2020, Sebastian_Review_AVS_2022}. They draw inspiration from the classical Otto cycle but leverage the unique properties of quantum systems to potentially achieve efficiencies beyond those attainable through classical means. Quantum heat engines incorporate ``quantumness” by utilizing quantum systems as working mediums~\cite{Scully_PRL_2002, Scully_Science_2003, Enrique_PRE_2012, Lutz_PRL_2014, Deffner_Entropy_2019, Maslennikov_Nature_Comm_2019, Singh_PRB_2021, Myers_NJP_2021, Sebastian_Review_AVS_2022, Myers_Nanomaterials_2023} and different types of reservoirs~\cite{Scully_PRL_2001, Huang_PRE_2012,  Zagoskin_PRB_2012, Lutz_PRL_2014, Ferdi_PRE_2014, Rui_PRE_2015, Jan_PRX_2017, Hardal_PRE_2017, Hardal_PRE_2017, Xiao_PLA_2018, Niedenzu_Nature_2018, Jackson_PRA_2018, Wang_PRE_2019, Assis_PRE_2020, Yanchao_PhysicaA_2020, Assis_JOP_2021, Rafael_PRB_2022, Rafael_PRB_2022} capable of generating quantum effects. 
The different types of reservoirs that can be used to enhance the efficiency of quantum Otto engines include
squeezed reservoirs~\cite{Scully_PRL_2001, Huang_PRE_2012, Zagoskin_PRB_2012, Lutz_PRL_2014, Ferdi_PRE_2014, Rui_PRE_2015, Jan_PRX_2017, Hardal_PRE_2017, Xiao_PLA_2018, Niedenzu_Nature_2018, Wang_PRE_2019, Assis_PRE_2020, Yanchao_PhysicaA_2020, Assis_JOP_2021,  Rafael_PRB_2022}, spin reservoirs~\cite{Jackson_PRA_2018}, superconducting reservoirs~\cite{Hardal_PRE_2017, Rafael_PRB_2022}, etc. The main motive behind this quantum reservoir engineering is to find ways to get better overall performance~\cite{Mendon_PRR_2020, Sebastian_Review_AVS_2022}.

The utilization of engineered quantum reservoirs, which are characterized by quantum coherence or correlations, has been shown to enhance the performance of quantum heat engines beyond the standard Carnot limit~\cite{Abah_EPL_2014}. Interestingly, this improvement of efficiency beyond the Carnot limit does not lead to a violation of the laws of thermodynamics, as some resources such as quantum coherence~\cite{Scully_Science_2003}, quantum correlations~\cite{Dillenschneider_EPL_2009, Llobet_PRX_2015}, and squeezed thermal reservoirs~\cite{ Scully_PRL_2001, Lutz_PRL_2014, Yanchao_PhysicaA_2020, Rafael_PRB_2022} are employed. A novel advancement in this reservoir engineering approach for enhancing the efficiency of a quantum heat engine involves the use of an effective negative temperature reservoir ~\cite{Assis_PRL_2019,Gabriella_2023,Arghya_arxiv_2023, Manab_arxiv_2023}. In a system with an effective negative temperature, the population distribution of energy levels is inverted, meaning that higher energy states are more probable than lower energy states. This is not a traditional negative temperature in the sense of being colder than absolute zero on the Kelvin scale. Instead, it is a mathematical description of a particular type of energy distribution. So, the system with the effective negative temperature has a higher average energy per particle compared to a system with a positive temperature~\cite{Struchtrup_PRL_2018, Ramsey_PR_1956}. In the context of heat exchange, when a positive-temperature system is in thermal contact with a system possessing an effective negative temperature, heat flows from the latter to the former. In this sense, the effective negative temperature is considered ``hotter" than the positive one. In Ref.~\cite{Purcell_PR_1951}, the authors generated spin states characterized by inverted population and regarded them as states at negative spin temperatures. 
See Refs.~\cite{Struchtrup_PRL_2018,Ramsey_PR_1956, Lincoln_Science_2013, Braun_Science_2013, Tacchino_PRL_2018} for more works in this regard. 
In Ref.~\cite{Assis_PRL_2019} Assis \textit{et al.} first conducted an experimental study employing such a reservoir in a quantum Otto engine. The operation of this engine was facilitated using a spin-$\frac{1}{2}$ particle of a $^{13}$C nucleus as the quantum fuel. During the heating stroke, this system was connected to an effective negative-temperature reservoir, while in the cooling stroke, it was coupled to a reservoir exhibiting a positive temperature. They considered that the two adiabatic strokes were governed by a unitary operation corresponding to a driving Hamiltonian described by a rotating magnetic field confined to the \(x,y\) plane. Notably, 
this effective negative temperature-based quantum Otto engine exhibits higher efficiency compared to a normal quantum Otto engine operating with both reservoirs at positive spin temperatures. 
Furthermore, in a recent research work~\cite{Arghya_arxiv_2023} on the effective negative temperature-based quantum Otto engine, the same setup was utilized, with the addition of an extra magnetic field introduced in the $z$ direction of the driving Hamiltonian. 
This study demonstrated that by adjusting the strength of the magnetic field in the \(z\) direction, the efficiency of the engine can be significantly enhanced in comparison to the scenario where no magnetic field was present in the $z$ direction.
In addition to improved efficiency, the effective negative temperature-based quantum Otto engine offers several other significant advantages. One notable advantage of this quantum Otto engine is that the faster the expansion and compression strokes are performed, the higher the efficiency. Therefore, the engine is not constrained to slow adiabatic processes to achieve superior performance~\cite{Assis_PRL_2019,Arghya_arxiv_2023}. Additionally, the effective temperature-based quantum Otto engine exhibits high robustness against disorder, which is typically a rare occurrence~\cite{Arghya_arxiv_2023}.

The effective negative temperature-based quantum Otto engines, studied previously in Refs.~\cite{Assis_PRL_2019, Arghya_arxiv_2023}, are predominantly focused on maintaining perfect thermalization, a condition where the working medium remains in contact with the reservoir for a sufficiently long time during the isochoric heating and cooling strokes. Such a long duration of contact allows the working substance to attain its canonical equilibrium state, corresponding to the hot and cold baths, 
respectively. However, quantum systems, by their very nature, are highly susceptible to their surrounding environments. Hence, prolonging an isochoric stroke for a long time can lead to undesirable consequences, including decoherence and fluctuations. These adverse effects can substantially deviate the engine from its ideal behavior. Additionally, in certain scenarios, achieving the canonical equilibrium state of the working substance after a prolonged period of evolution is not guaranteed. Therefore, it is necessary to explore the study of quantum Otto engines where the two isochoric strokes are prematurely truncated during the transient regime. This approach acknowledges the practical challenges posed by the sensitivity of quantum systems to their environment and the potential benefits of finite-time operations compared to steady-state operations. 
See, e.g., Refs.~\cite{Brask_PRE_2015,Mitchison_NJP_2015,Sreetama_EPL_2019,Ghosh_PRA_2021,Saha_arxiv_2023} in this regard.  
In this paper, we deal with an effective negative temperature-based quantum Otto engine operating in the transient regime of the two isochoric strokes. Throughout the remainder of this paper, we will refer to this engine as an effective negative temperature-based transient quantum Otto engine.

For the finite-time evolution during the two isochoric strokes, the behavior of the working substance is governed by an open quantum dynamics~\cite{Petruccione_book,Alicki_2007,Rivas_Huelga_book,Subhashish_Banerjee_book,Lidar_2020_lecture}. The simplest form of this dynamics can be derived when there exists a weak coupling between the system and the reservoir, and when we take into consideration the Born-Markov and secular approximations
~\cite{Sudarshan_JMP_1976,Lindblad_CMP_1976,Petruccione_book,Alicki_2007,Rivas_Huelga_book,Subhashish_Banerjee_book,Lidar_2020_lecture}. These approximations are applicable when the correlation function of the heat bath exhibits a faster decay rate compared to the relaxation time of the evolution of the quantum working medium. Within the framework of the Born-Markov approximations, information flows only in one direction, from the system to the reservoir. However, there are scenarios in which information can flow in the opposite direction, from the environment back to the system, leading to what is known as non-Markovian evolution~\cite{Esposito_NJP_2010, Breuer_RMP_2016, Seifert_PRL_2016, Strasberg_PRX_2017, Rivas_PRL_2020, Landi_RMP_2021}. Given that the Born-Markov approximations impose strict constraints on the evolution of the system and the characteristics of the heat baths, which are rarely found in natural settings, it becomes more practical to consider non-Markovian effects on the evolution of the working substance. In recent years, several significant studies have emerged, all of which revolve around the exploration of non-Markovian effects in the context of a quantum Otto cycle~\cite{Zhang_JOP_2014, Sibasish-Ghosh_PRE_2018, Pozas_NJP_2018, Pezzutto_QST_2019, Wiedmann_NJP_2020, Victor_CP_2020, Segal_PRL_2021, Sagnik_PRE_2022, Mao_PRE_2023, Miku_PRR_2023}. 

In this paper, 
we show that the effective negative temperature-based transient quantum Otto engines, comprising a spin-$\frac{1}{2}$ working substance, can achieve higher efficiency compared to an effective negative temperature-based quantum Otto engine operating in the steady-state regime of the isochoric strokes. Our investigation delves deeply into the Markovian as well as the non-Markovian dynamics of the working substance and is conducted within the framework of weak coupling between the system and the baths. In this context, non-Markovianity represents the unique phenomenon of information backflow from the environment to the system. In this comprehensive analysis, we assess the performance of the effective negative temperature-based transient quantum Otto engines across the spectrum from the Markovian to the non-Markovian regime. 
Our evaluation employs two key performance indicators. The first one is the maximum achievable efficiency at a certain time and the second one is the overall engine performance over an extended period encompassing the transient phase of the heating stroke. We find that the maximum efficiency experiences a significant boost with an increase in non-Markovianity.
However, intriguingly, the overall engine performance shows a decline as non-Markovianity increases. Moreover, our investigation leads to the discovery of a unique class of engines: the effective negative temperature-based necessarily transient quantum Otto engines. In scenarios where traditional effective negative temperature-based quantum Otto engines, dependent on either perfect thermalization or infinite-time protocols, are rendered ineffective due to temperature constraints, these engines extend their operational reach into previously inaccessible temperature ranges~\cite{Brask_PRE_2015, Mitchison_NJP_2015, Valerio_Quantum_2017, Sreetama_EPL_2019}.
Furthermore, we observe that the operational range of these effective negative temperature-based necessarily transient quantum Otto engines expands significantly with increasing non-Markovianity.

The remainder of the paper is structured as follows. In Sec.~\ref{sec:Protocol}, we introduce the framework of four-stroke effective negative temperature-based transient quantum Otto engines. Section~\ref{Sec:non_marko} briefly discusses the identification of information backflow and the canonical measure of non-Markovianity utilized in this study. In Sec.~\ref{sec:Reslt}, we investigate the efficiency of an effective negative temperature-based transient quantum Otto engine and explore the influence of information backflow on its performance. The existence and characteristics of effective negative temperature-based necessarily transient quantum Otto engines are demonstrated in Sec.~\ref{sec:Nec_Trans}. Finally, we provide our concluding remarks in Sec.~\ref{sec:Con}.

\section{Effective negative temperature-based transient quantum Otto engine}
\label{sec:Protocol}
Similar to a conventional quantum Otto engine, an effective negative temperature-based transient quantum Otto engine consists of four distinct strokes. 
The engine initiates with an expansion stroke, followed by an isochoric heating stroke, then progresses to a compression stroke, and finally concludes with an isochoric cooling stroke. The expansion and compression strokes are adiabatic, indicating no heat exchange between the working medium and the reservoir. While eigenenergies might vary during these strokes, the populations remain constant, thereby preserving the von Neumann entropy of the system. The unitary evolution of the expansion stroke is characterized by a change in Hamiltonian from $\mathcal{H}_{\text{cold}}$ to $\mathcal{H}_{\text{hot}}$ wherein the frequency $\nu_{\text{hot}}$, corresponding to $\mathcal{H}_{\text{hot}}$, is greater than the frequency $\nu_{\text{cold}}$, corresponding to $\mathcal{H}_{\text{cold}}$, resulting in an expansion of the energy gap, enabling the system to perform work. Conversely in the compression stroke, the driving Hamiltonian drives the system $\mathcal{H}_{\text{hot}}$ to again $\mathcal{H}_{\text{cold}}$ and there is a compression in the energy gap, allowing work to be done on the system by the surroundings. 
It is imperative that the ideal adiabatic strokes within the quantum Otto cycle are executed quasistatically~\cite{Sebastian_book, Sebastian_Review_AVS_2022}. On the other hand, in the isochoric stroke, the eigenenergies remain constant, yet the populations associated with each energy level undergo change. Consequently, this population shift leads to a variation in the von Neumann entropy of the system. In adherence to the first law of thermodynamics, the alteration in a system's internal energy throughout a thermodynamic process can be delineated into contributions from heat and work. 
The distinct nature of each stroke within the quantum Otto cycle, involving an exchange of either work or heat but never both simultaneously, renders it an appealing prospect for practical implementation.
We will now provide a detailed examination of each of these strokes.
\par
The initial state of the spin-$\frac{1}{2}$ working medium is prepared in a thermal state, specifically the Gibbs state denoted as $\rho_{\text{in}}^{\text{th}} = e^{-\beta_{\text{cold}}(\mathcal{H}_{\text{cold}} -\mu_{c} \hat{N}_{c})}/\mathcal{Z}_{\text{cold}}$, corresponding to the temperature $T_{\text{cold}}$ of the cold bath. The cold bath is characterized as having a positive temperature, with $\beta_{\text{cold}}=1/k_BT_{\text{cold}}$ representing the inverse temperature, where $k_B$ symbolizes the Boltzmann constant. Here, $\mu_c$ represents the chemical potential, accompanied by the number operator $\hat{N}_c$ and $\mathcal{Z}_{\text{cold}}= \Tr[e^{-\beta_{\text{cold}}( \mathcal{H}_{\text{cold}}-\mu_c \hat{N}_c)}] $ denotes the corresponding partition function.
 The Hamiltonian $\mathcal{H}_{\text{cold}}$ is taken as $\mathcal{H}_{\text{cold}} = -\hbar\pi \nu_{\text{cold}} \sigma_x + \hbar \frac{\Tilde{\omega}}{2} \sigma_z $, where $\hbar$ is the reduced Planck constant,  $\nu_{\text{cold}}$ represents the fixed frequency determined by the physical system, and  $\sigma_{i}$ \((i=x,y,z)\) signifies the Pauli matrices. The parameter $\Tilde{\omega}$ corresponds to the magnetic field strength along the $z$ direction.
\par
In the \textit{expansion stroke} 
the system undergoes a unitary evolution from time $t=0$ to $t=\tau$ governed by the Hamiltonian
\begin{equation}
\label{Equ:driven_Hamiltonian}
\mathcal{H}_{\text{exp}}(t) = - \hbar \pi \nu(t)\Big[\sigma_x \cos \omega t + \sigma_y \sin \omega t \Big ] + \hbar \sigma_z \frac{\Tilde{\omega}}{2}.
\end{equation}
Here, $\omega=\frac{\pi}{2\tau}$, ensuring a complete rotation of the magnetic field from the $x$ to $y$ direction \cite{Peterson_PRL_2019, Denzler_PRR_2020}. The function $\nu(t) = \nu_{\text{cold}}\Big[ 1- \left (\frac{t}{\tau} \right )\Big ] + \nu_{\text{hot}} \left (\frac{t}{\tau} \right )$ encapsulates how the energy spacing widens from $\nu_{\text{cold}}$ at time $t=0$ to $\nu_{\text{hot}}$ at $t=\tau$. Additionally, we incorporate a constant magnetic field along the $z$ direction with a strength of $\frac{\Tilde{\omega}}{2}$, where $\Tilde{\omega} =g \omega$ and $g$ is a dimensionless constant as taken in~\cite{Arghya_arxiv_2023}.
 The final duration of evolution, denoted as $\tau$, is determined by the specific experiment conducted~\cite{T.B.Batalhao_PRL_2014, T.B.Batalhao_PRL_2015}. 
The unitary operator by which the system evolves during this expansion stroke is denoted as $ U_{\tau,0} = \mathcal{T} e^{-\frac{i}{\hbar} \int_{t=0}^{\tau} \mathcal{H}_{\text{exp}}(t) dt}$, where $\mathcal{T}$ represents the time-ordering operator. 
At the end of the expansion stroke, i.e., at time $\tau$, the Hamiltonian of Eq.~(\ref{Equ:driven_Hamiltonian}) reduces to $\mathcal{H}_{\text{hot}}= - \hbar \pi \nu_{\text{hot}} \sigma_y + \hbar \frac{\Tilde{\omega}}{2} \sigma_z$.
Consequently, the resulting state transforms into $\rho_{\text{exp}} = U_{\tau,0} \rho_{\text{in}}^{\text{th}} U_{\tau,0}^{\dagger}$.
The transition probability between the eigenstates of the Hamiltonians $\mathcal{H}_{\text{cold}}$ and $\mathcal{H}_{\text{hot}}$ at time $\tau$, is defined as $\xi(\tau) = \big| \bra{\Psi_{\pm}^{\text{hot}}}U_{\tau,0}\ket{\Psi_{\mp}^{\text{cold}}} \big|^{2} = \big| \bra{\Psi_{\pm}^{\text{cold}}}U^{\dagger}_{\tau,0}\ket{\Psi_{\mp}^{\text{hot}}} \big|^{2}$.
Here, $\ket{\Psi_{\pm}^{\text{hot}}}$ and $\ket{\Psi_{\pm}^{\text{cold}}}$ correspond to the eigenstates of $\mathcal{H}_{\text{hot}}$ and $\mathcal{H}_{\text{cold}}$, respectively. The 
plus sign signifies the eigenstates associated with higher energy levels, while 
the minus sign represents those associated with lower energy levels. 
Precisely, this $\xi(\tau) $ is the adiabaticity parameter and for an ideal adiabatic process, i.e., for infinitely large $\tau$, it becomes zero. As shown in~\cite{Assis_PRL_2019,Arghya_arxiv_2023}, it is evident that $\xi(\tau)$ exhibits a higher value for shorter driving times $\tau$. As $\tau$ increases, the amplitude of oscillations in $\xi(\tau)$ diminishes, and for larger values of $\tau$, it nearly vanishes. 
The efficiency of the engine is directly related to the transition probability $\xi(\tau)$, and a higher value of the transition probability corresponds to a higher efficiency. So for our purpose, we limit the duration of the driving time to a small value, $\tau=100 \mu$s. Furthermore, by selecting an appropriate strength of magnetic field in the $z$ direction ($g$), the transition probability can be significantly enhanced. Thus, for a chosen driving time of $\tau=100 \mu$s, the value of $g$ is set to $0.2$ \cite{Arghya_arxiv_2023}.
This choices allows us to achieve a significant transition probability and hence a higher efficiency. Note that during this stroke, the working substance is isolated from its surroundings, and as a result, no heat exchange with the environment takes place. Consequently, the overall change in internal energy results from the work done in this phase.
\par
In the next phase, referred to as the \textit{isochoric heating stroke}, 
the working medium comes into contact with a reservoir kept at an effective negative temperature $T_{\text{hot}}$, denoted as the hot reservoir. The conventional quantum Otto engines typically operate within the steady-state regime during this isochoric process. However, as previously mentioned, our study is focused on examining the performance of an effective negative temperature-based quantum Otto engine within the transient regime of the isochoric processes. So, in contrast to the usual practice of allowing the system to reach thermal equilibrium with the connected thermal bath, we opt to terminate the stroke within a finite-time duration $\tilde{t}$, where $\tilde{t}<t_{\text{eq}}$. Here, $t_{\text{eq}}$ stands for the duration of the heating stroke at which the working substance reaches equilibrium with the hot bath. 
During this stroke, the Hamiltonian of the system remains constant at $\mathcal{H}_{\text{hot}}$, and as a result, no work is performed. Therefore, any alteration in the internal energy of the system is indicative of the heat exchange with the reservoir. 
The total Hamiltonian of the composite system-bath setup hence can be written as,
\begin{equation}
    H = H_{\text{sys}} + H_{\text{bath}} + H_{\text{int}}, 
\end{equation}
where $H_{\text{sys}}$ is the free Hamiltonian of the system and for this stroke, $H_{\text{sys}}=\mathcal{H}_{\text{hot}}$. $H_{\text{bath}}$ is the free Hamiltonian of the reservoir, and $H_{\text{int}}$ is the interaction Hamiltonian between the system and the reservoir, respectively, given as 
\begin{align}
&H_{\text{bath}} = 
\hbar \Omega\int^{\infty}_{0} d\omega^{\prime} b^{\dagger}_{\omega^{\prime}} b_{\omega^{\prime}}, \nonumber\\
&H_{\text{int}}=\hbar \sqrt{\Omega} \int_{0}^{\infty} d\omega^{\prime} \zeta(\omega^{\prime}) \Big( \sigma_{+} b_{\omega^{\prime}} + \sigma_{-} b^{\dagger}_{\omega^{\prime}}  \Big).
\label{Hamil}
\end{align}
Here we define $\sigma_{\pm}$ as $(\sigma_{x} \pm i \sigma_{y}) / 2$, and $\Omega$ as a constant having the unit of frequency. $b_{\omega^{\prime}} (b_{\omega^{\prime}}^{\dagger})$ represents the annihilation (creation) operators associated with each mode, and $\omega^{\prime}$ signifies the frequency of these individual modes. These operators have units of $\frac{1}{\sqrt{\omega^{\prime}}}$ and adhere to either the anticommutation or commutation relation, depending on whether they correspond to a fermionic or bosonic bath. For fermionic baths, the relation is expressed as $\{b_{\omega^{\prime}},b_{\omega^{\prime\prime}}^{\dagger}\}=\delta(\omega^{\prime}-\omega^{\prime\prime})$, where the curly braces denote anticommutation. Conversely, for bosonic baths, the relation takes the form $[b_{\omega^{\prime}},b_{\omega^{\prime\prime}}^{\dagger}]=\delta(\omega^{\prime}-\omega^{\prime\prime})$, where the square brackets signify commutation. 
The function $\zeta(\omega^{\prime})$ serves as a dimensionless coupling parameter between the system and the reservoir. To be more precise, $2\pi\Omega|\zeta(\omega^{\prime})|^2=\mathcal{J}(\omega^{\prime})$, where $\mathcal{J}(\omega^{\prime})$ represents the spectral density function of the reservoir that is coupled to the system. With this composite setup, the reduced dynamics of the system within the time interval 0 to $\tilde{t}$ is governed by the open quantum dynamics,
and the resulting output state at time $\tilde{t}$ comes out as 
$
\rho_{\text{heat}} = \Phi_{h}(\tilde{t},0)\rho_{\text{exp}} 
$. Here $\Phi_h(\tilde{t},0)$ is the dynamical map associated with the evolution of the system during this stroke, given by,
$
 \Phi_{h} (\tilde{t},0) = \mathcal{T} \exp \Big ( \int_{0}^{\tilde{t}} dt \mathcal{L}(\rho(t)) \Big )
$, where 
\begin{equation}
    \label{dyn_equ}
    \mathcal{L}(\rho(t))=\frac{d\rho(t)}{dt}=-\frac{i}{\hbar}\Big[H_{\text{sys}}, \rho(t)\Big ] +\mathcal{D}(t).
\end{equation}
Here  $\rho(t)$ refers to the reduced density matrix of the system, and $\mathcal{D}(t)$ represents the dissipative term that arises due to the influence of the environment at time $t$. The simplest form of this dissipative term $\mathcal{D}(t)$ is obtained when considering the weak coupling between the system and the bath, within the framework of the Born-Markov and secular approximations. In this scenario, Eq.~(\ref{dyn_equ}) simplifies to the Gorini-Kossakowski-Sudarshan-Lindblad (GKSL) master equation~\cite{Sudarshan_JMP_1976,Lindblad_CMP_1976,Petruccione_book,Alicki_2007,Rivas_Huelga_book,Lidar_2020_lecture}. These assumptions imply that information does not flow back from the environment to the system, thereby eliminating any memory effects. Since these assumptions confine the dynamics of the system to a highly specific scenario, and such situations are quite uncommon in nature, our work explores a more comprehensive quantum evolution of the system that takes into account the flow of information from the environment back to the system. We assume that initially the reservoir is in its thermal equilibrium state, $\rho_{\text{bath}}=\frac{e^{-\beta_{\text{hot}}H_{\text{bath}}}}{\text{Tr}\big(e^{-\beta_{\text{hot}}H_{\text{bath}}}\big)}$, where $\beta_{\text{hot}}$ is the inverse temperature of the hot bath defined as $\beta_{\text{hot}}=1/k_B T_{\text{hot}}$. Moreover, at the initial point of this stroke, the reservoir is uncorrelated with the system, i.e., the composite system-bath initial state can be written as $\rho_{SB}(0)=\rho_{\text{exp}}\otimes \rho_{\text{bath}}$. Taking these initial assumptions into account 
we can derive the master equation for open quantum dynamics in the form of Eq.~(\ref{dyn_equ}), with the dissipative term being~\cite{Tu_PRB_2008, Tu_QIP_2009, Xiong_PRA_2010, Wu_OE_2010, Jin_NJP_2010, Lei_PRA_2011, Lei_AP_2012, Nori_PRL_2012},
\begin{align}
\label{eq:dissi}
    \mathcal{D}(t)&= \sum_{\varepsilon \ge 0}2 \gamma_{\varepsilon}(t) \Big[ A(\varepsilon) \rho(t) A^{\dagger}(\varepsilon) -\frac{1}{2}\big\{ A^{\dagger}(\varepsilon) A(\varepsilon), \rho(t)\big\} 
    \Big] \nonumber \\
    &\phantom{amr mat}+ \widetilde{\gamma}_{\varepsilon}(t) \Big[ A^{\dagger}(\varepsilon) \rho (t) A (\varepsilon) 
     \mp  A(\varepsilon)\rho (t) A^{\dagger}(\varepsilon) \nonumber \\
     &\phantom{matha Kha}\pm A^{\dagger}(\varepsilon) A(\varepsilon) \rho(t) - \rho(t) A(\varepsilon) A^{\dagger}(\varepsilon) \Big]. 
\end{align}
Here the first part, associated with the coefficient $\gamma_{\varepsilon}(t)$, and the second part, associated with the coefficient $\widetilde{\gamma}_{\varepsilon}(t)$, provide non-unitary dissipation and fluctuation contributions, respectively, with their dynamics being linked through the intrinsic fluctuation-dissipation theorem~\cite{Weber_PR_1956, Kubo_1966, Felderhof_JOP_1978, Mucio_book_2021}. The symbols $\mp$ in the second term and $\pm$ in the third term of the fluctuation part are subsequently associated with fermions and bosons, respectively. 
The time-dependent dissipation coefficient $\gamma_{\varepsilon}(t)$ and the fluctuation coefficient $\widetilde{\gamma}_{\varepsilon}(t)$ can be formulated using the nonequilibrium Green's function of the system. See Refs.~\cite{Nori_PRL_2012, Lei_PRA_2011, Xiong_PRA_2010, Wu_OE_2010, Jin_NJP_2010, Tu_QIP_2009, Tu_PRB_2008, Lei_AP_2012} for a detailed discussion on this matter. The operators denoted as $A(\varepsilon)$ represent the Lindblad or jump operators associated with the transition energy $\varepsilon$. If we have two eigenvectors, $\ket{i}$ and $\ket{j}$, of the system Hamiltonian $H_{\text{sys}}$ corresponding to the eigenvalues $e_i$ and $e_j$, respectively, then the transition energy can be expressed as $\hbar \varepsilon = e_j - e_i$, and the Lindblad operators for the interaction Hamiltonian $H_{\text{int}}$ [Eq.~(\ref{Hamil})] can be expressed as
\begin{equation}
    A(\varepsilon)=\sum_{\varepsilon=\frac{1}{\hbar}(e_j-e_i)}\ket{i}\bra{i}\sigma_{x}\ket{j}\bra{j}.
\end{equation}
The master equation described in Eq.~(\ref{dyn_equ}) with the dissipative term given in Eq.~(\ref{eq:dissi}) can be reformulated in the conventional Lindblad form as
\begin{eqnarray}
\label{Equ:Master_Equ}
\mathcal{L}(\rho(t))
&=& -\frac{i}{\hbar} \Big[H_{\text{sys}}, \rho(t)\Big ] +  \sum_{\varepsilon \ge 0} \Gamma_{\varepsilon}(t) L_{A(\varepsilon), A^{\dagger}(\varepsilon)} \bigl[  \rho(t) \bigr]\nonumber\\  &&\phantom{dhur valo lagena}+ \widetilde{\gamma}_{\varepsilon}(t) L_{A^{\dagger}(\varepsilon), A(\varepsilon)} \bigl[  \rho(t) \bigr],
\end{eqnarray}
where $\Gamma_{\varepsilon}(t)=2\gamma_{\varepsilon}(t) \mp \widetilde{\gamma}_{\varepsilon}(t)$, with $-/+$ sign are associated with fermionic/bosonic bath, and 
\begin{equation}
    L_{A(\varepsilon), A^{\dagger}(\varepsilon)} \bigl[  \rho(t) \bigr] = A(\varepsilon) \rho (t) A^{\dagger}(\varepsilon) - \frac{1}{2}\big\{ A^{\dagger}(\varepsilon)A(\varepsilon), \rho (t)\big\}.
\end{equation}
If we now assume that the coupling between the system and the reservoir is weak and adopt a perturbative approach, considering terms only up to the second order of the coupling parameter $\zeta(\omega^{\prime})$, the dissipation and fluctuation coefficients are determined as respectively~\cite{Nori_PRL_2012,Xiong_PRA_2010},
\begin{align}
    \gamma_{\varepsilon}(t) &\approx \int_{0}^{t} ds \int \frac{d\omega^{\prime}}{2\pi} \mathcal{J}(\omega^{\prime}) \cos\Big[\big( \omega^{\prime} -\varepsilon \big) \big( t - s \big) \Big] \nonumber\\
    \widetilde{\gamma}_{\varepsilon}(t) &\approx 2 \int_{0}^{t} ds \int \frac{d\omega^{\prime}}{2\pi} \mathcal{J}(\omega^{\prime}) \overline{n}(\omega^{\prime}) \cos\Big[\big( \omega^{\prime} -\varepsilon \big) \big( t - s \big) \Big],
    \label{gamma}
\end{align}
where $\overline{n}(\omega^{\prime})=[\exp(\beta_{\text{hot}}(\hbar\omega^{\prime}-\mu_h))\pm 1]^{-1}$, with $\mu_h$ being the chemical potential of the hot reservoir and the sign $+ (-)$ corresponds to fermionic (bosonic) reservoir. For a fermionic reservoir the chemical potential $\mu_h>0$ and for a bosonic one $\mu_h<0$. Note that, in this stroke, as the reservoir exhibits an effective negative temperature, we specifically opt for a fermionic bath composed of two-level systems. This choice is essential because a bosonic bath is not suitable for describing a reservoir with an effective negative temperature~\cite{Gabriella_2023}. Throughout this paper, we will remain within the regime characterized by weak coupling between the system and the bath. Therefore, we confine our analysis to situations where both $\Gamma_{\varepsilon}(t)$ and $\widetilde{\gamma}_{\varepsilon}(t)$ are considerably smaller than the corresponding transition frequency $\varepsilon$. It is important to note that, we have omitted the effects of the frequency shift~\cite{Nori_PRL_2012, Lei_PRA_2011, Xiong_PRA_2010, Wu_OE_2010, Jin_NJP_2010, Tu_QIP_2009, Tu_PRB_2008, Lei_AP_2012} which arises in the first term of the quantum master equation presented in Eq.~(\ref{dyn_equ}) [as well as in Eq.~(\ref{Equ:Master_Equ})]. In the long-time Markov limit, Eq.~(\ref{Equ:Master_Equ}) simplifies to the GKSL quantum master equation, wherein the fluctuation and dissipation coefficients remain constant over time, as demonstrated in~\cite{Xiong_PRA_2010}.
In this work, we adopt a Lorentzian-type spectral density function for the hot reservoir, a choice commonly associated with the fermionic reservoirs~\cite{Tu_PRB_2008, Jin_NJP_2010, Lee_PRL_1993}. Specifically, we utilize the Lorentz-Drude form of the spectral density~\cite{Petruccione_book, Wei_PRA_2008, Denis_PRA_2020, Zhang_FP_2021}, which is described as
\begin{equation}
    \mathcal{J}(\omega^{\prime}) = \alpha \omega^{\prime} \frac{\omega^{2}_c}{\omega^{2}_c + \omega^{\prime^2}},
\end{equation}
where $\omega_c$ represents the spectral width of the bath, often referred to as the environment cutoff frequency, which is associated with memory effects. Additionally, the parameter $\alpha$ characterizes the strength of the interaction between the system and the reservoir.
\par
Now, let us move on to the next stroke, the  \textit{compression stroke}, 
which is essentially the reverse of the expansion stroke, wherein we achieve the compression of the energy gap. As a result, a time-reversed protocol of the expansion stroke is implemented, allowing us to express the driving Hamiltonian as $H_{\text{comp}}(t)=-H_{\text{exp}}(\tau - t)$. Similar to the expansion stroke, we can also assume it to be a unitary operation by the unitary $U^{\dagger}_{\tau,0}$. Hence, the compression stroke is performed on the output of the heating stroke, i.e., $\rho_{\text{heat}}$ and results in the final state $\rho_{\text{comp}} = U^{\dagger}_{\tau,0} \rho_{\text{heat}} U_{\tau,0}$. During this stroke, the Hamiltonian of the system changes from $\mathcal{H}_{\text{hot}}$ to $\mathcal{H}_{\text{cold}}$. Since the working substance is again isolated from the environment, there is no heat exchange involved. Instead, the change in the system's internal energy results from the work done during this stroke.
\par
Afterwards, we proceed to the \textit{cooling stroke} 
which marks the final step in completing one cycle of our designed Otto cycle framework.
In this phase, the working medium interacts with the cold reservoir, having the temperature $T_{\text{cold}}$. During this interaction, 
the Hamiltonian of the system is kept fixed at $\mathcal{H}_{\text{cold}}$, and
the evolution of the working medium follows the similar dynamics as the heating stroke. The only differences are that we will begin this stroke with the initial state of the working substance set as $\rho_{\text{comp}}$, and all the parameters related to the hot bath in Eqs.~(\ref{Equ:Master_Equ}) and~(\ref{gamma}) will be now replaced with the cold bath parameters. Also, in this case $H_{\text{sys}}=\mathcal{H}_{\text{cold}}$. In scenarios where we intend to investigate a perfect thermalization protocol, the system remains in contact with the cold reservoir until it achieves thermal equilibrium, effectively returning to its initial Gibbs state $\rho_{\text{in}}^{\text{th}}$. However, in cases involving imperfect thermalization or finite-time protocols, this stroke concludes before the working substance reaches its canonical equilibrium state corresponding to the temperature $T_{\text{cold}}$, and hence the cycle remains incomplete. In this study, we focus on these incomplete Otto cycles and specifically investigate the operation of an effective negative temperature-based transient Otto engine during the first cycle. In Sec.~\ref{sec:Reslt}, we will demonstrate that when assessing the efficiency of a quantum Otto engine for a fixed cycle, the heat exchanged in this cooling stroke is not involved in efficiency calculation, and there is no necessity for knowledge of the final state resulting from the cooling stroke. The final state is required only to initiate a new cycle, and this state is associated with the efficiency calculation of the next cycle. Hence, the final cooling stroke has no impact on the efficiency of the engine during the first cycle, and given our focus on the first cycle exclusively, we need not concern ourselves with the output state of the cooling stroke. Moreover, in this work, as we consider $T_{\text{cold}}$ as a positive temperature, we have the flexibility to employ both fermionic and bosonic baths during this stroke. Nevertheless, we choose to utilize a fermionic reservoir as the cold bath.

Let us now introduce the concept of effective negative temperature for a fermionic reservoir by defining the local inverse temperature in terms of population as
\begin{equation}
\label{Equ:relation_beta_population}
    \beta_{\text{cold(hot)}} = \frac{1}{2\pi\hbar\sqrt{\nu^{2}_{\text{cold(hot)}}+ (\frac{\Tilde{\omega}}{2\pi})^2}} \ln \left ( \frac{1- p^{+}_{\text{cold(hot)}}}{p^{+}_{\text{cold(hot)}}}     \right ),
\end{equation}
with
$p^{+}_{\text{cold(hot)}}=\bra{\Psi^{\text{cold(hot)}}_{+}}\rho^{\text{th}}_{\text{in}}(\rho_{\text{heat}}^{\text{th}})\ket{\Psi^{\text{cold(hot)}}_{+}}$, where $\rho_{\text{heat}}^{\text{th}}=e^{-\beta_{\text{hot}}(\mathcal{H}_{\text{hot}} -\mu_{h} \hat{N}_{h})}/\mathcal{Z}_{\text{hot}}$. Here $\mathcal{Z}_{\text{hot}}= \Tr[e^{-\beta_{\text{hot}}( \mathcal{H}_{\text{hot}}-\mu_h \hat{N}_h)}]$.
When the population of higher energy levels exceeds that of lower energy levels, the effective temperature becomes negative. From Eq. (\ref{Equ:relation_beta_population}), we can see that $p^{+}_{\text{cold(hot)}} \in [0, 0.5)$ results in a positive $\beta_{\text{cold(hot)}}$, while $p^{+}_{\text{cold(hot)}} \in [0.5, 1.0)$ yields a negative $\beta_{\text{cold(hot)}}$. 
We are interested in studying our designed Otto engine protocol when 
the cold reservoir is maintained at a positive temperature and the hot reservoir is set at an effective negative temperature. 
Therefore, in our analysis, we vary $p^{+}_{\text{hot}}$ within the range corresponding to effective negative temperatures, specifically from $0.5$ to $1.0$, while keeping $p^{+}_{\text{cold}}$ fixed at $0.261$. 
To ensure direct comparability with the results of the earlier studies~\cite{Assis_PRL_2019,Arghya_arxiv_2023} and maintain consistency across parameter values, we have taken $p^{+}_{\text{cold}}=0.261$, which aligns with experimental realization~\cite{Assis_PRL_2019}. However, it is essential to note that any value of $p^{+}_{\text{cold}} \in [0, 0.5)$ is theoretically valid.

\section{Information backflow and canonical measure of non-Markovianity}
\label{Sec:non_marko}
In a Markovian scenario, where the open quantum dynamics of a system adhere to the Born-Markov and secular approximations, it is assumed that the system-environment interaction is weak and the bath correlation functions quickly dissipate, leading to a loss of memory about the past states of the system. On the other hand, non-Markovian dynamics presents a more complex and realistic description of open quantum systems, allowing for the backflow of information from the environment to the system. In these scenarios, memory effects are considered, and the evolution of the system can depend on its past states, leading to intricate and often nonlocal dynamics. In general, non-Markovian effects become prominent when the system-environment coupling is strong.
However, non-Markovianity is not solely confined to strong system-environment coupling scenarios; it can also manifest itself in systems characterized by weak coupling with their environment. In weak coupling regimes, non-Markovian effects can emerge 
when the environment is structured in a way that allows for long-lasting correlations with the system.
In such cases, the presence of negative decoherence rates is a strong indicator of non-Markovian behavior. These negative rates signal that quantum processes can exhibit ``backward" evolution, suggesting that the system's future states are influenced by its past dynamics. This departure from the conventional assumption of memoryless evolution underscores the significance of accounting for non-Markovianity in the quantum description of weakly coupled systems, where the interplay between memory effects and negative decoherence rates can lead to unique and potentially exploitable quantum phenomena. In this study, our primary focus is on examining the consequences of non-Markovianity, stemming from the backflow of information from the environment to the system, on the performance of an effective negative temperature-based quantum Otto engine. While there exist several methods for detecting information backflow and the resulting non-Markovian behavior in system dynamics~\cite{Wolf_PRL_2008,  Breuer_PRL_2009, Rivas_PRL_2010, Lu_PRA_2010, Lorenzo_PRA_2011, Luo_PRA_2012, Zhong_PRA_2013, Liu_PRA_2013, Lorenzo_PRA_2013, Rivas_RPP_2014, Debarba_PRA_2017, Strasberg_PRL_2018, Huang_PRA_2021, Ahana_arxiv_2022}, our approach in this work focuses on identifying non-Markovian behavior through the observation of negative decoherence rates.

As outlined in the preceding section, the quantum master equation introduced in Eq.~(\ref{Equ:Master_Equ}) portrays the open quantum dynamics of a system, assuming a state of weak coupling between the system and its environment. The key distinction from the conventional GKSL master equation lies in the fact that, in this equation, the decoherence rates $\Gamma_{\varepsilon}(t)$ and $\widetilde{\gamma}_{\varepsilon}(t)$ are time-dependent, whereas for the Markovian GKSL equation, these rates are time independent.  Additionally, in the GKSL quantum master equation, the decoherence rates are always positive. However, it has been demonstrated in previous literature that under specific circumstances, either $\Gamma_{\varepsilon}(t)$ or $\widetilde{\gamma}_{\varepsilon}(t)$, or even both, can take negative values~\cite{Breuer1999,Maniscalco_2004,Andersson2007,Piilo2008,Anderson_PRA_2014,Siudzinska_2020}.
This observation serves as a clear indication of the presence of memory effects within the dynamics of the system, leading to the intriguing phenomenon of information backflow from the environment to the system. To explore the impact of non-Markovianity on the operation of a negative temperature-based transient quantum Otto engine, let us begin by determining the parameter regimes, characterized by $\omega_c$, in which we can detect evidence of non-Markovianity. 
\begin{figure*}
\includegraphics[
width=8.5cm, height=5.8cm ]{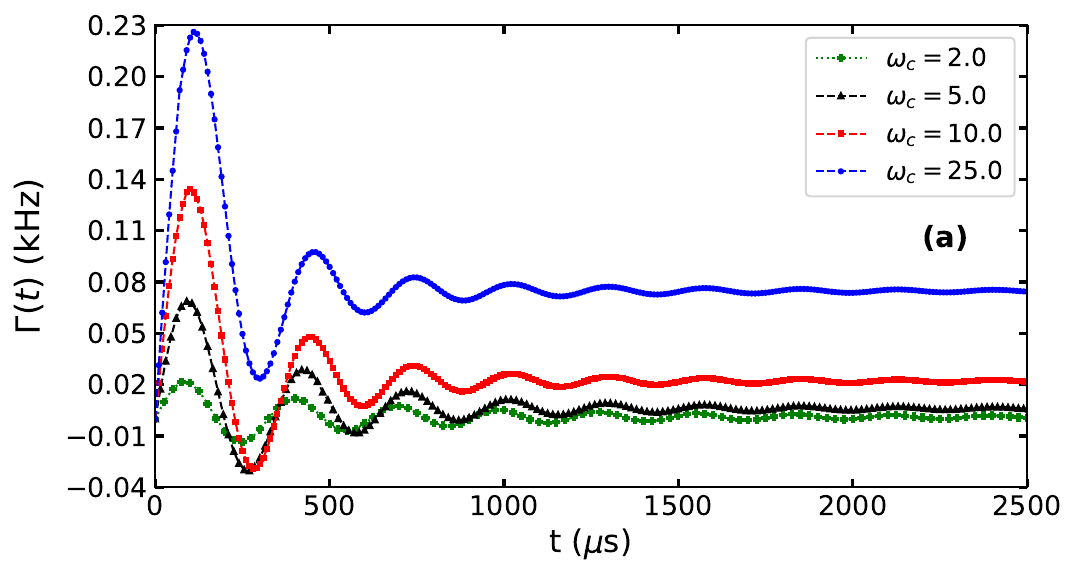}
\includegraphics[
width=8.5cm, height=5.8cm]{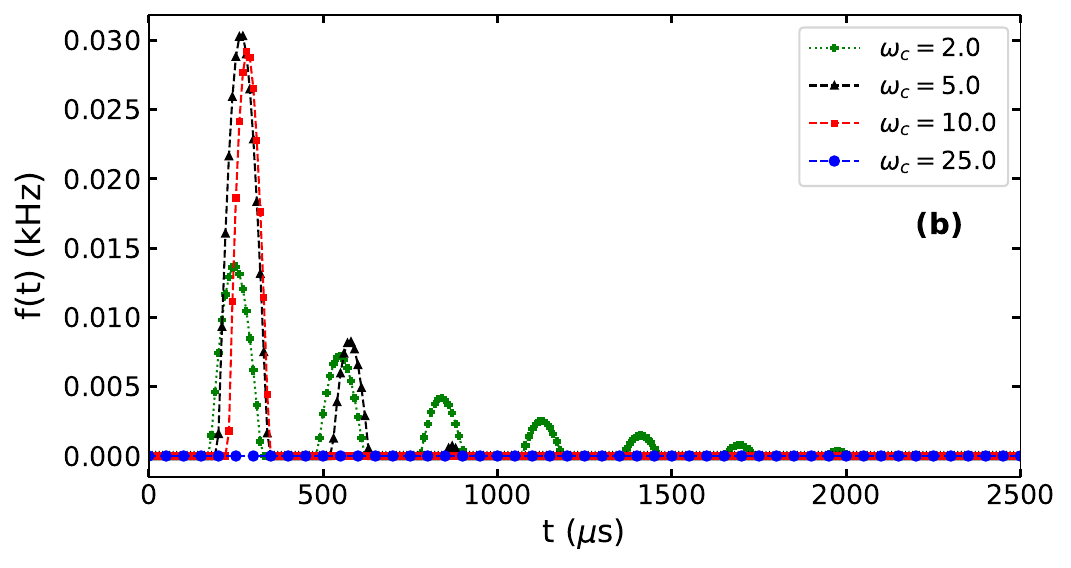}
\includegraphics[width=8.5cm,height=6.5cm]{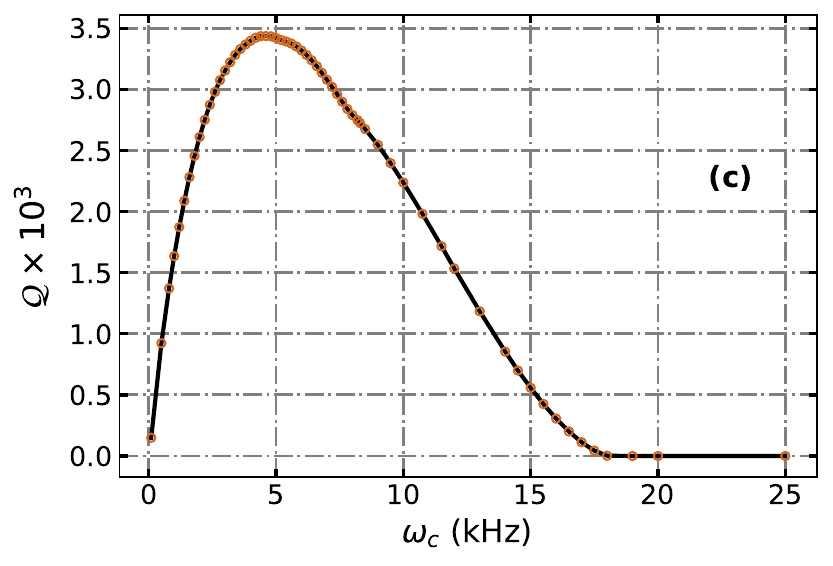}
\caption{Evidence of non-Markovianity in the open quantum evolution of the working substance during the heating stroke. Here in panel (a) we demonstrate the nature of the
decoherence rate $\Gamma(t)$ (in kHz) with time $t$ (in $\mu$s) 
for different cutoff frequencies $\omega_c$ (in kHz). Panel (b) displays the time dynamics of the witness of non-Markovianity $f(t)$ (in kHz) for different $\omega_c$, and panel (c) presents the variation of the quantifier of non-Markovianity $\mathcal{Q}$ with the change in $\omega_c$. 
Here we take $\nu_{\text{cold}} = 2.0$ kHz and $\nu_{\text{hot}} = 3.6 $ kHz. The driving time of both the expansion and compression stroke is taken as $\tau =100\; \mu$s. The strength of the \(z\)-directional magnetic field is set as $g =0.2$ and $p^{+}_{hot}=0.99$.
The coupling strength in the spectral density function, $\alpha =0.6$. The units of the parameters are written in the corresponding labels of $x$ and $y$ axes. The quantifier $\mathcal{Q}$, plotted here is a dimensionless quantity.
}
\label{Fig:Gamma_witness_quantifier}
\end{figure*}

In the heating stroke of the Otto engine setup, as discussed in Sec.~\ref{sec:Protocol}, the Hamiltonian of the system remains fixed at $\mathcal{H}_{\text{hot}}$. The two eigenvalues of $\mathcal{H}_{\text{hot}}$ are denoted as $e_{\pm}=\pm \pi\hbar\sqrt{\nu^{2}_{\text{hot}}+ (\frac{\Tilde{\omega}}{2\pi})^2}$. So, we have only two possible values of $\varepsilon$, for $\varepsilon\ge 0$: $\varepsilon=0$ and
$\varepsilon=e_+-e_-$. In this specific scenario, the transition for $\varepsilon=0$ is not allowed and hence 
we have only $\varepsilon=e_+-e_-$. For ease of notation, in the further discussions of this paper we will utilize $\Gamma_{\varepsilon}(t)=\Gamma(t)$, $\gamma_{\varepsilon}(t)=\gamma(t)$, and $\widetilde{\gamma}_{\varepsilon}(t)=\widetilde{\gamma}(t)$.
In Fig.~\ref{Fig:Gamma_witness_quantifier}-(a), the time dynamics of the decoherence rate $\Gamma(t)$ during the transient regime of the heating stroke is depicted, by varying the cutoff frequency $\omega_c$ of the hot reservoir. We observe that for a higher value of $\omega_c$, specifically when $\omega_c=25$ kHz, $\Gamma(t)$ remains positive throughout the entire duration of the evolution. However, as we reduce the value of $\omega_c$, a time interval emerges during which $\Gamma(t)$ takes on negative values. Initially, as $\omega_c$ decreases, the magnitude of these negative values increases. However, beyond a certain threshold of $\omega_c$, the magnitude of the negative values of $\Gamma(t)$ once again diminishes. See the curve, representing the situation where $\omega_c=2$ kHz, of Fig.~\ref{Fig:Gamma_witness_quantifier}-(a). Conversely, the other decoherence rate $\widetilde{\gamma}(t)$, defined in Eq.~(\ref{gamma}),  maintains a positive value throughout the entire evolution period. The presence of negative values in $\Gamma(t)$ provides an evidence of information backflow from the environment to the system. As mentioned earlier, given that the cooling stroke does not affect the efficiency of the first cycle of a quantum Otto engine we do not delve into the examination of information backflow during the cooling stroke. Note that in the heating stroke, we have employed a fermionic bath as the environment, and conventionally, it should have a chemical potential $\mu_h$ greater than zero. However, for the sake of convenience, we have opted to set $\mu_h$ equal to zero throughout this paper. We have conducted tests and confirmed that for values of $\mu_h$ within the range of $0$ to $\varepsilon$, where $\varepsilon=e_+-e_-$, all the quantities studied in this paper exhibit qualitatively similar behavior. Therefore, selecting $\mu_h=0$ does not introduce any significant deviations or issues in our analysis. Similarly, we also set $\mu_c=0$ in the initial state $\rho_{\text{in}}^{\text{th}}$. 

We now try to quantify the degree of information backflow 
by employing a suitable measure designed for this purpose. Several distinct measures of non-Markovianity have been proposed in the literature~\cite{Rivas_RPP_2014, Wolf_PRL_2008,  Breuer_PRL_2009, Rivas_PRL_2010, Lu_PRA_2010, Lorenzo_PRA_2011, Luo_PRA_2012, Zhong_PRA_2013, Liu_PRA_2013, Lorenzo_PRA_2013, Debarba_PRA_2017, Strasberg_PRL_2018, Huang_PRA_2021, Ahana_arxiv_2022}, each capturing, different aspects of non-Markovianity. Note that most of these measures do not necessarily agree with each another.
In this work, we use a witness and quantifier of non-Markovianity that relies on the presence of negative decoherence rates associated with the Lindblad-type master equation given in Eq.~(\ref{Equ:Master_Equ}).
A local-in-time master equation, applicable to a quantum system in a $d$-dimensional Hilbert space, can be expressed in a canonical form as~\cite{Anderson_PRA_2014}
\begin{align}
\label{Equ:canonical}
&\dot{\rho}(t)=-\frac{i}{\hbar}[H(t),\rho(t)] \nonumber\\
&+ \sum_{k=1}^{d^2-1} \Lambda_{k}(t) \Big[\mathcal{A}_{k}(t)\rho(t)\mathcal{A}_k^{\dagger}(t)-\frac{1}{2}\big\{\mathcal{A}_k^{\dagger}(t)\mathcal{A}_{k}(t),\rho(t)\big\}\Big].
\end{align}
The  jump operators $\mathcal{A}_{k}(t)$ are subject to the following conditions:
$\Tr[\mathcal{A}_{k}(t)] = 0$ and $\Tr[\mathcal{A}^{\dagger}_{j}(t) \mathcal{A}_{k}(t)] = \delta_{jk}$. Non-Markovianity in the time evolution of a system is identified when there exists at least one canonical decoherence rate $\Lambda_{k}(t)$ with a strictly negative value. Hence, a function depending on the occurrence of negative decoherence rates at time $t$ can be defined as
\begin{equation}
    f_{k}(t) := \max \big[ 0, - \Lambda_{k}(t) \big] \geq 0.
    \label{eq:12}
\end{equation}
This function characterizes the presence of non-Markovian behavior within individual decoherence channels. We can also introduce a function as
\begin{equation}
    f(t) = \sum_{k=1}^{d^2 -1 } f_{k} (t).
    \label{eq:13}
\end{equation}
The function $f(t)$ will take a value greater than 0 if there is at least one decoherence rate with negative values for any $k$; otherwise, it remains at $0$. Hence, we can use $f(t)$ as a witness of non-Markovianity.
Consequently, the total amount of non-Markovianity, or the quantifier of non-Markovianity, can be defined over a time interval $[t_0 , t^{\prime}]$ as,
\begin{equation}
    \mathcal{Q} (t_0, t^{\prime}) =  \int_{t_0}^{t^{\prime}} ds ~ f(s).
    \end{equation}
If the quantifier, $\mathcal{Q} (t_0, t^{\prime})$ is greater than $0$, it signifies that the evolution of the system exhibits non-Markovian behavior. If we now compare Eq.~(\ref{Equ:Master_Equ}) with the canonical form of master equations given in Eq.~(\ref{Equ:canonical}), we can see that Eq.~(\ref{Equ:Master_Equ}) is already written in the canonical form with the decoherence rates $\Lambda_1(t)=2\gamma_0(t)$ for $\varepsilon=0$, and  $\Lambda_2(t)=\Gamma(t)$, $\Lambda_3(t)=\widetilde{\gamma}(t)$ for $\varepsilon=e_+-e_-$, for the heating stroke discussed above. As $\varepsilon=0$ does not affect the evolution of the system, to evaluate the quantifier and witness of non-Markovianity, we only have $k=2$ and $3$ in Eqs.~(\ref{Equ:canonical}),~(\ref{eq:12}), and~(\ref{eq:13}).
The behavior of the witness of non-Markovianity $f(t)$ for the heating stroke is demonstrated in 
Fig.~\ref{Fig:Gamma_witness_quantifier}-(b). This reveals that when $\omega_c$ is set at a higher value, say at $25$ kHz, the witness of non-Markovianity does not exhibit any positive values, whereas for certain lower values of  $\omega_c$, the witness yields positive values, indicating the presence of non-Markovian characteristics. Notably, this pattern aligns with the trend illustrated in Fig.~\ref{Fig:Gamma_witness_quantifier}-(a). By examining 
Figs.~\ref{Fig:Gamma_witness_quantifier}-(a) and~\ref{Fig:Gamma_witness_quantifier}-(b), it becomes apparent that the negative values of $\Gamma(t)$ and the maximum of $f(t)$ increase as $\omega_c$ decreases. However, beyond a certain threshold value of $\omega_c$, these values decrease. In order to comprehend the relationship between the non-Markovian behavior of the system and the cutoff frequency $\omega_c$, we delve into an investigation of how the degree of non-Markovianity, $\mathcal{Q}(t_0,t^{\prime})$, evolves in response to changes in $\omega_c$ in Fig.~\ref{Fig:Gamma_witness_quantifier}-(c). Here we take $t_0=0$ and $t^{\prime}=10^4$. 
The study indicates that for the cutoff frequency range $\omega_c \gtrapprox 18$ kHz the system does not show any non-Markovian nature, i.e.,  $\mathcal{Q} =0$.
Later, when the cutoff frequency $\omega_c \lessapprox 18 $ kHz, non-Markovianity appears with $\mathcal{Q} > 0$ and the quantifier of non-Markovianity ($\mathcal{Q}$) enhances with further reductions in the cutoff frequency, and this nature of $\mathcal{Q}$ is consistent in the range of cutoff frequency $18$ kHz $\gtrapprox \omega_c \gtrapprox 4$ kHz. 
Notably, this increase is not consistently monotonic; rather, below a critical value of $\omega_c \lessapprox 4 $ kHz, the degree of non-Markovianity, $\mathcal{Q}$ starts to decrease.

\section{
Performance of effective negative temperature-based transient quantum Otto engine}
\label{sec:Reslt}
In this section, we will study the impact of information backflow from the environment to the system on the efficiency of a negative temperature-based transient quantum Otto engine.
In this context, we first compute the engine's efficiency ($\eta$) as a function of the degree of non-Markovianity ($\mathcal{Q}$). Subsequently, by examining the overall trend of the efficiency graph with the time duration of heating stroke, we introduce another quantity, the \textit{overall performance of an effective negative temperature-based quantum Otto engine}. This quantity not only accounts for the maximum efficiency achieved but also assesses the engine's ability to sustain a substantial amount of efficiency over time.

As previously mentioned, in each stroke of the quantum Otto engine, there is either an exchange of heat or work, but never both simultaneously. During the expansion phase ($\rho^{\text{th}}_{\text{in}} \rightarrow \rho_{\text{exp}} $) and the compression phase ($\rho_{\text{heat}} \rightarrow \rho_{\text{comp}}$), the change in internal energy occurs as work is done. In contrast, during the heat exchange processes between the working medium and the reservoirs in the heating and cooling strokes, energy transfer takes place in the form of heat. The efficiency of a quantum Otto engine is defined as the ratio of the net work to the absorbed heat. So, in the context of this paper, the work output results from the expansion stroke as $\langle W_1\rangle=\Tr[\rho_{\text{exp}} \mathcal{H}_{\text{hot}}] - \Tr[\rho^{\text{th}}_{\text{in}} \mathcal{H}_{\text{cold}}]$, and from the compression stroke as $\langle W_2\rangle=\Tr[\rho_{\text{comp}} \mathcal{H}_{\text{cold}}] - \Tr[\rho_{\text{heat}} \mathcal{H}_{\text{hot}}]$, with the total work output as 
\begin{equation}
    \langle W\rangle=\langle W_1\rangle +\langle W_2 \rangle.
\end{equation}
The energy input is given by the heat intake in the heating stroke, which is
\begin{equation}
    \langle Q_{\text{hot}} \rangle = \Tr[\rho_{\text{heat}} \mathcal{H}_{\text{hot}}] - \Tr[\rho_{\text{exp}} \mathcal{H}_{\text{hot}}]. 
\end{equation}
Therefore, the efficiency of a negative temperature-based transient quantum Otto engine is given by
\begin{equation}
    \eta = -  \langle W \rangle / \langle Q_{\text{hot}} \rangle .
\end{equation}
In the scope of our definition for work and heat, within the valid operational range of a quantum Otto engine $\langle W \rangle <0$ and $\langle Q_{\text{hot}}\rangle >0$. Hence, a minus sign is incorporated into the definition of the engine's efficiency to account for this convention.

\begin{figure}
    \centering
    \includegraphics[width=8.5cm, height=5.8cm ]{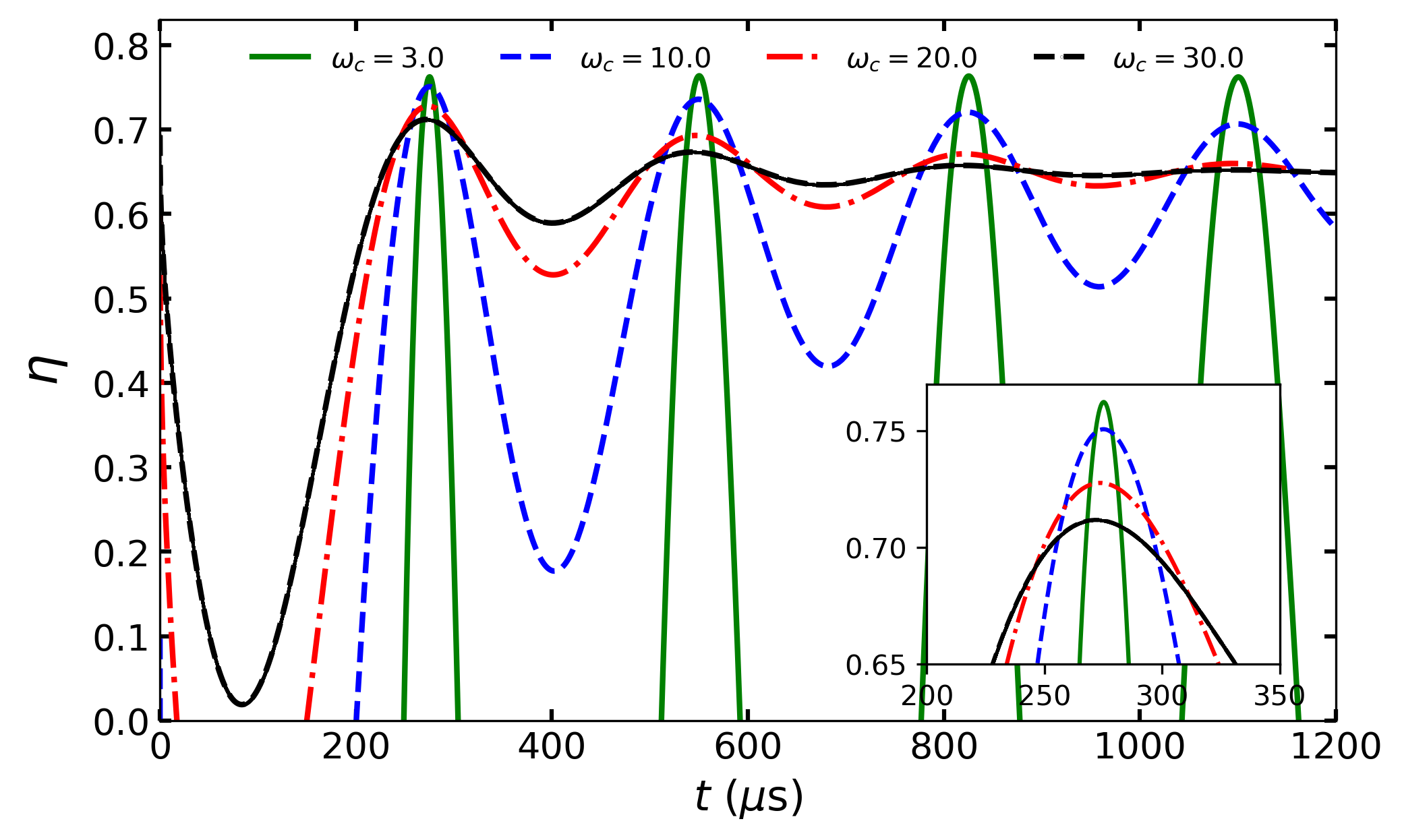}
    \caption{Efficiency of an effective negative temperature-based transient quantum Otto engine. Here we plot the efficiency $\eta$ vs. the evolution time $t$ of the system during the heating stroke of the engine
    for different values of $\omega_c$. A magnified version of the efficiency for $t=200$ to $350 \mu$s is presented in the inset. All the parameters are same as in Fig.~\ref{Fig:Gamma_witness_quantifier}. Here $t$ is in the unit of $\mu$s and $\eta$ is a dimensionless quantity.}
    \label{Fig:Efficiency}
\end{figure}

\begin{figure*}
\includegraphics[width=8.5cm, height=5.8cm ]{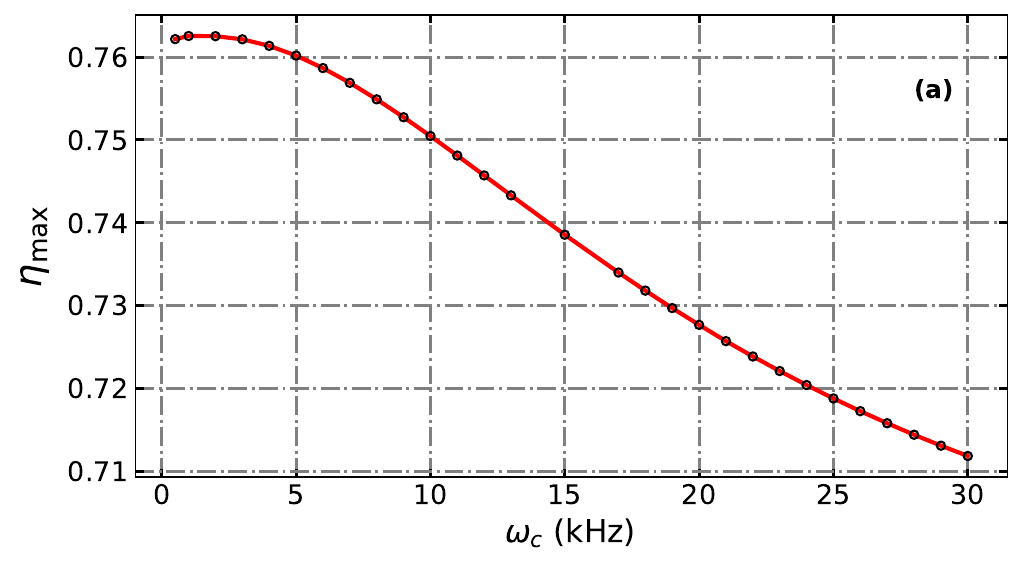}
\includegraphics[width=8.5cm, height=5.8cm ]{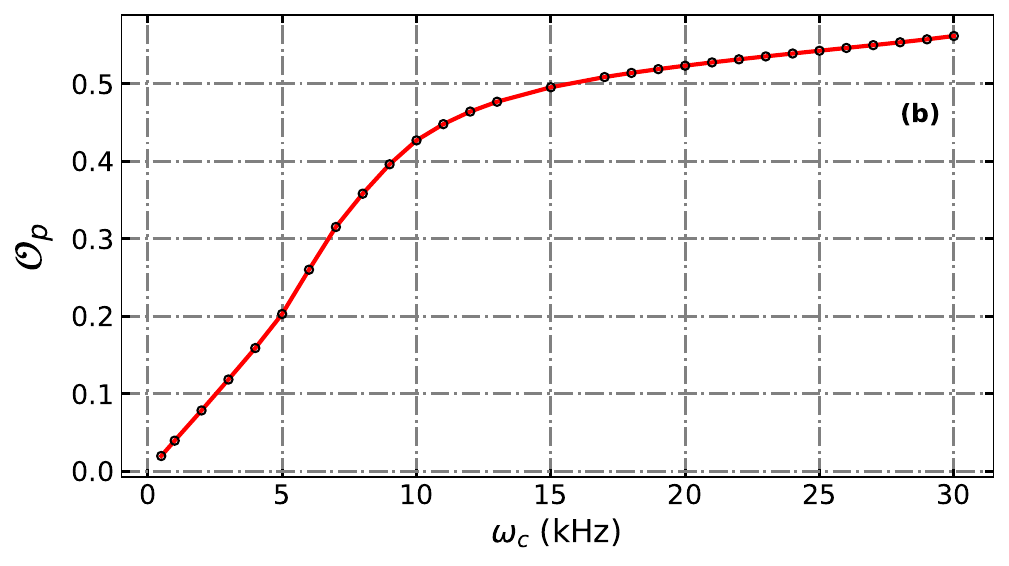}
\caption{Dependence of the maximum efficiency and the overall performance of an effective negative temperature-based quantum Otto engine on the non-Markovianity. 
Here we plot the maximum efficiency $\eta_{\text{max}}$ by varying $\omega_c$ in panel (a), and the overall performance of the engine $\mathcal{O}_p$ with the same in panel (b). For panel (b) we take $t_{f}=10^3 \mu$s.
All other considerations are same as in Fig.~\ref{Fig:Gamma_witness_quantifier}. The $\omega_c$ plotted along the $x$ axes of the both the panels are in kHz, and the quantities depicted along the $y$ axes are dimensionless.
}
\label{Fig:Eff_OPE_eta-max}
\end{figure*}

We now investigate the behavior of efficiency of the engine with respect to the finite time duration of the heating stroke.
Also, we examine how the maximum efficiency obtained at a fixed time during the heating stroke changes with variations in the cutoff frequency ($\omega_c$) of the hot reservoir.  From the depiction of Fig.~\ref{Fig:Efficiency}, we observe that the efficiency $\eta$ exhibits oscillations with respect to the evolution time of the system during the heating stroke, 
and eventually, it reaches a constant value $\eta_{\text{sat}}$ after a sufficiently long time $t_{\text{eq}}$.
For all values of $\omega_c$ the oscillating nature of the efficiency with $t$ is qualitatively the same, but as we approach the lower cutoff frequency, 
the fluctuations in efficiency become more pronounced, and it takes a longer time for the system to reach the stable value ($\eta_{\text{sat}}$). 
Note that for all $\omega_c$, the efficiency curves eventually converge to the same value of $\eta_{\text{sat}}$ and this stabilized final efficiency precisely matches the efficiency achieved for an infinite-time protocol, $\eta_{\text{INF}}=0.649$, as reported in Ref.~\cite{Arghya_arxiv_2023}.
Furthermore, it is worth noting that there are specific time intervals during which a negative temperature-based transient quantum Otto engine can achieve a higher efficiency than the one operating under the infinite-time protocol, i.e., for certain values of $t$, $\eta$ can be greater than $\eta_{\text{INF}}$.
For instance, with $\omega_c = 30$ kHz, the highest efficiency achieved is $\eta_{\text{max}} = 0.712$. This maximum efficiency, $\eta_{\text{max}}$, is attained within the time interval $t \in [271.71, 272.08] \mu$s. Following that, a second-highest peak of efficiency, $\eta=0.6733$, can be achieved within the time interval $t \in [543.62, 544.29] \mu$s, and subsequently, after a series of peaks, the efficiency gradually stabilizes at $\eta_{\text{sat}} = 0.649$. Hence, by selecting an appropriate termination time of the heating stroke, denoted as $\tilde{t}$, within the interval $[271.71, 272.08] \mu$s, we can attain the maximum efficiency associated with $\omega_c=30$ kHz. This demonstrates that a negative temperature-based transient quantum Otto engine can achieve a higher efficiency than $\eta_{\text{INF}}$ by strategically terminating the heating stroke within a specific time frame. Therefore, such an engine can offer advantages over conventional engines operating under infinite-time protocols. For the other values of $\omega_c$ considered in Fig.~\ref{Fig:Efficiency}, $\eta_{\text{max}}$ is higher, but the time window in which we achieve the maximum efficiency $\eta_{\text{max}}$ decreases as $\omega_c$ decreases. 


Let us now turn our attention to the behavior of individual efficiency curves in relation to information backflow from the environment to the system. 
Building upon our earlier examination of non-Markovianity (as illustrated in Fig.~\ref{Fig:Gamma_witness_quantifier}), it has been ascertained that non-Markovianity increases as $\omega_c$ decreases up to a threshold value. However, with a further decrease in $\omega_c$, non-Markovianity begins to decrease.
So, there might be a relation between the efficiency of an effective negative temperature-based transient quantum Otto engine and the non-Markovianity. 
To delve into this relation we first divide our study of efficiency into two aspects.\\
\\
(i) The first one is the maximum efficiency $\eta_{\text{max}}$, achieved by an effective negative temperature-based quantum Otto engine operating with a finite-time protocol.\\
\\
(ii) The second one is the overall performance of the engine throughout a sufficiently extensive transient period of the heating stroke.\\ 

We present the behavior of the first figure of merit $\eta_{\text{max}}$,
by varying the cutoff frequency of the hot reservoir in Fig.~\ref{Fig:Eff_OPE_eta-max}-(a). It is evident from the depiction that the maximum efficiency of the engine, $\eta_{\text{max}}$, increases as $\omega_c$ decreases. However, the rate of this increase differs between the non-Markovian and Markovian regimes. In the Markovian regime, for $\omega_c \gtrapprox 18$ kHz, where the quantifier of non-Markovianity remains at $0$, the increase in $\eta_{\text{max}}$ is less pronounced compared to the non-Markovian regime within the range $5$ kHz $\lessapprox \omega_c \lessapprox 18$ kHz. As $\omega_c$ drops to $\lessapprox 5$ kHz, $\eta_{\text{max}}$ starts to saturate, and finally, for a very small value of $\omega_c$ ($\omega_c=0.5$ kHz), $\eta_{\text{max}}$ tends to decrease. 
On the other hand, the quantifier of non-Markovianity, $\mathcal{Q}$, decreases abruptly in this specific regime. See Fig.~\ref{Fig:Gamma_witness_quantifier}-(c). 

 

To investigate the second figure of merit, let us first highlight an interesting aspect of the efficiency curves for different values of $\omega_c$. From Fig.~\ref{Fig:Efficiency}, we can see that 
$\eta_{\text{max}}$ increases with decreasing $\omega_c$, and the influence of non-Markovianity aids in achieving a higher maximum efficiency.  However, it is important to notice that although the maximum efficiency is higher for the non-Markovian evolution of the system, the operational range of the engine, where a positive efficiency is attained ($0 \le \eta <1$), is more extensive in the Markovian scenarios for higher values of $\omega_c$. Moreover, as non-Markovianity decreases (corresponding to an increase in $\omega_c$), this operational range expands. To quantify the operational range of an effective negative temperature-based quantum Otto engine throughout the finite-time duration of the heating stroke, we introduce the \textit{overall performance of an effective negative temperature-based quantum Otto engine} as
\begin{equation}
    \mathcal{O}_p=\frac{1}{t_{f}}\int_0^{t_{f}} \eta(t) dt.
\end{equation}
Here, 
$t_{f}$ can be any time between 
$0<t_{f}\le t_{\text{eq}}$.
We now illustrate the behavior of the overall performance $\mathcal{O}_p$ with the change of the cutoff frequency of the hot reservoir depicted in Fig.~\ref{Fig:Eff_OPE_eta-max}-(b). 
The overall performance of the engine clearly diminishes as the cutoff frequency $\omega_c$ decreases, signifying a reduction in performance as the system transitions from the Markovian to the non-Markovian regime. Notably, the rate of this performance decline differs between the Markovian and non-Markovian regimes. In the Markovian range ($\omega_c \gtrapprox 18$ kHz), the rate of decrease in $\mathcal{O}_p$ is relatively gradual. However, as the system shifts from the Markovian to the non-Markovian regime (for $\omega_c \lessapprox 18$ kHz), the rate of decrease in $\mathcal{O}_p$ accelerates. Once again, for small values of $\omega_c$ within the range of $5$ kHz to $0.5$ kHz, the rate of decline in $\mathcal{O}_p$ diminishes.
In contrast, the maximum attainable efficiency $\eta_{\text{max}}$, depicted in Fig.~\ref{Fig:Eff_OPE_eta-max}-(b), exhibits an inverse behavior compared to $\mathcal{O}_p$, as discussed earlier. Thus, it can be concluded that if we define our figure of merit as the overall performance of the engine, then the engine operating within the Markovian range outperforms its non-Markovian counterparts. The reason behind this is that, in the non-Markovian regime, we achieve higher efficiencies but for less time, and as non-Markovianity increases this phenomenon gets prominent. This results in an overall reduction in the engine's performance with an increase in non-Markovianity, as the operational window during which the model functions as an engine decreases. So, to comprehensively assess the engine's behavior, we employ two figures of merit: the maximum achievable efficiency at a specific finite time during the isochoric heating stroke, and the overall performance of the engine over an extended period during the transient phase of this stroke. This dual metric approach is adopted not only for completeness but also to emphasize that achieving higher efficiency at some particular instances of time is insufficient for realizing the full capability of an engine.

 \section{Effective negative temperature-based necessarily transient quantum Otto engine}
\label{sec:Nec_Trans}
Up until this point, our examination has revolved around the performance of an effective negative temperature-based quantum Otto engine depending on the finite-time duration of the heating stroke, while maintaining fixed populations $p^{+}_{\text{hot}}=0.99$ and $p^{+}_{\text{cold}}=0.261$. Our findings have revealed that the maximum efficiency of the engine improves with an augmented influx of information from the environment to the system. Additionally, the overall performance of these engines, as defined previously, increases as the system's behavior approaches a Markovian regime. In this section, we will investigate another significant advantage linked to our finite-time or transient quantum engine concept by posing a question of whether there exists a scenario within this effective negative temperature-based quantum Otto engine where the transient heating or cooling stroke becomes the ``sole" viable option for the engine's operation. 

To answer this question, we investigate the efficiency of an effective negative temperature-based transient quantum Otto engine at a fixed duration of the heating stroke $\tilde{t}$, by varying the effective negative temperature of the hot bath, defined by the population $p^{+}_{\text{hot}}$. As previously discussed, when the values of $p^{+}_{\text{cold(hot)}}$ fall within the range of $[0.5, 1.0)$, $\beta_{\text{cold(hot)}}$ takes negative values. Therefore, in our analysis, we vary $p^{+}_{\text{hot}}$ from $0.5$ to $0.99$ while maintaining $p^{+}_{\text{cold}}$ fixed at a value of $0.261$.
The duration of the expansion and compression strokes is fixed at $\tau=100 \mu$s. Furthermore, for this investigation, we determine the duration of the heating stroke as the time at which we reach the maximum efficiency ($\eta_{\text{max}}$) for a specific cutoff frequency ($\omega_c$). This specific time is denoted as $\tilde{t}_{\text{max}}$, and these values for different $\omega_c$ have already been obtained in the analysis presented in Fig.~\ref{Fig:Efficiency}.
\begin{figure}
\includegraphics[width=8.5cm, height=5.8cm ]{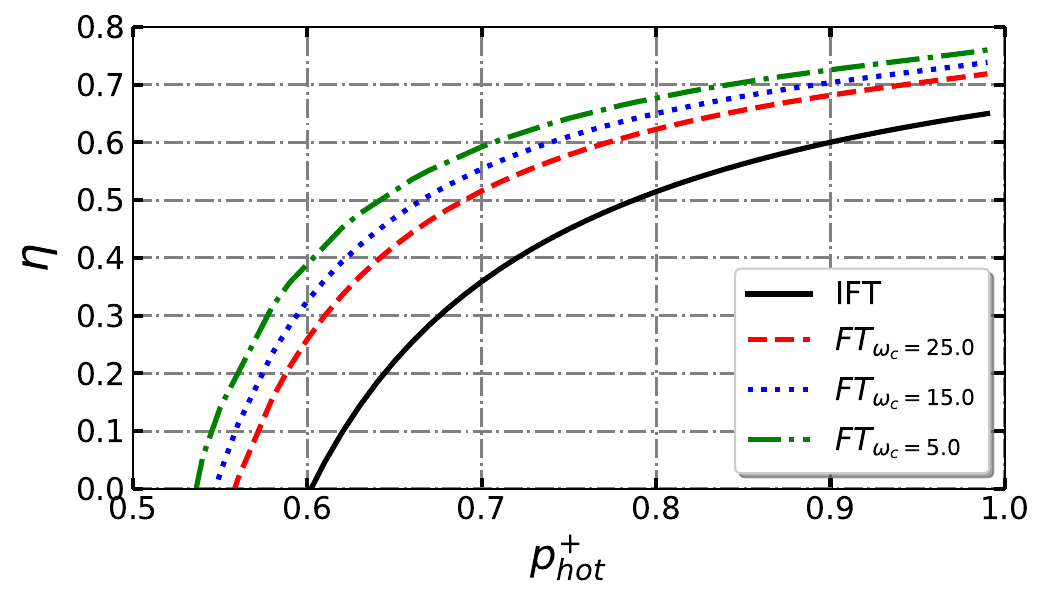}
\caption{Efficiency of an effective negative temperature-based necessarily transient quantum Otto engine. Here we depict the efficiency $\eta$ vs. the population denoted as $p^{+}_{\text{hot}}$. The solid black curve presents the efficiency of an effective negative temperature-based quantum Otto engine which operates in the equilibrium regime of the heating stroke (IFT). The efficiencies for the same operating in the transient regime of the heating stroke are plotted by colored lines for different values of $\omega_c$. For $\omega_c=25$ kHz, $\tilde{t}_{\text{max}}=272.7 \mu$s, for $\omega_c=15$ kHz, $\tilde{t}_{\text{max}}=274.55 \mu$s, and for $\omega_c=5$ kHz, $\tilde{t}_{\text{max}}=275.2 \mu$s. All other considerations are same as in Fig.~\ref{Fig:Gamma_witness_quantifier}. The quantities plotted here are dimensionless.
}
\label{Fig:ph_eff}
\end{figure}
In Fig.~\ref{Fig:ph_eff}, the behavior of efficiency of the engine is depicted as we vary the effective negative temperature of the hot bath through population, denoted as $p^{+}_{\text{hot}}$. The black solid curve corresponds to an infinite-time protocol (IFT) with $g=0.2$. In this case, the engine commences its operation when $p^{+}_{\text{hot}}$ reaches $0.61$ and obtain the maximum efficiency $\eta_{\text{INF}}=0.649$ at $p^{+}_{\text{hot}}=0.99$. This investigation has previously been conducted 
with $g=0$ in Ref.~\cite{Assis_PRL_2019} and with $g\ne 0$ in Ref.~\cite{Arghya_arxiv_2023}. Let us now shift our focus to the same investigation for finite-time protocols. In Fig.~\ref{Fig:ph_eff}, we present the efficiency of the effective negative temperature-based transient quantum Otto engine while varying $p^{+}_{\text{hot}}$ from $0.5$ to $0.99$, with the heating stroke duration fixed at $\tilde{t}_{\text{max}}$ for different $\omega_c$, each denoted by distinct colors and symbols. It is important to note that $\tilde{t}_{\text{max}}$ varies for different $\omega_c$, and in this analysis, we maintain $g=0.2$.
A notable observation is that, in contrast to the IFT scenarios, the engine can commence operation at lower values of $p^{+}_{\text{hot}}$, which correspond to lower values of $\beta_{\text{hot}}$, depending on the chosen cutoff frequency. Additionally, the efficiency obtained for the finite-time protocol (FT) at each $p^{+}_{\text{hot}}$ and for different $\omega_c$ surpasses that achieved by the IFT. Furthermore, as $\omega_c$ decreases, signifying an increase in the degree of non-Markovianity $\mathcal{Q}$, both the operational range and efficiency of the effective negative temperature-based transient engine expand. However, it is important to mention that this operational range and efficiency cannot be indefinitely increased by lowering the cutoff frequency of the reservoir. As we have previously demonstrated, for very small values of $\omega_c$, the quantifier of non-Markovianity saturates and then decreases. Similarly, in this context, beyond a certain value of $\omega_c$, the efficiency $\eta$ as a function of $p^{+}_{\text{hot}}$ also saturates and subsequently declines. 
Hence, we can infer the existence of effective negative temperature-based transient quantum Otto engines that can operate in the region of $\beta_{\text{hot}}$ where conventional negative temperature-based quantum Otto engines employing IFT cannot function. In this specific regime of effective negative temperature of the hot bath, we refer to these engines as effective negative temperature-based necessarily transient quantum Otto engines.


\section{Conclusion}
\label{sec:Con}
It is evident, based on both theoretical predictions~\cite{Assis_PRL_2019,Arghya_arxiv_2023} and experimental observation~\cite{Assis_PRL_2019}, that a quantum Otto engine operating with a reservoir characterized by an effective negative temperature offers numerous advantages. The primary advantage lies in its higher efficiency when compared to a typical quantum Otto engine operating with both reservoirs at positive spin temperatures. However, previous studies have been confined to scenarios involving perfect thermalization or an infinite-time protocol, where the final states of the isochoric strokes always converge to a thermal state.
In this paper, we have presented an effective negative temperature-based quantum Otto engine that has been operated within the transient or finite-time domain of the isochoric strokes. We have found that these transient engines, employing a spin-$\frac{1}{2}$ working substance, exhibit superior efficiency compared to their steady-state counterparts. 
We have studied the performance of these effective negative temperature-based transient quantum Otto engines as they transition from the Markovian to the non-Markovian regime. This study has been conducted within the framework of weak coupling between the system and the bath, allowing for the unique phenomenon of information backflow from the environment to the system. 
We have shown that the maximum achievable efficiency at a specific time of the heating stroke experiences a significant boost with an increase in non-Markovianity. On the contrary, the overall engine performance over an extended period of this stroke declines as non-Markovianity increases. Furthermore, our investigation has led to the discovery of a unique class of engines: the effective negative temperature-based necessarily transient quantum Otto engines. In scenarios where the conventional infinite-time effective negative temperature-based Otto-cycle protocol fails to function as an engine, the effective negative temperature-based necessarily transient quantum Otto engines emerge as a pragmatic alternative for operation within this particular realm. Moreover, we observed that the operational range of these effective negative temperature-based necessarily transient quantum Otto engines expands significantly with increasing non-Markovianity.
\vspace{0.2in}
\section*{Acknowledgment}
\label{sec:Ack}
We acknowledge computations performed using Armadillo~\cite{Sanderson,Sanderson1} on the cluster computing facility of the Harish-Chandra Research Institute, India. This research was supported in part by the `INFOSYS scholarship for senior students'. A. Maity acknowledges the support from the Interdisciplinary Cyber-Physical Systems (ICPS) program of the Department of Science and Technology (DST), India, Grant No.: DST/ICPS/QuST/Theme- 1/2019/23. A. Ghoshal acknowledges support from the Alexander von Humboldt Foundation.

\bibliography{FTO}

\begin{thebibliography}{115}%
\makeatletter
\providecommand \@ifxundefined [1]{%
 \@ifx{#1\undefined}
}%
\providecommand \@ifnum [1]{%
 \ifnum #1\expandafter \@firstoftwo
 \else \expandafter \@secondoftwo
 \fi
}%
\providecommand \@ifx [1]{%
 \ifx #1\expandafter \@firstoftwo
 \else \expandafter \@secondoftwo
 \fi
}%
\providecommand \natexlab [1]{#1}%
\providecommand \enquote  [1]{``#1''}%
\providecommand \bibnamefont  [1]{#1}%
\providecommand \bibfnamefont [1]{#1}%
\providecommand \citenamefont [1]{#1}%
\providecommand \href@noop [0]{\@secondoftwo}%
\providecommand \href [0]{\begingroup \@sanitize@url \@href}%
\providecommand \@href[1]{\@@startlink{#1}\@@href}%
\providecommand \@@href[1]{\endgroup#1\@@endlink}%
\providecommand \@sanitize@url [0]{\catcode `\\12\catcode `\$12\catcode
  `\&12\catcode `\#12\catcode `\^12\catcode `\_12\catcode `\%12\relax}%
\providecommand \@@startlink[1]{}%
\providecommand \@@endlink[0]{}%
\providecommand \url  [0]{\begingroup\@sanitize@url \@url }%
\providecommand \@url [1]{\endgroup\@href {#1}{\urlprefix }}%
\providecommand \urlprefix  [0]{URL }%
\providecommand \Eprint [0]{\href }%
\providecommand \doibase [0]{https://doi.org/}%
\providecommand \selectlanguage [0]{\@gobble}%
\providecommand \bibinfo  [0]{\@secondoftwo}%
\providecommand \bibfield  [0]{\@secondoftwo}%
\providecommand \translation [1]{[#1]}%
\providecommand \BibitemOpen [0]{}%
\providecommand \bibitemStop [0]{}%
\providecommand \bibitemNoStop [0]{.\EOS\space}%
\providecommand \EOS [0]{\spacefactor3000\relax}%
\providecommand \BibitemShut  [1]{\csname bibitem#1\endcsname}%
\let\auto@bib@innerbib\@empty
\bibitem [{\citenamefont {Scovil}\ and\ \citenamefont
  {Schulz-DuBois}(1959)}]{Scovil_PRL_1959}%
  \BibitemOpen
  \bibfield  {author} {\bibinfo {author} {\bibfnamefont {H.~E.~D.}\
  \bibnamefont {Scovil}}\ and\ \bibinfo {author} {\bibfnamefont {E.~O.}\
  \bibnamefont {Schulz-DuBois}},\ }\bibfield  {title} {\bibinfo {title}
  {Three-level masers as heat engines},\ }\href
  {https://doi.org/10.1103/PhysRevLett.2.262} {\bibfield  {journal} {\bibinfo
  {journal} {Phys. Rev. Lett.}\ }\textbf {\bibinfo {volume} {2}},\ \bibinfo
  {pages} {262} (\bibinfo {year} {1959})}\BibitemShut {NoStop}%
\bibitem [{\citenamefont {Kosloff}(1984)}]{Kosloff_JOCP_1984}%
  \BibitemOpen
  \bibfield  {author} {\bibinfo {author} {\bibfnamefont {R.}~\bibnamefont
  {Kosloff}},\ }\bibfield  {title} {\bibinfo {title} {{A quantum mechanical
  open system as a model of a heat engine}},\ }\href
  {https://doi.org/10.1063/1.446862} {\bibfield  {journal} {\bibinfo  {journal}
  {The Journal of Chemical Physics}\ }\textbf {\bibinfo {volume} {80}},\
  \bibinfo {pages} {1625} (\bibinfo {year} {1984})}\BibitemShut {NoStop}%
\bibitem [{\citenamefont {Scully}\ \emph {et~al.}(2003)\citenamefont {Scully},
  \citenamefont {Zubairy}, \citenamefont {Agarwal},\ and\ \citenamefont
  {Walther}}]{Scully_Science_2003}%
  \BibitemOpen
  \bibfield  {author} {\bibinfo {author} {\bibfnamefont {M.~O.}\ \bibnamefont
  {Scully}}, \bibinfo {author} {\bibfnamefont {M.~S.}\ \bibnamefont {Zubairy}},
  \bibinfo {author} {\bibfnamefont {G.~S.}\ \bibnamefont {Agarwal}},\ and\
  \bibinfo {author} {\bibfnamefont {H.}~\bibnamefont {Walther}},\ }\bibfield
  {title} {\bibinfo {title} {Extracting work from a single heat bath via
  vanishing quantum coherence},\ }\href
  {https://doi.org/10.1126/science.1078955} {\bibfield  {journal} {\bibinfo
  {journal} {Science}\ }\textbf {\bibinfo {volume} {299}},\ \bibinfo {pages}
  {862} (\bibinfo {year} {2003})}\BibitemShut {NoStop}%
\bibitem [{\citenamefont {Jochen~Gemmer}(2009)}]{Gemmer_QT_book}%
  \BibitemOpen
  \bibfield  {author} {\bibinfo {author} {\bibfnamefont {G.~M.}\ \bibnamefont
  {Jochen~Gemmer}, \bibfnamefont {M.~Michel}},\ }\href
  {https://doi.org/https://doi.org/10.1007/978-3-540-70510-9} {\emph {\bibinfo
  {title} {Quantum Thermodynamics}}}\ (\bibinfo  {publisher} {Springer Berlin,
  Heidelberg},\ \bibinfo {year} {2009})\BibitemShut {NoStop}%
\bibitem [{\citenamefont {Deffner}\ and\ \citenamefont
  {Campbell}(2019)}]{Sebastian_book}%
  \BibitemOpen
  \bibfield  {author} {\bibinfo {author} {\bibfnamefont {S.}~\bibnamefont
  {Deffner}}\ and\ \bibinfo {author} {\bibfnamefont {S.}~\bibnamefont
  {Campbell}},\ }\href {https://doi.org/10.1088/2053-2571/ab21c6} {\emph
  {\bibinfo {title} {Quantum Thermodynamics}}},\ 2053-2571\ (\bibinfo
  {publisher} {Morgan \& Claypool Publishers, San Rafael},\ \bibinfo {year}
  {2019})\BibitemShut {NoStop}%
\bibitem [{\citenamefont {Pe\~{n}a}\ \emph {et~al.}(2020)\citenamefont
  {Pe\~{n}a}, \citenamefont {Negrete}, \citenamefont {Cort\'{e}s},\ and\
  \citenamefont {Vargas}}]{Francisco_Entropy_2020}%
  \BibitemOpen
  \bibfield  {author} {\bibinfo {author} {\bibfnamefont {F.~J.}\ \bibnamefont
  {Pe\~{n}a}}, \bibinfo {author} {\bibfnamefont {O.}~\bibnamefont {Negrete}},
  \bibinfo {author} {\bibfnamefont {N.}~\bibnamefont {Cort\'{e}s}},\ and\
  \bibinfo {author} {\bibfnamefont {P.}~\bibnamefont {Vargas}},\ }\bibfield
  {title} {\bibinfo {title} {Otto engine: Classical and quantum approach},\
  }\href {https://doi.org/10.3390/e22070755} {\bibfield  {journal} {\bibinfo
  {journal} {Entropy}\ }\textbf {\bibinfo {volume} {22}},\ \bibinfo {pages}
  {755} (\bibinfo {year} {2020})}\BibitemShut {NoStop}%
\bibitem [{\citenamefont {Myers}\ \emph {et~al.}(2022)\citenamefont {Myers},
  \citenamefont {Abah},\ and\ \citenamefont
  {Deffner}}]{Sebastian_Review_AVS_2022}%
  \BibitemOpen
  \bibfield  {author} {\bibinfo {author} {\bibfnamefont {N.~M.}\ \bibnamefont
  {Myers}}, \bibinfo {author} {\bibfnamefont {O.}~\bibnamefont {Abah}},\ and\
  \bibinfo {author} {\bibfnamefont {S.}~\bibnamefont {Deffner}},\ }\bibfield
  {title} {\bibinfo {title} {Quantum thermodynamic devices: From theoretical
  proposals to experimental reality},\ }\href
  {https://doi.org/10.1116/5.0083192} {\bibfield  {journal} {\bibinfo
  {journal} {AVS Quantum Science}\ }\textbf {\bibinfo {volume} {4}},\ \bibinfo
  {pages} {027101} (\bibinfo {year} {2022})}\BibitemShut {NoStop}%
\bibitem [{\citenamefont {Scully}(2002)}]{Scully_PRL_2002}%
  \BibitemOpen
  \bibfield  {author} {\bibinfo {author} {\bibfnamefont {M.~O.}\ \bibnamefont
  {Scully}},\ }\bibfield  {title} {\bibinfo {title} {Quantum afterburner:
  Improving the efficiency of an ideal heat engine},\ }\href
  {https://doi.org/10.1103/PhysRevLett.88.050602} {\bibfield  {journal}
  {\bibinfo  {journal} {Phys. Rev. Lett.}\ }\textbf {\bibinfo {volume} {88}},\
  \bibinfo {pages} {050602} (\bibinfo {year} {2002})}\BibitemShut {NoStop}%
\bibitem [{\citenamefont {Mu\~noz}\ and\ \citenamefont
  {Pe\~na}(2012)}]{Enrique_PRE_2012}%
  \BibitemOpen
  \bibfield  {author} {\bibinfo {author} {\bibfnamefont {E.}~\bibnamefont
  {Mu\~noz}}\ and\ \bibinfo {author} {\bibfnamefont {F.~J.}\ \bibnamefont
  {Pe\~na}},\ }\bibfield  {title} {\bibinfo {title} {Quantum heat engine in the
  relativistic limit: The case of a dirac particle},\ }\href
  {https://doi.org/10.1103/PhysRevE.86.061108} {\bibfield  {journal} {\bibinfo
  {journal} {Phys. Rev. E}\ }\textbf {\bibinfo {volume} {86}},\ \bibinfo
  {pages} {061108} (\bibinfo {year} {2012})}\BibitemShut {NoStop}%
\bibitem [{\citenamefont {Ro\ss{}nagel}\ \emph {et~al.}(2014)\citenamefont
  {Ro\ss{}nagel}, \citenamefont {Abah}, \citenamefont {Schmidt-Kaler},
  \citenamefont {Singer},\ and\ \citenamefont {Lutz}}]{Lutz_PRL_2014}%
  \BibitemOpen
  \bibfield  {author} {\bibinfo {author} {\bibfnamefont {J.}~\bibnamefont
  {Ro\ss{}nagel}}, \bibinfo {author} {\bibfnamefont {O.}~\bibnamefont {Abah}},
  \bibinfo {author} {\bibfnamefont {F.}~\bibnamefont {Schmidt-Kaler}}, \bibinfo
  {author} {\bibfnamefont {K.}~\bibnamefont {Singer}},\ and\ \bibinfo {author}
  {\bibfnamefont {E.}~\bibnamefont {Lutz}},\ }\bibfield  {title} {\bibinfo
  {title} {Nanoscale heat engine beyond the carnot limit},\ }\href
  {https://doi.org/10.1103/PhysRevLett.112.030602} {\bibfield  {journal}
  {\bibinfo  {journal} {Phys. Rev. Lett.}\ }\textbf {\bibinfo {volume} {112}},\
  \bibinfo {pages} {030602} (\bibinfo {year} {2014})}\BibitemShut {NoStop}%
\bibitem [{\citenamefont {Cherubim}\ \emph {et~al.}(2019)\citenamefont
  {Cherubim}, \citenamefont {Brito},\ and\ \citenamefont
  {Deffner}}]{Deffner_Entropy_2019}%
  \BibitemOpen
  \bibfield  {author} {\bibinfo {author} {\bibfnamefont {C.}~\bibnamefont
  {Cherubim}}, \bibinfo {author} {\bibfnamefont {F.}~\bibnamefont {Brito}},\
  and\ \bibinfo {author} {\bibfnamefont {S.}~\bibnamefont {Deffner}},\
  }\bibfield  {title} {\bibinfo {title} {Non-thermal quantum engine in transmon
  qubits},\ }\href {https://doi.org/10.3390/e21060545} {\bibfield  {journal}
  {\bibinfo  {journal} {Entropy}\ }\textbf {\bibinfo {volume} {21}},\ \bibinfo
  {pages} {545} (\bibinfo {year} {2019})}\BibitemShut {NoStop}%
\bibitem [{\citenamefont {Maslennikov}\ \emph {et~al.}(2019)\citenamefont
  {Maslennikov}, \citenamefont {Ding}, \citenamefont {Habl{\"u}tzel},
  \citenamefont {Gan}, \citenamefont {Roulet}, \citenamefont {Nimmrichter},
  \citenamefont {Dai}, \citenamefont {Scarani},\ and\ \citenamefont
  {Matsukevich}}]{Maslennikov_Nature_Comm_2019}%
  \BibitemOpen
  \bibfield  {author} {\bibinfo {author} {\bibfnamefont {G.}~\bibnamefont
  {Maslennikov}}, \bibinfo {author} {\bibfnamefont {S.}~\bibnamefont {Ding}},
  \bibinfo {author} {\bibfnamefont {R.}~\bibnamefont {Habl{\"u}tzel}}, \bibinfo
  {author} {\bibfnamefont {J.}~\bibnamefont {Gan}}, \bibinfo {author}
  {\bibfnamefont {A.}~\bibnamefont {Roulet}}, \bibinfo {author} {\bibfnamefont
  {S.}~\bibnamefont {Nimmrichter}}, \bibinfo {author} {\bibfnamefont
  {J.}~\bibnamefont {Dai}}, \bibinfo {author} {\bibfnamefont {V.}~\bibnamefont
  {Scarani}},\ and\ \bibinfo {author} {\bibfnamefont {D.}~\bibnamefont
  {Matsukevich}},\ }\bibfield  {title} {\bibinfo {title} {Quantum absorption
  refrigerator with trapped ions},\ }\href
  {https://doi.org/10.1038/s41467-018-08090-0} {\bibfield  {journal} {\bibinfo
  {journal} {Nature Communications}\ }\textbf {\bibinfo {volume} {10}},\
  \bibinfo {pages} {202} (\bibinfo {year} {2019})}\BibitemShut {NoStop}%
\bibitem [{\citenamefont {Singh}\ and\ \citenamefont
  {Benjamin}(2021)}]{Singh_PRB_2021}%
  \BibitemOpen
  \bibfield  {author} {\bibinfo {author} {\bibfnamefont {A.}~\bibnamefont
  {Singh}}\ and\ \bibinfo {author} {\bibfnamefont {C.}~\bibnamefont
  {Benjamin}},\ }\bibfield  {title} {\bibinfo {title} {Magic angle twisted
  bilayer graphene as a highly efficient quantum otto engine},\ }\href
  {https://doi.org/10.1103/PhysRevB.104.125445} {\bibfield  {journal} {\bibinfo
   {journal} {Phys. Rev. B}\ }\textbf {\bibinfo {volume} {104}},\ \bibinfo
  {pages} {125445} (\bibinfo {year} {2021})}\BibitemShut {NoStop}%
\bibitem [{\citenamefont {Myers}\ \emph {et~al.}(2021)\citenamefont {Myers},
  \citenamefont {Abah},\ and\ \citenamefont {Deffner}}]{Myers_NJP_2021}%
  \BibitemOpen
  \bibfield  {author} {\bibinfo {author} {\bibfnamefont {N.~M.}\ \bibnamefont
  {Myers}}, \bibinfo {author} {\bibfnamefont {O.}~\bibnamefont {Abah}},\ and\
  \bibinfo {author} {\bibfnamefont {S.}~\bibnamefont {Deffner}},\ }\bibfield
  {title} {\bibinfo {title} {Quantum otto engines at relativistic energies},\
  }\href {https://doi.org/10.1088/1367-2630/ac2756} {\bibfield  {journal}
  {\bibinfo  {journal} {New Journal of Physics}\ }\textbf {\bibinfo {volume}
  {23}},\ \bibinfo {pages} {105001} (\bibinfo {year} {2021})}\BibitemShut
  {NoStop}%
\bibitem [{\citenamefont {Myers}\ \emph {et~al.}(2023)\citenamefont {Myers},
  \citenamefont {Pe\~{n}a}, \citenamefont {Cort\'{e}s},\ and\ \citenamefont
  {Vargas}}]{Myers_Nanomaterials_2023}%
  \BibitemOpen
  \bibfield  {author} {\bibinfo {author} {\bibfnamefont {N.~M.}\ \bibnamefont
  {Myers}}, \bibinfo {author} {\bibfnamefont {F.~J.}\ \bibnamefont {Pe\~{n}a}},
  \bibinfo {author} {\bibfnamefont {N.}~\bibnamefont {Cort\'{e}s}},\ and\
  \bibinfo {author} {\bibfnamefont {P.}~\bibnamefont {Vargas}},\ }\bibfield
  {title} {\bibinfo {title} {Multilayer graphene as an endoreversible otto
  engine},\ }\href {https://www.mdpi.com/2079-4991/13/9/1548} {\bibfield
  {journal} {\bibinfo  {journal} {Nanomaterials}\ }\textbf {\bibinfo {volume}
  {13}} (\bibinfo {year} {2023})}\BibitemShut {NoStop}%
\bibitem [{\citenamefont {Scully}(2001)}]{Scully_PRL_2001}%
  \BibitemOpen
  \bibfield  {author} {\bibinfo {author} {\bibfnamefont {M.~O.}\ \bibnamefont
  {Scully}},\ }\bibfield  {title} {\bibinfo {title} {Extracting work from a
  single thermal bath via quantum negentropy},\ }\href
  {https://doi.org/10.1103/PhysRevLett.87.220601} {\bibfield  {journal}
  {\bibinfo  {journal} {Phys. Rev. Lett.}\ }\textbf {\bibinfo {volume} {87}},\
  \bibinfo {pages} {220601} (\bibinfo {year} {2001})}\BibitemShut {NoStop}%
\bibitem [{\citenamefont {Huang}\ \emph {et~al.}(2012)\citenamefont {Huang},
  \citenamefont {Wang},\ and\ \citenamefont {Yi}}]{Huang_PRE_2012}%
  \BibitemOpen
  \bibfield  {author} {\bibinfo {author} {\bibfnamefont {X.~L.}\ \bibnamefont
  {Huang}}, \bibinfo {author} {\bibfnamefont {T.}~\bibnamefont {Wang}},\ and\
  \bibinfo {author} {\bibfnamefont {X.~X.}\ \bibnamefont {Yi}},\ }\bibfield
  {title} {\bibinfo {title} {Effects of reservoir squeezing on quantum systems
  and work extraction},\ }\href {https://doi.org/10.1103/PhysRevE.86.051105}
  {\bibfield  {journal} {\bibinfo  {journal} {Phys. Rev. E}\ }\textbf {\bibinfo
  {volume} {86}},\ \bibinfo {pages} {051105} (\bibinfo {year}
  {2012})}\BibitemShut {NoStop}%
\bibitem [{\citenamefont {Zagoskin}\ \emph {et~al.}(2012)\citenamefont
  {Zagoskin}, \citenamefont {Savel'ev}, \citenamefont {Nori},\ and\
  \citenamefont {Kusmartsev}}]{Zagoskin_PRB_2012}%
  \BibitemOpen
  \bibfield  {author} {\bibinfo {author} {\bibfnamefont {A.~M.}\ \bibnamefont
  {Zagoskin}}, \bibinfo {author} {\bibfnamefont {S.}~\bibnamefont {Savel'ev}},
  \bibinfo {author} {\bibfnamefont {F.}~\bibnamefont {Nori}},\ and\ \bibinfo
  {author} {\bibfnamefont {F.~V.}\ \bibnamefont {Kusmartsev}},\ }\bibfield
  {title} {\bibinfo {title} {Squeezing as the source of inefficiency in the
  quantum otto cycle},\ }\href {https://doi.org/10.1103/PhysRevB.86.014501}
  {\bibfield  {journal} {\bibinfo  {journal} {Phys. Rev. B}\ }\textbf {\bibinfo
  {volume} {86}},\ \bibinfo {pages} {014501} (\bibinfo {year}
  {2012})}\BibitemShut {NoStop}%
\bibitem [{\citenamefont {Altintas}\ \emph {et~al.}(2014)\citenamefont
  {Altintas}, \citenamefont {Hardal},\ and\ \citenamefont
  {M\"ustecapl{\i}o\~{g}lu}}]{Ferdi_PRE_2014}%
  \BibitemOpen
  \bibfield  {author} {\bibinfo {author} {\bibfnamefont {F.}~\bibnamefont
  {Altintas}}, \bibinfo {author} {\bibfnamefont {A.~{\"{U}}.~C.}\ \bibnamefont
  {Hardal}},\ and\ \bibinfo {author} {\bibfnamefont {O.~E.}\ \bibnamefont
  {M\"ustecapl{\i}o\~{g}lu}},\ }\bibfield  {title} {\bibinfo {title} {Quantum
  correlated heat engine with spin squeezing},\ }\href
  {https://doi.org/10.1103/PhysRevE.90.032102} {\bibfield  {journal} {\bibinfo
  {journal} {Phys. Rev. E}\ }\textbf {\bibinfo {volume} {90}},\ \bibinfo
  {pages} {032102} (\bibinfo {year} {2014})}\BibitemShut {NoStop}%
\bibitem [{\citenamefont {Long}\ and\ \citenamefont
  {Liu}(2015)}]{Rui_PRE_2015}%
  \BibitemOpen
  \bibfield  {author} {\bibinfo {author} {\bibfnamefont {R.}~\bibnamefont
  {Long}}\ and\ \bibinfo {author} {\bibfnamefont {W.}~\bibnamefont {Liu}},\
  }\bibfield  {title} {\bibinfo {title} {Performance of quantum otto
  refrigerators with squeezing},\ }\href
  {https://doi.org/10.1103/PhysRevE.91.062137} {\bibfield  {journal} {\bibinfo
  {journal} {Phys. Rev. E}\ }\textbf {\bibinfo {volume} {91}},\ \bibinfo
  {pages} {062137} (\bibinfo {year} {2015})}\BibitemShut {NoStop}%
\bibitem [{\citenamefont {Klaers}\ \emph {et~al.}(2017)\citenamefont {Klaers},
  \citenamefont {Faelt}, \citenamefont {Imamoglu},\ and\ \citenamefont
  {Togan}}]{Jan_PRX_2017}%
  \BibitemOpen
  \bibfield  {author} {\bibinfo {author} {\bibfnamefont {J.}~\bibnamefont
  {Klaers}}, \bibinfo {author} {\bibfnamefont {S.}~\bibnamefont {Faelt}},
  \bibinfo {author} {\bibfnamefont {A.}~\bibnamefont {Imamoglu}},\ and\
  \bibinfo {author} {\bibfnamefont {E.}~\bibnamefont {Togan}},\ }\bibfield
  {title} {\bibinfo {title} {Squeezed thermal reservoirs as a resource for a
  nanomechanical engine beyond the carnot limit},\ }\href
  {https://doi.org/10.1103/PhysRevX.7.031044} {\bibfield  {journal} {\bibinfo
  {journal} {Phys. Rev. X}\ }\textbf {\bibinfo {volume} {7}},\ \bibinfo {pages}
  {031044} (\bibinfo {year} {2017})}\BibitemShut {NoStop}%
\bibitem [{\citenamefont {Hardal}\ \emph {et~al.}(2017)\citenamefont {Hardal},
  \citenamefont {Aslan}, \citenamefont {Wilson},\ and\ \citenamefont
  {M\"ustecapl\ifmmode \imath \else \i \fi{}o\ifmmode~\breve{g}\else
  \u{g}\fi{}lu}}]{Hardal_PRE_2017}%
  \BibitemOpen
  \bibfield  {author} {\bibinfo {author} {\bibfnamefont {A.~{\"U}.~C.}\
  \bibnamefont {Hardal}}, \bibinfo {author} {\bibfnamefont {N.}~\bibnamefont
  {Aslan}}, \bibinfo {author} {\bibfnamefont {C.~M.}\ \bibnamefont {Wilson}},\
  and\ \bibinfo {author} {\bibfnamefont {O.~E.}\ \bibnamefont
  {M\"ustecapl\ifmmode \imath \else \i \fi{}o\ifmmode~\breve{g}\else
  \u{g}\fi{}lu}},\ }\bibfield  {title} {\bibinfo {title} {Quantum heat engine
  with coupled superconducting resonators},\ }\href
  {https://doi.org/10.1103/PhysRevE.96.062120} {\bibfield  {journal} {\bibinfo
  {journal} {Phys. Rev. E}\ }\textbf {\bibinfo {volume} {96}},\ \bibinfo
  {pages} {062120} (\bibinfo {year} {2017})}\BibitemShut {NoStop}%
\bibitem [{\citenamefont {Xiao}\ and\ \citenamefont
  {Li}(2018)}]{Xiao_PLA_2018}%
  \BibitemOpen
  \bibfield  {author} {\bibinfo {author} {\bibfnamefont {B.}~\bibnamefont
  {Xiao}}\ and\ \bibinfo {author} {\bibfnamefont {R.}~\bibnamefont {Li}},\
  }\bibfield  {title} {\bibinfo {title} {Finite time thermodynamic analysis of
  quantum otto heat engine with squeezed thermal bath},\ }\href
  {https://doi.org/https://doi.org/10.1016/j.physleta.2018.07.033} {\bibfield
  {journal} {\bibinfo  {journal} {Physics Letters A}\ }\textbf {\bibinfo
  {volume} {382}},\ \bibinfo {pages} {3051} (\bibinfo {year}
  {2018})}\BibitemShut {NoStop}%
\bibitem [{\citenamefont {Niedenzu}\ \emph {et~al.}(2018)\citenamefont
  {Niedenzu}, \citenamefont {Mukherjee}, \citenamefont {Ghosh}, \citenamefont
  {Kofman},\ and\ \citenamefont {Kurizki}}]{Niedenzu_Nature_2018}%
  \BibitemOpen
  \bibfield  {author} {\bibinfo {author} {\bibfnamefont {W.}~\bibnamefont
  {Niedenzu}}, \bibinfo {author} {\bibfnamefont {V.}~\bibnamefont {Mukherjee}},
  \bibinfo {author} {\bibfnamefont {A.}~\bibnamefont {Ghosh}}, \bibinfo
  {author} {\bibfnamefont {A.~G.}\ \bibnamefont {Kofman}},\ and\ \bibinfo
  {author} {\bibfnamefont {G.}~\bibnamefont {Kurizki}},\ }\bibfield  {title}
  {\bibinfo {title} {Quantum engine efficiency bound beyond the second law of
  thermodynamics},\ }\href {https://doi.org/10.1038/s41467-017-01991-6}
  {\bibfield  {journal} {\bibinfo  {journal} {Nature Communications}\ }\textbf
  {\bibinfo {volume} {9}},\ \bibinfo {pages} {165} (\bibinfo {year}
  {2018})}\BibitemShut {NoStop}%
\bibitem [{\citenamefont {Wright}\ \emph {et~al.}(2018)\citenamefont {Wright},
  \citenamefont {Gould}, \citenamefont {Carvalho}, \citenamefont {Bedkihal},\
  and\ \citenamefont {Vaccaro}}]{Jackson_PRA_2018}%
  \BibitemOpen
  \bibfield  {author} {\bibinfo {author} {\bibfnamefont {J.~S. S.~T.}\
  \bibnamefont {Wright}}, \bibinfo {author} {\bibfnamefont {T.}~\bibnamefont
  {Gould}}, \bibinfo {author} {\bibfnamefont {A.~R.~R.}\ \bibnamefont
  {Carvalho}}, \bibinfo {author} {\bibfnamefont {S.}~\bibnamefont {Bedkihal}},\
  and\ \bibinfo {author} {\bibfnamefont {J.~A.}\ \bibnamefont {Vaccaro}},\
  }\bibfield  {title} {\bibinfo {title} {Quantum heat engine operating between
  thermal and spin reservoirs},\ }\href
  {https://doi.org/10.1103/PhysRevA.97.052104} {\bibfield  {journal} {\bibinfo
  {journal} {Phys. Rev. A}\ }\textbf {\bibinfo {volume} {97}},\ \bibinfo
  {pages} {052104} (\bibinfo {year} {2018})}\BibitemShut {NoStop}%
\bibitem [{\citenamefont {Wang}\ \emph {et~al.}(2019)\citenamefont {Wang},
  \citenamefont {He},\ and\ \citenamefont {Ma}}]{Wang_PRE_2019}%
  \BibitemOpen
  \bibfield  {author} {\bibinfo {author} {\bibfnamefont {J.}~\bibnamefont
  {Wang}}, \bibinfo {author} {\bibfnamefont {J.}~\bibnamefont {He}},\ and\
  \bibinfo {author} {\bibfnamefont {Y.}~\bibnamefont {Ma}},\ }\bibfield
  {title} {\bibinfo {title} {Finite-time performance of a quantum heat engine
  with a squeezed thermal bath},\ }\href
  {https://doi.org/10.1103/PhysRevE.100.052126} {\bibfield  {journal} {\bibinfo
   {journal} {Phys. Rev. E}\ }\textbf {\bibinfo {volume} {100}},\ \bibinfo
  {pages} {052126} (\bibinfo {year} {2019})}\BibitemShut {NoStop}%
\bibitem [{\citenamefont {de~Assis}\ \emph {et~al.}(2020)\citenamefont
  {de~Assis}, \citenamefont {Sales}, \citenamefont {da~Cunha},\ and\
  \citenamefont {de~Almeida}}]{Assis_PRE_2020}%
  \BibitemOpen
  \bibfield  {author} {\bibinfo {author} {\bibfnamefont {R.~J.}\ \bibnamefont
  {de~Assis}}, \bibinfo {author} {\bibfnamefont {J.~S.}\ \bibnamefont {Sales}},
  \bibinfo {author} {\bibfnamefont {J.~A.~R.}\ \bibnamefont {da~Cunha}},\ and\
  \bibinfo {author} {\bibfnamefont {N.~G.}\ \bibnamefont {de~Almeida}},\
  }\bibfield  {title} {\bibinfo {title} {Universal two-level quantum otto
  machine under a squeezed reservoir},\ }\href
  {https://doi.org/10.1103/PhysRevE.102.052131} {\bibfield  {journal} {\bibinfo
   {journal} {Phys. Rev. E}\ }\textbf {\bibinfo {volume} {102}},\ \bibinfo
  {pages} {052131} (\bibinfo {year} {2020})}\BibitemShut {NoStop}%
\bibitem [{\citenamefont {Zhang}(2020)}]{Yanchao_PhysicaA_2020}%
  \BibitemOpen
  \bibfield  {author} {\bibinfo {author} {\bibfnamefont {Y.}~\bibnamefont
  {Zhang}},\ }\bibfield  {title} {\bibinfo {title} {Optimization performance of
  quantum otto heat engines and refrigerators with squeezed thermal
  reservoirs},\ }\href
  {https://doi.org/https://doi.org/10.1016/j.physa.2020.125083} {\bibfield
  {journal} {\bibinfo  {journal} {Physica A: Statistical Mechanics and its
  Applications}\ }\textbf {\bibinfo {volume} {559}},\ \bibinfo {pages} {125083}
  (\bibinfo {year} {2020})}\BibitemShut {NoStop}%
\bibitem [{\citenamefont {de~Assis}\ \emph {et~al.}(2021)\citenamefont
  {de~Assis}, \citenamefont {Sales}, \citenamefont {Mendes},\ and\
  \citenamefont {de~Almeida}}]{Assis_JOP_2021}%
  \BibitemOpen
  \bibfield  {author} {\bibinfo {author} {\bibfnamefont {R.~J.}\ \bibnamefont
  {de~Assis}}, \bibinfo {author} {\bibfnamefont {J.~S.}\ \bibnamefont {Sales}},
  \bibinfo {author} {\bibfnamefont {U.~C.}\ \bibnamefont {Mendes}},\ and\
  \bibinfo {author} {\bibfnamefont {N.~G.}\ \bibnamefont {de~Almeida}},\
  }\bibfield  {title} {\bibinfo {title} {Two-level quantum otto heat engine
  operating with unit efficiency far from the quasi-static regime under a
  squeezed reservoir},\ }\href {https://doi.org/10.1088/1361-6455/abcfd9}
  {\bibfield  {journal} {\bibinfo  {journal} {Journal of Physics B: Atomic,
  Molecular and Optical Physics}\ }\textbf {\bibinfo {volume} {54}},\ \bibinfo
  {pages} {095501} (\bibinfo {year} {2021})}\BibitemShut {NoStop}%
\bibitem [{\citenamefont {Tabatabaei}\ \emph {et~al.}(2022)\citenamefont
  {Tabatabaei}, \citenamefont {S\'anchez}, \citenamefont {Yeyati},\ and\
  \citenamefont {S\'anchez}}]{Rafael_PRB_2022}%
  \BibitemOpen
  \bibfield  {author} {\bibinfo {author} {\bibfnamefont {S.~M.}\ \bibnamefont
  {Tabatabaei}}, \bibinfo {author} {\bibfnamefont {D.}~\bibnamefont
  {S\'anchez}}, \bibinfo {author} {\bibfnamefont {A.~L.}\ \bibnamefont
  {Yeyati}},\ and\ \bibinfo {author} {\bibfnamefont {R.}~\bibnamefont
  {S\'anchez}},\ }\bibfield  {title} {\bibinfo {title} {Nonlocal quantum heat
  engines made of hybrid superconducting devices},\ }\href
  {https://doi.org/10.1103/PhysRevB.106.115419} {\bibfield  {journal} {\bibinfo
   {journal} {Phys. Rev. B}\ }\textbf {\bibinfo {volume} {106}},\ \bibinfo
  {pages} {115419} (\bibinfo {year} {2022})}\BibitemShut {NoStop}%
\bibitem [{\citenamefont {Mendon\ifmmode~\mbox{\c{c}}\else \c{c}\fi{}a}\ \emph
  {et~al.}(2020)\citenamefont {Mendon\ifmmode~\mbox{\c{c}}\else \c{c}\fi{}a},
  \citenamefont {Souza}, \citenamefont {de~Assis}, \citenamefont {de~Almeida},
  \citenamefont {Sarthour}, \citenamefont {Oliveira},\ and\ \citenamefont
  {Villas-Boas}}]{Mendon_PRR_2020}%
  \BibitemOpen
  \bibfield  {author} {\bibinfo {author} {\bibfnamefont {T.~M.}\ \bibnamefont
  {Mendon\ifmmode~\mbox{\c{c}}\else \c{c}\fi{}a}}, \bibinfo {author}
  {\bibfnamefont {A.~M.}\ \bibnamefont {Souza}}, \bibinfo {author}
  {\bibfnamefont {R.~J.}\ \bibnamefont {de~Assis}}, \bibinfo {author}
  {\bibfnamefont {N.~G.}\ \bibnamefont {de~Almeida}}, \bibinfo {author}
  {\bibfnamefont {R.~S.}\ \bibnamefont {Sarthour}}, \bibinfo {author}
  {\bibfnamefont {I.~S.}\ \bibnamefont {Oliveira}},\ and\ \bibinfo {author}
  {\bibfnamefont {C.~J.}\ \bibnamefont {Villas-Boas}},\ }\bibfield  {title}
  {\bibinfo {title} {Reservoir engineering for maximally efficient quantum
  engines},\ }\href {https://doi.org/10.1103/PhysRevResearch.2.043419}
  {\bibfield  {journal} {\bibinfo  {journal} {Phys. Rev. Res.}\ }\textbf
  {\bibinfo {volume} {2}},\ \bibinfo {pages} {043419} (\bibinfo {year}
  {2020})}\BibitemShut {NoStop}%
\bibitem [{\citenamefont {Abah}\ and\ \citenamefont
  {Lutz}(2014)}]{Abah_EPL_2014}%
  \BibitemOpen
  \bibfield  {author} {\bibinfo {author} {\bibfnamefont {O.}~\bibnamefont
  {Abah}}\ and\ \bibinfo {author} {\bibfnamefont {E.}~\bibnamefont {Lutz}},\
  }\bibfield  {title} {\bibinfo {title} {Efficiency of heat engines coupled to
  nonequilibrium reservoirs},\ }\href
  {https://doi.org/10.1209/0295-5075/106/20001} {\bibfield  {journal} {\bibinfo
   {journal} {Europhysics Letters}\ }\textbf {\bibinfo {volume} {106}},\
  \bibinfo {pages} {20001} (\bibinfo {year} {2014})}\BibitemShut {NoStop}%
\bibitem [{\citenamefont {Dillenschneider}\ and\ \citenamefont
  {Lutz}(2009)}]{Dillenschneider_EPL_2009}%
  \BibitemOpen
  \bibfield  {author} {\bibinfo {author} {\bibfnamefont {R.}~\bibnamefont
  {Dillenschneider}}\ and\ \bibinfo {author} {\bibfnamefont {E.}~\bibnamefont
  {Lutz}},\ }\bibfield  {title} {\bibinfo {title} {Energetics of quantum
  correlations},\ }\href {https://doi.org/10.1209/0295-5075/88/50003}
  {\bibfield  {journal} {\bibinfo  {journal} {Europhysics Letters}\ }\textbf
  {\bibinfo {volume} {88}},\ \bibinfo {pages} {50003} (\bibinfo {year}
  {2009})}\BibitemShut {NoStop}%
\bibitem [{\citenamefont {Perarnau-Llobet}\ \emph {et~al.}(2015)\citenamefont
  {Perarnau-Llobet}, \citenamefont {Hovhannisyan}, \citenamefont {Huber},
  \citenamefont {Skrzypczyk}, \citenamefont {Brunner},\ and\ \citenamefont
  {Ac\'{\i}n}}]{Llobet_PRX_2015}%
  \BibitemOpen
  \bibfield  {author} {\bibinfo {author} {\bibfnamefont {M.}~\bibnamefont
  {Perarnau-Llobet}}, \bibinfo {author} {\bibfnamefont {K.~V.}\ \bibnamefont
  {Hovhannisyan}}, \bibinfo {author} {\bibfnamefont {M.}~\bibnamefont {Huber}},
  \bibinfo {author} {\bibfnamefont {P.}~\bibnamefont {Skrzypczyk}}, \bibinfo
  {author} {\bibfnamefont {N.}~\bibnamefont {Brunner}},\ and\ \bibinfo {author}
  {\bibfnamefont {A.}~\bibnamefont {Ac\'{\i}n}},\ }\bibfield  {title} {\bibinfo
  {title} {Extractable work from correlations},\ }\href
  {https://doi.org/10.1103/PhysRevX.5.041011} {\bibfield  {journal} {\bibinfo
  {journal} {Phys. Rev. X}\ }\textbf {\bibinfo {volume} {5}},\ \bibinfo {pages}
  {041011} (\bibinfo {year} {2015})}\BibitemShut {NoStop}%
\bibitem [{\citenamefont {de~Assis}\ \emph {et~al.}(2019)\citenamefont
  {de~Assis}, \citenamefont {de~Mendon\ifmmode~\mbox{\c{c}}\else \c{c}\fi{}a},
  \citenamefont {Villas-Boas}, \citenamefont {de~Souza}, \citenamefont
  {Sarthour}, \citenamefont {Oliveira},\ and\ \citenamefont
  {de~Almeida}}]{Assis_PRL_2019}%
  \BibitemOpen
  \bibfield  {author} {\bibinfo {author} {\bibfnamefont {R.~J.}\ \bibnamefont
  {de~Assis}}, \bibinfo {author} {\bibfnamefont {T.~M.}\ \bibnamefont
  {de~Mendon\ifmmode~\mbox{\c{c}}\else \c{c}\fi{}a}}, \bibinfo {author}
  {\bibfnamefont {C.~J.}\ \bibnamefont {Villas-Boas}}, \bibinfo {author}
  {\bibfnamefont {A.~M.}\ \bibnamefont {de~Souza}}, \bibinfo {author}
  {\bibfnamefont {R.~S.}\ \bibnamefont {Sarthour}}, \bibinfo {author}
  {\bibfnamefont {I.~S.}\ \bibnamefont {Oliveira}},\ and\ \bibinfo {author}
  {\bibfnamefont {N.~G.}\ \bibnamefont {de~Almeida}},\ }\bibfield  {title}
  {\bibinfo {title} {Efficiency of a quantum otto heat engine operating under a
  reservoir at effective negative temperatures},\ }\href
  {https://doi.org/10.1103/PhysRevLett.122.240602} {\bibfield  {journal}
  {\bibinfo  {journal} {Phys. Rev. Lett.}\ }\textbf {\bibinfo {volume} {122}},\
  \bibinfo {pages} {240602} (\bibinfo {year} {2019})}\BibitemShut {NoStop}%
\bibitem [{\citenamefont {Damas}\ \emph {et~al.}(2023)\citenamefont {Damas},
  \citenamefont {{de Assis}},\ and\ \citenamefont {{de
  Almeida}}}]{Gabriella_2023}%
  \BibitemOpen
  \bibfield  {author} {\bibinfo {author} {\bibfnamefont {G.~G.}\ \bibnamefont
  {Damas}}, \bibinfo {author} {\bibfnamefont {R.~J.}\ \bibnamefont {{de
  Assis}}},\ and\ \bibinfo {author} {\bibfnamefont {N.~G.}\ \bibnamefont {{de
  Almeida}}},\ }\bibfield  {title} {\bibinfo {title} {Negative temperature is
  cool for cooling},\ }\href
  {https://doi.org/https://doi.org/10.1016/j.physleta.2023.129038} {\bibfield
  {journal} {\bibinfo  {journal} {Physics Letters A}\ }\textbf {\bibinfo
  {volume} {482}},\ \bibinfo {pages} {129038} (\bibinfo {year}
  {2023})}\BibitemShut {NoStop}%
\bibitem [{\citenamefont {Maity}\ and\ \citenamefont
  {Sen(De)}()}]{Arghya_arxiv_2023}%
  \BibitemOpen
  \bibfield  {author} {\bibinfo {author} {\bibfnamefont {A.}~\bibnamefont
  {Maity}}\ and\ \bibinfo {author} {\bibfnamefont {A.}~\bibnamefont
  {Sen(De)}},\ }\href@noop {} {\bibinfo {title} {Enhanced efficiency in quantum
  otto engine via additional magnetic field and effective negative
  temperature}},\ \Eprint {https://arxiv.org/abs/2304.10420} {arXiv:2304.10420
  [quant-ph]} \BibitemShut {NoStop}%
\bibitem [{\citenamefont {Bera}\ \emph {et~al.}(2023)\citenamefont {Bera},
  \citenamefont {Pandit}, \citenamefont {Chatterjee}, \citenamefont {Singh},
  \citenamefont {Lewenstein}, \citenamefont {Bhattacharya},\ and\ \citenamefont
  {Bera}}]{Manab_arxiv_2023}%
  \BibitemOpen
  \bibfield  {author} {\bibinfo {author} {\bibfnamefont {M.~L.}\ \bibnamefont
  {Bera}}, \bibinfo {author} {\bibfnamefont {T.}~\bibnamefont {Pandit}},
  \bibinfo {author} {\bibfnamefont {K.}~\bibnamefont {Chatterjee}}, \bibinfo
  {author} {\bibfnamefont {V.}~\bibnamefont {Singh}}, \bibinfo {author}
  {\bibfnamefont {M.}~\bibnamefont {Lewenstein}}, \bibinfo {author}
  {\bibfnamefont {U.}~\bibnamefont {Bhattacharya}},\ and\ \bibinfo {author}
  {\bibfnamefont {M.~N.}\ \bibnamefont {Bera}},\ }\href@noop {} {\bibinfo
  {title} {Steady-state quantum thermodynamics with synthetic negative
  temperatures}} (\bibinfo {year} {2023}),\ \Eprint
  {https://arxiv.org/abs/2305.01215} {arXiv:2305.01215 [quant-ph]} \BibitemShut
  {NoStop}%
\bibitem [{\citenamefont {Struchtrup}(2018)}]{Struchtrup_PRL_2018}%
  \BibitemOpen
  \bibfield  {author} {\bibinfo {author} {\bibfnamefont {H.}~\bibnamefont
  {Struchtrup}},\ }\bibfield  {title} {\bibinfo {title} {Work storage in states
  of apparent negative thermodynamic temperature},\ }\href
  {https://doi.org/10.1103/PhysRevLett.120.250602} {\bibfield  {journal}
  {\bibinfo  {journal} {Phys. Rev. Lett.}\ }\textbf {\bibinfo {volume} {120}},\
  \bibinfo {pages} {250602} (\bibinfo {year} {2018})}\BibitemShut {NoStop}%
\bibitem [{\citenamefont {Ramsey}(1956)}]{Ramsey_PR_1956}%
  \BibitemOpen
  \bibfield  {author} {\bibinfo {author} {\bibfnamefont {N.~F.}\ \bibnamefont
  {Ramsey}},\ }\bibfield  {title} {\bibinfo {title} {Thermodynamics and
  statistical mechanics at negative absolute temperatures},\ }\href
  {https://doi.org/10.1103/PhysRev.103.20} {\bibfield  {journal} {\bibinfo
  {journal} {Phys. Rev.}\ }\textbf {\bibinfo {volume} {103}},\ \bibinfo {pages}
  {20} (\bibinfo {year} {1956})}\BibitemShut {NoStop}%
\bibitem [{\citenamefont {Purcell}\ and\ \citenamefont
  {Pound}(1951)}]{Purcell_PR_1951}%
  \BibitemOpen
  \bibfield  {author} {\bibinfo {author} {\bibfnamefont {E.~M.}\ \bibnamefont
  {Purcell}}\ and\ \bibinfo {author} {\bibfnamefont {R.~V.}\ \bibnamefont
  {Pound}},\ }\bibfield  {title} {\bibinfo {title} {A nuclear spin system at
  negative temperature},\ }\href {https://doi.org/10.1103/PhysRev.81.279}
  {\bibfield  {journal} {\bibinfo  {journal} {Phys. Rev.}\ }\textbf {\bibinfo
  {volume} {81}},\ \bibinfo {pages} {279} (\bibinfo {year} {1951})}\BibitemShut
  {NoStop}%
\bibitem [{\citenamefont {Carr}(2013)}]{Lincoln_Science_2013}%
  \BibitemOpen
  \bibfield  {author} {\bibinfo {author} {\bibfnamefont {L.~D.}\ \bibnamefont
  {Carr}},\ }\bibfield  {title} {\bibinfo {title} {Negative temperatures?},\
  }\href {https://doi.org/10.1126/science.1232558} {\bibfield  {journal}
  {\bibinfo  {journal} {Science}\ }\textbf {\bibinfo {volume} {339}},\ \bibinfo
  {pages} {42} (\bibinfo {year} {2013})}\BibitemShut {NoStop}%
\bibitem [{\citenamefont {Braun}\ \emph {et~al.}(2013)\citenamefont {Braun},
  \citenamefont {Ronzheimer}, \citenamefont {Schreiber}, \citenamefont
  {Hodgman}, \citenamefont {Rom}, \citenamefont {Bloch},\ and\ \citenamefont
  {Schneider}}]{Braun_Science_2013}%
  \BibitemOpen
  \bibfield  {author} {\bibinfo {author} {\bibfnamefont {S.}~\bibnamefont
  {Braun}}, \bibinfo {author} {\bibfnamefont {J.~P.}\ \bibnamefont
  {Ronzheimer}}, \bibinfo {author} {\bibfnamefont {M.}~\bibnamefont
  {Schreiber}}, \bibinfo {author} {\bibfnamefont {S.~S.}\ \bibnamefont
  {Hodgman}}, \bibinfo {author} {\bibfnamefont {T.}~\bibnamefont {Rom}},
  \bibinfo {author} {\bibfnamefont {I.}~\bibnamefont {Bloch}},\ and\ \bibinfo
  {author} {\bibfnamefont {U.}~\bibnamefont {Schneider}},\ }\bibfield  {title}
  {\bibinfo {title} {Negative absolute temperature for motional degrees of
  freedom},\ }\href {https://doi.org/10.1126/science.1227831} {\bibfield
  {journal} {\bibinfo  {journal} {Science}\ }\textbf {\bibinfo {volume}
  {339}},\ \bibinfo {pages} {52} (\bibinfo {year} {2013})}\BibitemShut
  {NoStop}%
\bibitem [{\citenamefont {Tacchino}\ \emph {et~al.}(2018)\citenamefont
  {Tacchino}, \citenamefont {Auff\`eves}, \citenamefont {Santos},\ and\
  \citenamefont {Gerace}}]{Tacchino_PRL_2018}%
  \BibitemOpen
  \bibfield  {author} {\bibinfo {author} {\bibfnamefont {F.}~\bibnamefont
  {Tacchino}}, \bibinfo {author} {\bibfnamefont {A.}~\bibnamefont
  {Auff\`eves}}, \bibinfo {author} {\bibfnamefont {M.~F.}\ \bibnamefont
  {Santos}},\ and\ \bibinfo {author} {\bibfnamefont {D.}~\bibnamefont
  {Gerace}},\ }\bibfield  {title} {\bibinfo {title} {Steady state entanglement
  beyond thermal limits},\ }\href
  {https://doi.org/10.1103/PhysRevLett.120.063604} {\bibfield  {journal}
  {\bibinfo  {journal} {Phys. Rev. Lett.}\ }\textbf {\bibinfo {volume} {120}},\
  \bibinfo {pages} {063604} (\bibinfo {year} {2018})}\BibitemShut {NoStop}%
\bibitem [{\citenamefont {Brask}\ and\ \citenamefont
  {Brunner}(2015)}]{Brask_PRE_2015}%
  \BibitemOpen
  \bibfield  {author} {\bibinfo {author} {\bibfnamefont {J.~B.}\ \bibnamefont
  {Brask}}\ and\ \bibinfo {author} {\bibfnamefont {N.}~\bibnamefont
  {Brunner}},\ }\bibfield  {title} {\bibinfo {title} {Small quantum absorption
  refrigerator in the transient regime: Time scales, enhanced cooling, and
  entanglement},\ }\href {https://doi.org/10.1103/PhysRevE.92.062101}
  {\bibfield  {journal} {\bibinfo  {journal} {Phys. Rev. E}\ }\textbf {\bibinfo
  {volume} {92}},\ \bibinfo {pages} {062101} (\bibinfo {year}
  {2015})}\BibitemShut {NoStop}%
\bibitem [{\citenamefont {Mitchison}\ \emph {et~al.}(2015)\citenamefont
  {Mitchison}, \citenamefont {Woods}, \citenamefont {Prior},\ and\
  \citenamefont {Huber}}]{Mitchison_NJP_2015}%
  \BibitemOpen
  \bibfield  {author} {\bibinfo {author} {\bibfnamefont {M.~T.}\ \bibnamefont
  {Mitchison}}, \bibinfo {author} {\bibfnamefont {M.~P.}\ \bibnamefont
  {Woods}}, \bibinfo {author} {\bibfnamefont {J.}~\bibnamefont {Prior}},\ and\
  \bibinfo {author} {\bibfnamefont {M.}~\bibnamefont {Huber}},\ }\bibfield
  {title} {\bibinfo {title} {Coherence-assisted single-shot cooling by quantum
  absorption refrigerators},\ }\href
  {https://doi.org/10.1088/1367-2630/17/11/115013} {\bibfield  {journal}
  {\bibinfo  {journal} {New Journal of Physics}\ }\textbf {\bibinfo {volume}
  {17}},\ \bibinfo {pages} {115013} (\bibinfo {year} {2015})}\BibitemShut
  {NoStop}%
\bibitem [{\citenamefont {Das}\ \emph {et~al.}(2019)\citenamefont {Das},
  \citenamefont {Misra}, \citenamefont {Pal}, \citenamefont {Sen(De)},\ and\
  \citenamefont {Sen}}]{Sreetama_EPL_2019}%
  \BibitemOpen
  \bibfield  {author} {\bibinfo {author} {\bibfnamefont {S.}~\bibnamefont
  {Das}}, \bibinfo {author} {\bibfnamefont {A.}~\bibnamefont {Misra}}, \bibinfo
  {author} {\bibfnamefont {A.~K.}\ \bibnamefont {Pal}}, \bibinfo {author}
  {\bibfnamefont {A.}~\bibnamefont {Sen(De)}},\ and\ \bibinfo {author}
  {\bibfnamefont {U.}~\bibnamefont {Sen}},\ }\bibfield  {title} {\bibinfo
  {title} {Necessarily transient quantum refrigerator},\ }\href
  {https://doi.org/10.1209/0295-5075/125/20007} {\bibfield  {journal} {\bibinfo
   {journal} {Europhysics Letters}\ }\textbf {\bibinfo {volume} {125}},\
  \bibinfo {pages} {20007} (\bibinfo {year} {2019})}\BibitemShut {NoStop}%
\bibitem [{\citenamefont {Ghosh}\ \emph {et~al.}(2021)\citenamefont {Ghosh},
  \citenamefont {Ghoshal},\ and\ \citenamefont {Sen}}]{Ghosh_PRA_2021}%
  \BibitemOpen
  \bibfield  {author} {\bibinfo {author} {\bibfnamefont {R.}~\bibnamefont
  {Ghosh}}, \bibinfo {author} {\bibfnamefont {A.}~\bibnamefont {Ghoshal}},\
  and\ \bibinfo {author} {\bibfnamefont {U.}~\bibnamefont {Sen}},\ }\bibfield
  {title} {\bibinfo {title} {Quantum thermal transistors: Operation
  characteristics in steady state versus transient regimes},\ }\href
  {https://doi.org/10.1103/PhysRevA.103.052613} {\bibfield  {journal} {\bibinfo
   {journal} {Phys. Rev. A}\ }\textbf {\bibinfo {volume} {103}},\ \bibinfo
  {pages} {052613} (\bibinfo {year} {2021})}\BibitemShut {NoStop}%
\bibitem [{\citenamefont {Saha}\ \emph {et~al.}()\citenamefont {Saha},
  \citenamefont {Ghoshal},\ and\ \citenamefont {Sen}}]{Saha_arxiv_2023}%
  \BibitemOpen
  \bibfield  {author} {\bibinfo {author} {\bibfnamefont {D.}~\bibnamefont
  {Saha}}, \bibinfo {author} {\bibfnamefont {A.}~\bibnamefont {Ghoshal}},\ and\
  \bibinfo {author} {\bibfnamefont {U.}~\bibnamefont {Sen}},\ }\href@noop {}
  {\bibinfo {title} {Temperature- and interaction-tweaked efficiency boost of
  finite-time robust quantum otto engines}},\ \Eprint
  {https://arxiv.org/abs/2309.11483} {arXiv:2309.11483 [quant-ph]} \BibitemShut
  {NoStop}%
\bibitem [{\citenamefont {Breuer}\ and\ \citenamefont
  {Petruccione}(2002)}]{Petruccione_book}%
  \BibitemOpen
  \bibfield  {author} {\bibinfo {author} {\bibfnamefont {H.~P.}\ \bibnamefont
  {Breuer}}\ and\ \bibinfo {author} {\bibfnamefont {F.}~\bibnamefont
  {Petruccione}},\ }\href@noop {} {\emph {\bibinfo {title} {The theory of open
  quantum systems}}}\ (\bibinfo  {publisher} {Oxford University Press},\
  \bibinfo {address} {New York},\ \bibinfo {year} {2002})\BibitemShut {NoStop}%
\bibitem [{\citenamefont {Alicki}\ and\ \citenamefont
  {Lendi}(2007)}]{Alicki_2007}%
  \BibitemOpen
  \bibfield  {author} {\bibinfo {author} {\bibfnamefont {R.}~\bibnamefont
  {Alicki}}\ and\ \bibinfo {author} {\bibfnamefont {K.}~\bibnamefont {Lendi}},\
  }\href@noop {} {\emph {\bibinfo {title} {Quantum Dynamical Semigroups and
  Applications}}}\ (\bibinfo  {publisher} {Springer, Berlin},\ \bibinfo {year}
  {2007})\BibitemShut {NoStop}%
\bibitem [{\citenamefont {Rivas}\ and\ \citenamefont
  {Huelga}(2011)}]{Rivas_Huelga_book}%
  \BibitemOpen
  \bibfield  {author} {\bibinfo {author} {\bibfnamefont {A.}~\bibnamefont
  {Rivas}}\ and\ \bibinfo {author} {\bibfnamefont {S.~F.}\ \bibnamefont
  {Huelga}},\ }\href
  {https://doi.org/https://doi.org/10.1007/978-3-642-23354-8} {\emph {\bibinfo
  {title} {Open Quantum Systems}}}\ (\bibinfo  {publisher} {Springer, Berlin},\
  \bibinfo {year} {2011})\BibitemShut {NoStop}%
\bibitem [{\citenamefont {Banerjee}(2018)}]{Subhashish_Banerjee_book}%
  \BibitemOpen
  \bibfield  {author} {\bibinfo {author} {\bibfnamefont {S.}~\bibnamefont
  {Banerjee}},\ }\href
  {https://doi.org/https://doi.org/10.1007/978-981-13-3182-4} {\emph {\bibinfo
  {title} {Open Quantum Systems:Dynamics of Nonclassical Evolution}}},\
  2366-8849\ (\bibinfo  {publisher} {Springer, Singapore},\ \bibinfo {year}
  {2018})\BibitemShut {NoStop}%
\bibitem [{\citenamefont {Lidar}()}]{Lidar_2020_lecture}%
  \BibitemOpen
  \bibfield  {author} {\bibinfo {author} {\bibfnamefont {D.~A.}\ \bibnamefont
  {Lidar}},\ }\href@noop {} {\bibinfo {title} {Lecture notes on the theory of
  open quantum systems}},\ \Eprint {https://arxiv.org/abs/1902.00967}
  {arXiv:1902.00967 [quant-ph]} \BibitemShut {NoStop}%
\bibitem [{\citenamefont {Gorini}\ \emph {et~al.}(1976)\citenamefont {Gorini},
  \citenamefont {Kossakowski},\ and\ \citenamefont
  {Sudarshan}}]{Sudarshan_JMP_1976}%
  \BibitemOpen
  \bibfield  {author} {\bibinfo {author} {\bibfnamefont {V.}~\bibnamefont
  {Gorini}}, \bibinfo {author} {\bibfnamefont {A.}~\bibnamefont
  {Kossakowski}},\ and\ \bibinfo {author} {\bibfnamefont {E.~C.~G.}\
  \bibnamefont {Sudarshan}},\ }\bibfield  {title} {\bibinfo {title}
  {{Completely positive dynamical semigroups of N‐level systems}},\ }\href
  {https://doi.org/10.1063/1.522979} {\bibfield  {journal} {\bibinfo  {journal}
  {Journal of Mathematical Physics}\ }\textbf {\bibinfo {volume} {17}},\
  \bibinfo {pages} {821} (\bibinfo {year} {1976})}\BibitemShut {NoStop}%
\bibitem [{\citenamefont {{Lindblad}}(1976)}]{Lindblad_CMP_1976}%
  \BibitemOpen
  \bibfield  {author} {\bibinfo {author} {\bibfnamefont {G.}~\bibnamefont
  {{Lindblad}}},\ }\bibfield  {title} {\bibinfo {title} {{On the generators of
  quantum dynamical semigroups}},\ }\href {https://doi.org/10.1007/BF01608499}
  {\bibfield  {journal} {\bibinfo  {journal} {Communications in Mathematical
  Physics}\ }\textbf {\bibinfo {volume} {48}},\ \bibinfo {pages} {119}
  (\bibinfo {year} {1976})}\BibitemShut {NoStop}%
\bibitem [{\citenamefont {Esposito}\ \emph {et~al.}(2010)\citenamefont
  {Esposito}, \citenamefont {Lindenberg},\ and\ \citenamefont {den
  Broeck}}]{Esposito_NJP_2010}%
  \BibitemOpen
  \bibfield  {author} {\bibinfo {author} {\bibfnamefont {M.}~\bibnamefont
  {Esposito}}, \bibinfo {author} {\bibfnamefont {K.}~\bibnamefont
  {Lindenberg}},\ and\ \bibinfo {author} {\bibfnamefont {C.~V.}\ \bibnamefont
  {den Broeck}},\ }\bibfield  {title} {\bibinfo {title} {Entropy production as
  correlation between system and reservoir},\ }\href
  {https://doi.org/10.1088/1367-2630/12/1/013013} {\bibfield  {journal}
  {\bibinfo  {journal} {New Journal of Physics}\ }\textbf {\bibinfo {volume}
  {12}},\ \bibinfo {pages} {013013} (\bibinfo {year} {2010})}\BibitemShut
  {NoStop}%
\bibitem [{\citenamefont {Breuer}\ \emph {et~al.}(2016)\citenamefont {Breuer},
  \citenamefont {Laine}, \citenamefont {Piilo},\ and\ \citenamefont
  {Vacchini}}]{Breuer_RMP_2016}%
  \BibitemOpen
  \bibfield  {author} {\bibinfo {author} {\bibfnamefont {H.-P.}\ \bibnamefont
  {Breuer}}, \bibinfo {author} {\bibfnamefont {E.-M.}\ \bibnamefont {Laine}},
  \bibinfo {author} {\bibfnamefont {J.}~\bibnamefont {Piilo}},\ and\ \bibinfo
  {author} {\bibfnamefont {B.}~\bibnamefont {Vacchini}},\ }\bibfield  {title}
  {\bibinfo {title} {Colloquium: Non-{M}arkovian dynamics in open quantum
  systems},\ }\href {https://doi.org/10.1103/RevModPhys.88.021002} {\bibfield
  {journal} {\bibinfo  {journal} {Rev. Mod. Phys.}\ }\textbf {\bibinfo {volume}
  {88}},\ \bibinfo {pages} {021002} (\bibinfo {year} {2016})}\BibitemShut
  {NoStop}%
\bibitem [{\citenamefont {Seifert}(2016)}]{Seifert_PRL_2016}%
  \BibitemOpen
  \bibfield  {author} {\bibinfo {author} {\bibfnamefont {U.}~\bibnamefont
  {Seifert}},\ }\bibfield  {title} {\bibinfo {title} {First and second law of
  thermodynamics at strong coupling},\ }\href
  {https://doi.org/10.1103/PhysRevLett.116.020601} {\bibfield  {journal}
  {\bibinfo  {journal} {Phys. Rev. Lett.}\ }\textbf {\bibinfo {volume} {116}},\
  \bibinfo {pages} {020601} (\bibinfo {year} {2016})}\BibitemShut {NoStop}%
\bibitem [{\citenamefont {Strasberg}\ \emph {et~al.}(2017)\citenamefont
  {Strasberg}, \citenamefont {Schaller}, \citenamefont {Brandes},\ and\
  \citenamefont {Esposito}}]{Strasberg_PRX_2017}%
  \BibitemOpen
  \bibfield  {author} {\bibinfo {author} {\bibfnamefont {P.}~\bibnamefont
  {Strasberg}}, \bibinfo {author} {\bibfnamefont {G.}~\bibnamefont {Schaller}},
  \bibinfo {author} {\bibfnamefont {T.}~\bibnamefont {Brandes}},\ and\ \bibinfo
  {author} {\bibfnamefont {M.}~\bibnamefont {Esposito}},\ }\bibfield  {title}
  {\bibinfo {title} {Quantum and information thermodynamics: A unifying
  framework based on repeated interactions},\ }\href
  {https://doi.org/10.1103/PhysRevX.7.021003} {\bibfield  {journal} {\bibinfo
  {journal} {Phys. Rev. X}\ }\textbf {\bibinfo {volume} {7}},\ \bibinfo {pages}
  {021003} (\bibinfo {year} {2017})}\BibitemShut {NoStop}%
\bibitem [{\citenamefont {Rivas}(2020)}]{Rivas_PRL_2020}%
  \BibitemOpen
  \bibfield  {author} {\bibinfo {author} {\bibfnamefont {A.}~\bibnamefont
  {Rivas}},\ }\bibfield  {title} {\bibinfo {title} {Strong coupling
  thermodynamics of open quantum systems},\ }\href
  {https://doi.org/10.1103/PhysRevLett.124.160601} {\bibfield  {journal}
  {\bibinfo  {journal} {Phys. Rev. Lett.}\ }\textbf {\bibinfo {volume} {124}},\
  \bibinfo {pages} {160601} (\bibinfo {year} {2020})}\BibitemShut {NoStop}%
\bibitem [{\citenamefont {Landi}\ and\ \citenamefont
  {Paternostro}(2021)}]{Landi_RMP_2021}%
  \BibitemOpen
  \bibfield  {author} {\bibinfo {author} {\bibfnamefont {G.~T.}\ \bibnamefont
  {Landi}}\ and\ \bibinfo {author} {\bibfnamefont {M.}~\bibnamefont
  {Paternostro}},\ }\bibfield  {title} {\bibinfo {title} {Irreversible entropy
  production: From classical to quantum},\ }\href
  {https://doi.org/10.1103/RevModPhys.93.035008} {\bibfield  {journal}
  {\bibinfo  {journal} {Rev. Mod. Phys.}\ }\textbf {\bibinfo {volume} {93}},\
  \bibinfo {pages} {035008} (\bibinfo {year} {2021})}\BibitemShut {NoStop}%
\bibitem [{\citenamefont {Zhang}\ \emph {et~al.}(2014)\citenamefont {Zhang},
  \citenamefont {Huang},\ and\ \citenamefont {Yi}}]{Zhang_JOP_2014}%
  \BibitemOpen
  \bibfield  {author} {\bibinfo {author} {\bibfnamefont {X.~Y.}\ \bibnamefont
  {Zhang}}, \bibinfo {author} {\bibfnamefont {X.~L.}\ \bibnamefont {Huang}},\
  and\ \bibinfo {author} {\bibfnamefont {X.~X.}\ \bibnamefont {Yi}},\
  }\bibfield  {title} {\bibinfo {title} {Quantum otto heat engine with a
  non-{M}arkovian reservoir},\ }\href
  {https://doi.org/10.1088/1751-8113/47/45/455002} {\bibfield  {journal}
  {\bibinfo  {journal} {Journal of Physics A: Mathematical and Theoretical}\
  }\textbf {\bibinfo {volume} {47}},\ \bibinfo {pages} {455002} (\bibinfo
  {year} {2014})}\BibitemShut {NoStop}%
\bibitem [{\citenamefont {Thomas}\ \emph {et~al.}(2018)\citenamefont {Thomas},
  \citenamefont {Siddharth}, \citenamefont {Banerjee},\ and\ \citenamefont
  {Ghosh}}]{Sibasish-Ghosh_PRE_2018}%
  \BibitemOpen
  \bibfield  {author} {\bibinfo {author} {\bibfnamefont {G.}~\bibnamefont
  {Thomas}}, \bibinfo {author} {\bibfnamefont {N.}~\bibnamefont {Siddharth}},
  \bibinfo {author} {\bibfnamefont {S.}~\bibnamefont {Banerjee}},\ and\
  \bibinfo {author} {\bibfnamefont {S.}~\bibnamefont {Ghosh}},\ }\bibfield
  {title} {\bibinfo {title} {Thermodynamics of non-{M}arkovian reservoirs and
  heat engines},\ }\href {https://doi.org/10.1103/PhysRevE.97.062108}
  {\bibfield  {journal} {\bibinfo  {journal} {Phys. Rev. E}\ }\textbf {\bibinfo
  {volume} {97}},\ \bibinfo {pages} {062108} (\bibinfo {year}
  {2018})}\BibitemShut {NoStop}%
\bibitem [{\citenamefont {Pozas-Kerstjens}\ \emph {et~al.}(2018)\citenamefont
  {Pozas-Kerstjens}, \citenamefont {Brown},\ and\ \citenamefont
  {Hovhannisyan}}]{Pozas_NJP_2018}%
  \BibitemOpen
  \bibfield  {author} {\bibinfo {author} {\bibfnamefont {A.}~\bibnamefont
  {Pozas-Kerstjens}}, \bibinfo {author} {\bibfnamefont {E.~G.}\ \bibnamefont
  {Brown}},\ and\ \bibinfo {author} {\bibfnamefont {K.~V.}\ \bibnamefont
  {Hovhannisyan}},\ }\bibfield  {title} {\bibinfo {title} {A quantum otto
  engine with finite heat baths: energy, correlations, and degradation},\
  }\href {https://doi.org/10.1088/1367-2630/aaba02} {\bibfield  {journal}
  {\bibinfo  {journal} {New Journal of Physics}\ }\textbf {\bibinfo {volume}
  {20}},\ \bibinfo {pages} {043034} (\bibinfo {year} {2018})}\BibitemShut
  {NoStop}%
\bibitem [{\citenamefont {Pezzutto}\ \emph {et~al.}(2019)\citenamefont
  {Pezzutto}, \citenamefont {Paternostro},\ and\ \citenamefont
  {Omar}}]{Pezzutto_QST_2019}%
  \BibitemOpen
  \bibfield  {author} {\bibinfo {author} {\bibfnamefont {M.}~\bibnamefont
  {Pezzutto}}, \bibinfo {author} {\bibfnamefont {M.}~\bibnamefont
  {Paternostro}},\ and\ \bibinfo {author} {\bibfnamefont {Y.}~\bibnamefont
  {Omar}},\ }\bibfield  {title} {\bibinfo {title} {An out-of-equilibrium
  non-{M}arkovian quantum heat engine},\ }\href
  {https://doi.org/10.1088/2058-9565/aaf5b4} {\bibfield  {journal} {\bibinfo
  {journal} {Quantum Science and Technology}\ }\textbf {\bibinfo {volume}
  {4}},\ \bibinfo {pages} {025002} (\bibinfo {year} {2019})}\BibitemShut
  {NoStop}%
\bibitem [{\citenamefont {Wiedmann}\ \emph {et~al.}(2020)\citenamefont
  {Wiedmann}, \citenamefont {Stockburger},\ and\ \citenamefont
  {Ankerhold}}]{Wiedmann_NJP_2020}%
  \BibitemOpen
  \bibfield  {author} {\bibinfo {author} {\bibfnamefont {M.}~\bibnamefont
  {Wiedmann}}, \bibinfo {author} {\bibfnamefont {J.~T.}\ \bibnamefont
  {Stockburger}},\ and\ \bibinfo {author} {\bibfnamefont {J.}~\bibnamefont
  {Ankerhold}},\ }\bibfield  {title} {\bibinfo {title} {Non-{M}arkovian
  dynamics of a quantum heat engine: out-of-equilibrium operation and thermal
  coupling control},\ }\href {https://doi.org/10.1088/1367-2630/ab725a}
  {\bibfield  {journal} {\bibinfo  {journal} {New Journal of Physics}\ }\textbf
  {\bibinfo {volume} {22}},\ \bibinfo {pages} {033007} (\bibinfo {year}
  {2020})}\BibitemShut {NoStop}%
\bibitem [{\citenamefont {Mukherjee}\ \emph {et~al.}(2020)\citenamefont
  {Mukherjee}, \citenamefont {Kofman},\ and\ \citenamefont
  {Kurizki}}]{Victor_CP_2020}%
  \BibitemOpen
  \bibfield  {author} {\bibinfo {author} {\bibfnamefont {V.}~\bibnamefont
  {Mukherjee}}, \bibinfo {author} {\bibfnamefont {A.~G.}\ \bibnamefont
  {Kofman}},\ and\ \bibinfo {author} {\bibfnamefont {G.}~\bibnamefont
  {Kurizki}},\ }\bibfield  {title} {\bibinfo {title} {Anti-{Z}eno quantum
  advantage in fast-driven heat machines},\ }\href
  {https://doi.org/10.1038/s42005-019-0272-z} {\bibfield  {journal} {\bibinfo
  {journal} {Communications Physics}\ }\textbf {\bibinfo {volume} {3}},\
  \bibinfo {pages} {8} (\bibinfo {year} {2020})}\BibitemShut {NoStop}%
\bibitem [{\citenamefont {Liu}\ \emph {et~al.}(2021)\citenamefont {Liu},
  \citenamefont {Jung},\ and\ \citenamefont {Segal}}]{Segal_PRL_2021}%
  \BibitemOpen
  \bibfield  {author} {\bibinfo {author} {\bibfnamefont {J.}~\bibnamefont
  {Liu}}, \bibinfo {author} {\bibfnamefont {K.~A.}\ \bibnamefont {Jung}},\ and\
  \bibinfo {author} {\bibfnamefont {D.}~\bibnamefont {Segal}},\ }\bibfield
  {title} {\bibinfo {title} {Periodically driven quantum thermal machines from
  warming up to limit cycle},\ }\href
  {https://doi.org/10.1103/PhysRevLett.127.200602} {\bibfield  {journal}
  {\bibinfo  {journal} {Phys. Rev. Lett.}\ }\textbf {\bibinfo {volume} {127}},\
  \bibinfo {pages} {200602} (\bibinfo {year} {2021})}\BibitemShut {NoStop}%
\bibitem [{\citenamefont {Chakraborty}\ \emph {et~al.}(2022)\citenamefont
  {Chakraborty}, \citenamefont {Das},\ and\ \citenamefont {Chru\ifmmode
  \acute{s}\else \'{s}\fi{}ci\ifmmode~\acute{n}\else
  \'{n}\fi{}ski}}]{Sagnik_PRE_2022}%
  \BibitemOpen
  \bibfield  {author} {\bibinfo {author} {\bibfnamefont {S.}~\bibnamefont
  {Chakraborty}}, \bibinfo {author} {\bibfnamefont {A.}~\bibnamefont {Das}},\
  and\ \bibinfo {author} {\bibfnamefont {D.}~\bibnamefont {Chru\ifmmode
  \acute{s}\else \'{s}\fi{}ci\ifmmode~\acute{n}\else \'{n}\fi{}ski}},\
  }\bibfield  {title} {\bibinfo {title} {Strongly coupled quantum otto cycle
  with single qubit bath},\ }\href
  {https://doi.org/10.1103/PhysRevE.106.064133} {\bibfield  {journal} {\bibinfo
   {journal} {Phys. Rev. E}\ }\textbf {\bibinfo {volume} {106}},\ \bibinfo
  {pages} {064133} (\bibinfo {year} {2022})}\BibitemShut {NoStop}%
\bibitem [{\citenamefont {Kaneyasu}\ and\ \citenamefont
  {Hasegawa}(2023)}]{Mao_PRE_2023}%
  \BibitemOpen
  \bibfield  {author} {\bibinfo {author} {\bibfnamefont {M.}~\bibnamefont
  {Kaneyasu}}\ and\ \bibinfo {author} {\bibfnamefont {Y.}~\bibnamefont
  {Hasegawa}},\ }\bibfield  {title} {\bibinfo {title} {Quantum otto cycle under
  strong coupling},\ }\href {https://doi.org/10.1103/PhysRevE.107.044127}
  {\bibfield  {journal} {\bibinfo  {journal} {Phys. Rev. E}\ }\textbf {\bibinfo
  {volume} {107}},\ \bibinfo {pages} {044127} (\bibinfo {year}
  {2023})}\BibitemShut {NoStop}%
\bibitem [{\citenamefont {Ishizaki}\ \emph {et~al.}(2023)\citenamefont
  {Ishizaki}, \citenamefont {Hatano},\ and\ \citenamefont
  {Tajima}}]{Miku_PRR_2023}%
  \BibitemOpen
  \bibfield  {author} {\bibinfo {author} {\bibfnamefont {M.}~\bibnamefont
  {Ishizaki}}, \bibinfo {author} {\bibfnamefont {N.}~\bibnamefont {Hatano}},\
  and\ \bibinfo {author} {\bibfnamefont {H.}~\bibnamefont {Tajima}},\
  }\bibfield  {title} {\bibinfo {title} {Switching the function of the quantum
  otto cycle in non-{M}arkovian dynamics: Heat engine, heater, and heat pump},\
  }\href {https://doi.org/10.1103/PhysRevResearch.5.023066} {\bibfield
  {journal} {\bibinfo  {journal} {Phys. Rev. Res.}\ }\textbf {\bibinfo {volume}
  {5}},\ \bibinfo {pages} {023066} (\bibinfo {year} {2023})}\BibitemShut
  {NoStop}%
\bibitem [{\citenamefont {Nimmrichter}\ \emph {et~al.}(2017)\citenamefont
  {Nimmrichter}, \citenamefont {Dai}, \citenamefont {Roulet},\ and\
  \citenamefont {Scarani}}]{Valerio_Quantum_2017}%
  \BibitemOpen
  \bibfield  {author} {\bibinfo {author} {\bibfnamefont {S.}~\bibnamefont
  {Nimmrichter}}, \bibinfo {author} {\bibfnamefont {J.}~\bibnamefont {Dai}},
  \bibinfo {author} {\bibfnamefont {A.}~\bibnamefont {Roulet}},\ and\ \bibinfo
  {author} {\bibfnamefont {V.}~\bibnamefont {Scarani}},\ }\bibfield  {title}
  {\bibinfo {title} {Quantum and classical dynamics of a three-mode absorption
  refrigerator},\ }\href {https://doi.org/10.22331/q-2017-12-11-37} {\bibfield
  {journal} {\bibinfo  {journal} {{Quantum}}\ }\textbf {\bibinfo {volume}
  {1}},\ \bibinfo {pages} {37} (\bibinfo {year} {2017})}\BibitemShut {NoStop}%
\bibitem [{\citenamefont {Peterson}\ \emph {et~al.}(2019)\citenamefont
  {Peterson}, \citenamefont {Batalh\~ao}, \citenamefont {Herrera},
  \citenamefont {Souza}, \citenamefont {Sarthour}, \citenamefont {Oliveira},\
  and\ \citenamefont {Serra}}]{Peterson_PRL_2019}%
  \BibitemOpen
  \bibfield  {author} {\bibinfo {author} {\bibfnamefont {J.~P.~S.}\
  \bibnamefont {Peterson}}, \bibinfo {author} {\bibfnamefont {T.~B.}\
  \bibnamefont {Batalh\~ao}}, \bibinfo {author} {\bibfnamefont
  {M.}~\bibnamefont {Herrera}}, \bibinfo {author} {\bibfnamefont {A.~M.}\
  \bibnamefont {Souza}}, \bibinfo {author} {\bibfnamefont {R.~S.}\ \bibnamefont
  {Sarthour}}, \bibinfo {author} {\bibfnamefont {I.~S.}\ \bibnamefont
  {Oliveira}},\ and\ \bibinfo {author} {\bibfnamefont {R.~M.}\ \bibnamefont
  {Serra}},\ }\bibfield  {title} {\bibinfo {title} {Experimental
  characterization of a spin quantum heat engine},\ }\href
  {https://doi.org/10.1103/PhysRevLett.123.240601} {\bibfield  {journal}
  {\bibinfo  {journal} {Phys. Rev. Lett.}\ }\textbf {\bibinfo {volume} {123}},\
  \bibinfo {pages} {240601} (\bibinfo {year} {2019})}\BibitemShut {NoStop}%
\bibitem [{\citenamefont {Denzler}\ and\ \citenamefont
  {Lutz}(2020)}]{Denzler_PRR_2020}%
  \BibitemOpen
  \bibfield  {author} {\bibinfo {author} {\bibfnamefont {T.}~\bibnamefont
  {Denzler}}\ and\ \bibinfo {author} {\bibfnamefont {E.}~\bibnamefont {Lutz}},\
  }\bibfield  {title} {\bibinfo {title} {Efficiency fluctuations of a quantum
  heat engine},\ }\href {https://doi.org/10.1103/PhysRevResearch.2.032062}
  {\bibfield  {journal} {\bibinfo  {journal} {Phys. Rev. Res.}\ }\textbf
  {\bibinfo {volume} {2}},\ \bibinfo {pages} {032062} (\bibinfo {year}
  {2020})}\BibitemShut {NoStop}%
\bibitem [{\citenamefont {Batalh\~ao}\ \emph {et~al.}(2014)\citenamefont
  {Batalh\~ao}, \citenamefont {Souza}, \citenamefont {Mazzola}, \citenamefont
  {Auccaise}, \citenamefont {Sarthour}, \citenamefont {Oliveira}, \citenamefont
  {Goold}, \citenamefont {De~Chiara}, \citenamefont {Paternostro},\ and\
  \citenamefont {Serra}}]{T.B.Batalhao_PRL_2014}%
  \BibitemOpen
  \bibfield  {author} {\bibinfo {author} {\bibfnamefont {T.~B.}\ \bibnamefont
  {Batalh\~ao}}, \bibinfo {author} {\bibfnamefont {A.~M.}\ \bibnamefont
  {Souza}}, \bibinfo {author} {\bibfnamefont {L.}~\bibnamefont {Mazzola}},
  \bibinfo {author} {\bibfnamefont {R.}~\bibnamefont {Auccaise}}, \bibinfo
  {author} {\bibfnamefont {R.~S.}\ \bibnamefont {Sarthour}}, \bibinfo {author}
  {\bibfnamefont {I.~S.}\ \bibnamefont {Oliveira}}, \bibinfo {author}
  {\bibfnamefont {J.}~\bibnamefont {Goold}}, \bibinfo {author} {\bibfnamefont
  {G.}~\bibnamefont {De~Chiara}}, \bibinfo {author} {\bibfnamefont
  {M.}~\bibnamefont {Paternostro}},\ and\ \bibinfo {author} {\bibfnamefont
  {R.~M.}\ \bibnamefont {Serra}},\ }\bibfield  {title} {\bibinfo {title}
  {Experimental reconstruction of work distribution and study of fluctuation
  relations in a closed quantum system},\ }\href
  {https://doi.org/10.1103/PhysRevLett.113.140601} {\bibfield  {journal}
  {\bibinfo  {journal} {Phys. Rev. Lett.}\ }\textbf {\bibinfo {volume} {113}},\
  \bibinfo {pages} {140601} (\bibinfo {year} {2014})}\BibitemShut {NoStop}%
\bibitem [{\citenamefont {Batalh\~ao}\ \emph {et~al.}(2015)\citenamefont
  {Batalh\~ao}, \citenamefont {Souza}, \citenamefont {Sarthour}, \citenamefont
  {Oliveira}, \citenamefont {Paternostro}, \citenamefont {Lutz},\ and\
  \citenamefont {Serra}}]{T.B.Batalhao_PRL_2015}%
  \BibitemOpen
  \bibfield  {author} {\bibinfo {author} {\bibfnamefont {T.~B.}\ \bibnamefont
  {Batalh\~ao}}, \bibinfo {author} {\bibfnamefont {A.~M.}\ \bibnamefont
  {Souza}}, \bibinfo {author} {\bibfnamefont {R.~S.}\ \bibnamefont {Sarthour}},
  \bibinfo {author} {\bibfnamefont {I.~S.}\ \bibnamefont {Oliveira}}, \bibinfo
  {author} {\bibfnamefont {M.}~\bibnamefont {Paternostro}}, \bibinfo {author}
  {\bibfnamefont {E.}~\bibnamefont {Lutz}},\ and\ \bibinfo {author}
  {\bibfnamefont {R.~M.}\ \bibnamefont {Serra}},\ }\bibfield  {title} {\bibinfo
  {title} {Irreversibility and the arrow of time in a quenched quantum
  system},\ }\href {https://doi.org/10.1103/PhysRevLett.115.190601} {\bibfield
  {journal} {\bibinfo  {journal} {Phys. Rev. Lett.}\ }\textbf {\bibinfo
  {volume} {115}},\ \bibinfo {pages} {190601} (\bibinfo {year}
  {2015})}\BibitemShut {NoStop}%
\bibitem [{\citenamefont {Tu}\ and\ \citenamefont {Zhang}(2008)}]{Tu_PRB_2008}%
  \BibitemOpen
  \bibfield  {author} {\bibinfo {author} {\bibfnamefont {M.~W.~Y.}\
  \bibnamefont {Tu}}\ and\ \bibinfo {author} {\bibfnamefont {W.-M.}\
  \bibnamefont {Zhang}},\ }\bibfield  {title} {\bibinfo {title}
  {Non-{M}arkovian decoherence theory for a double-dot charge qubit},\ }\href
  {https://doi.org/10.1103/PhysRevB.78.235311} {\bibfield  {journal} {\bibinfo
  {journal} {Phys. Rev. B}\ }\textbf {\bibinfo {volume} {78}},\ \bibinfo
  {pages} {235311} (\bibinfo {year} {2008})}\BibitemShut {NoStop}%
\bibitem [{\citenamefont {Tu}\ \emph {et~al.}(2009)\citenamefont {Tu},
  \citenamefont {Lee},\ and\ \citenamefont {Zhang}}]{Tu_QIP_2009}%
  \BibitemOpen
  \bibfield  {author} {\bibinfo {author} {\bibfnamefont {M.~W.-Y.}\
  \bibnamefont {Tu}}, \bibinfo {author} {\bibfnamefont {M.-T.}\ \bibnamefont
  {Lee}},\ and\ \bibinfo {author} {\bibfnamefont {W.-M.}\ \bibnamefont
  {Zhang}},\ }\bibfield  {title} {\bibinfo {title} {Exact master equation and
  non-{M}arkovian decoherence for quantum dot quantum computing},\ }\href
  {https://doi.org/10.1007/s11128-009-0143-8} {\bibfield  {journal} {\bibinfo
  {journal} {Quantum Information Processing}\ }\textbf {\bibinfo {volume}
  {8}},\ \bibinfo {pages} {631} (\bibinfo {year} {2009})}\BibitemShut {NoStop}%
\bibitem [{\citenamefont {Xiong}\ \emph {et~al.}(2010)\citenamefont {Xiong},
  \citenamefont {Zhang}, \citenamefont {Wang},\ and\ \citenamefont
  {Wu}}]{Xiong_PRA_2010}%
  \BibitemOpen
  \bibfield  {author} {\bibinfo {author} {\bibfnamefont {H.-N.}\ \bibnamefont
  {Xiong}}, \bibinfo {author} {\bibfnamefont {W.-M.}\ \bibnamefont {Zhang}},
  \bibinfo {author} {\bibfnamefont {X.}~\bibnamefont {Wang}},\ and\ \bibinfo
  {author} {\bibfnamefont {M.-H.}\ \bibnamefont {Wu}},\ }\bibfield  {title}
  {\bibinfo {title} {Exact non-{M}arkovian cavity dynamics strongly coupled to
  a reservoir},\ }\href {https://doi.org/10.1103/PhysRevA.82.012105} {\bibfield
   {journal} {\bibinfo  {journal} {Phys. Rev. A}\ }\textbf {\bibinfo {volume}
  {82}},\ \bibinfo {pages} {012105} (\bibinfo {year} {2010})}\BibitemShut
  {NoStop}%
\bibitem [{\citenamefont {Wu}\ \emph {et~al.}(2010)\citenamefont {Wu},
  \citenamefont {Lei}, \citenamefont {Zhang},\ and\ \citenamefont
  {Xiong}}]{Wu_OE_2010}%
  \BibitemOpen
  \bibfield  {author} {\bibinfo {author} {\bibfnamefont {M.-H.}\ \bibnamefont
  {Wu}}, \bibinfo {author} {\bibfnamefont {C.~U.}\ \bibnamefont {Lei}},
  \bibinfo {author} {\bibfnamefont {W.-M.}\ \bibnamefont {Zhang}},\ and\
  \bibinfo {author} {\bibfnamefont {H.-N.}\ \bibnamefont {Xiong}},\ }\bibfield
  {title} {\bibinfo {title} {Non-{M}arkovian dynamics of a microcavity coupled
  to a waveguide in photonic crystals},\ }\href
  {https://doi.org/10.1364/OE.18.018407} {\bibfield  {journal} {\bibinfo
  {journal} {Opt. Express}\ }\textbf {\bibinfo {volume} {18}},\ \bibinfo
  {pages} {18407} (\bibinfo {year} {2010})}\BibitemShut {NoStop}%
\bibitem [{\citenamefont {Jin}\ \emph {et~al.}(2010)\citenamefont {Jin},
  \citenamefont {Tu}, \citenamefont {Zhang},\ and\ \citenamefont
  {Yan}}]{Jin_NJP_2010}%
  \BibitemOpen
  \bibfield  {author} {\bibinfo {author} {\bibfnamefont {J.}~\bibnamefont
  {Jin}}, \bibinfo {author} {\bibfnamefont {M.~W.-Y.}\ \bibnamefont {Tu}},
  \bibinfo {author} {\bibfnamefont {W.-M.}\ \bibnamefont {Zhang}},\ and\
  \bibinfo {author} {\bibfnamefont {Y.}~\bibnamefont {Yan}},\ }\bibfield
  {title} {\bibinfo {title} {Non-equilibrium quantum theory for nanodevices
  based on the {F}eynman–{V}ernon influence functional},\ }\href
  {https://doi.org/10.1088/1367-2630/12/8/083013} {\bibfield  {journal}
  {\bibinfo  {journal} {New Journal of Physics}\ }\textbf {\bibinfo {volume}
  {12}},\ \bibinfo {pages} {083013} (\bibinfo {year} {2010})}\BibitemShut
  {NoStop}%
\bibitem [{\citenamefont {Lei}\ and\ \citenamefont
  {Zhang}(2011)}]{Lei_PRA_2011}%
  \BibitemOpen
  \bibfield  {author} {\bibinfo {author} {\bibfnamefont {C.~U.}\ \bibnamefont
  {Lei}}\ and\ \bibinfo {author} {\bibfnamefont {W.-M.}\ \bibnamefont
  {Zhang}},\ }\bibfield  {title} {\bibinfo {title} {Decoherence suppression of
  open quantum systems through a strong coupling to non-markovian reservoirs},\
  }\href {https://doi.org/10.1103/PhysRevA.84.052116} {\bibfield  {journal}
  {\bibinfo  {journal} {Phys. Rev. A}\ }\textbf {\bibinfo {volume} {84}},\
  \bibinfo {pages} {052116} (\bibinfo {year} {2011})}\BibitemShut {NoStop}%
\bibitem [{\citenamefont {Lei}\ and\ \citenamefont
  {Zhang}(2012)}]{Lei_AP_2012}%
  \BibitemOpen
  \bibfield  {author} {\bibinfo {author} {\bibfnamefont {C.~U.}\ \bibnamefont
  {Lei}}\ and\ \bibinfo {author} {\bibfnamefont {W.-M.}\ \bibnamefont
  {Zhang}},\ }\bibfield  {title} {\bibinfo {title} {A quantum photonic
  dissipative transport theory},\ }\href
  {https://doi.org/https://doi.org/10.1016/j.aop.2012.02.005} {\bibfield
  {journal} {\bibinfo  {journal} {Annals of Physics}\ }\textbf {\bibinfo
  {volume} {327}},\ \bibinfo {pages} {1408} (\bibinfo {year}
  {2012})}\BibitemShut {NoStop}%
\bibitem [{\citenamefont {Zhang}\ \emph {et~al.}(2012)\citenamefont {Zhang},
  \citenamefont {Lo}, \citenamefont {Xiong}, \citenamefont {Tu},\ and\
  \citenamefont {Nori}}]{Nori_PRL_2012}%
  \BibitemOpen
  \bibfield  {author} {\bibinfo {author} {\bibfnamefont {W.-M.}\ \bibnamefont
  {Zhang}}, \bibinfo {author} {\bibfnamefont {P.-Y.}\ \bibnamefont {Lo}},
  \bibinfo {author} {\bibfnamefont {H.-N.}\ \bibnamefont {Xiong}}, \bibinfo
  {author} {\bibfnamefont {M.~W.-Y.}\ \bibnamefont {Tu}},\ and\ \bibinfo
  {author} {\bibfnamefont {F.}~\bibnamefont {Nori}},\ }\bibfield  {title}
  {\bibinfo {title} {General non-markovian dynamics of open quantum systems},\
  }\href {https://doi.org/10.1103/PhysRevLett.109.170402} {\bibfield  {journal}
  {\bibinfo  {journal} {Phys. Rev. Lett.}\ }\textbf {\bibinfo {volume} {109}},\
  \bibinfo {pages} {170402} (\bibinfo {year} {2012})}\BibitemShut {NoStop}%
\bibitem [{\citenamefont {Weber}(1956)}]{Weber_PR_1956}%
  \BibitemOpen
  \bibfield  {author} {\bibinfo {author} {\bibfnamefont {J.}~\bibnamefont
  {Weber}},\ }\bibfield  {title} {\bibinfo {title} {Fluctuation dissipation
  theorem},\ }\href {https://doi.org/10.1103/PhysRev.101.1620} {\bibfield
  {journal} {\bibinfo  {journal} {Phys. Rev.}\ }\textbf {\bibinfo {volume}
  {101}},\ \bibinfo {pages} {1620} (\bibinfo {year} {1956})}\BibitemShut
  {NoStop}%
\bibitem [{\citenamefont {Kubo}(1966)}]{Kubo_1966}%
  \BibitemOpen
  \bibfield  {author} {\bibinfo {author} {\bibfnamefont {R.}~\bibnamefont
  {Kubo}},\ }\bibfield  {title} {\bibinfo {title} {The fluctuation-dissipation
  theorem},\ }\href {https://doi.org/10.1088/0034-4885/29/1/306} {\bibfield
  {journal} {\bibinfo  {journal} {Reports on Progress in Physics}\ }\textbf
  {\bibinfo {volume} {29}},\ \bibinfo {pages} {255} (\bibinfo {year}
  {1966})}\BibitemShut {NoStop}%
\bibitem [{\citenamefont {Felderhof}(1978)}]{Felderhof_JOP_1978}%
  \BibitemOpen
  \bibfield  {author} {\bibinfo {author} {\bibfnamefont {B.~U.}\ \bibnamefont
  {Felderhof}},\ }\bibfield  {title} {\bibinfo {title} {On the derivation of
  the fluctuation-dissipation theorem},\ }\href
  {https://doi.org/10.1088/0305-4470/11/5/021} {\bibfield  {journal} {\bibinfo
  {journal} {Journal of Physics A: Mathematical and General}\ }\textbf
  {\bibinfo {volume} {11}},\ \bibinfo {pages} {921} (\bibinfo {year}
  {1978})}\BibitemShut {NoStop}%
\bibitem [{\citenamefont {Continentino}(2021)}]{Mucio_book_2021}%
  \BibitemOpen
  \bibfield  {author} {\bibinfo {author} {\bibfnamefont {M.~A.}\ \bibnamefont
  {Continentino}},\ }\href {https://doi.org/10.1088/978-0-7503-3395-5ch1}
  {\emph {\bibinfo {title} {Key Methods and Concepts in Condensed Matter
  Physics}}}\ (\bibinfo  {publisher} {IOP Publishing},\ \bibinfo {year}
  {2021})\BibitemShut {NoStop}%
\bibitem [{\citenamefont {Meir}\ \emph {et~al.}(1993)\citenamefont {Meir},
  \citenamefont {Wingreen},\ and\ \citenamefont {Lee}}]{Lee_PRL_1993}%
  \BibitemOpen
  \bibfield  {author} {\bibinfo {author} {\bibfnamefont {Y.}~\bibnamefont
  {Meir}}, \bibinfo {author} {\bibfnamefont {N.~S.}\ \bibnamefont {Wingreen}},\
  and\ \bibinfo {author} {\bibfnamefont {P.~A.}\ \bibnamefont {Lee}},\
  }\bibfield  {title} {\bibinfo {title} {Low-temperature transport through a
  quantum dot: The anderson model out of equilibrium},\ }\href
  {https://doi.org/10.1103/PhysRevLett.70.2601} {\bibfield  {journal} {\bibinfo
   {journal} {Phys. Rev. Lett.}\ }\textbf {\bibinfo {volume} {70}},\ \bibinfo
  {pages} {2601} (\bibinfo {year} {1993})}\BibitemShut {NoStop}%
\bibitem [{\citenamefont {Cui}\ \emph {et~al.}(2008)\citenamefont {Cui},
  \citenamefont {Xi},\ and\ \citenamefont {Pan}}]{Wei_PRA_2008}%
  \BibitemOpen
  \bibfield  {author} {\bibinfo {author} {\bibfnamefont {W.}~\bibnamefont
  {Cui}}, \bibinfo {author} {\bibfnamefont {Z.~R.}\ \bibnamefont {Xi}},\ and\
  \bibinfo {author} {\bibfnamefont {Y.}~\bibnamefont {Pan}},\ }\bibfield
  {title} {\bibinfo {title} {Optimal decoherence control in non-{M}arkovian
  open dissipative quantum systems},\ }\href
  {https://doi.org/10.1103/PhysRevA.77.032117} {\bibfield  {journal} {\bibinfo
  {journal} {Phys. Rev. A}\ }\textbf {\bibinfo {volume} {77}},\ \bibinfo
  {pages} {032117} (\bibinfo {year} {2008})}\BibitemShut {NoStop}%
\bibitem [{\citenamefont {Lacroix}\ \emph {et~al.}(2020)\citenamefont
  {Lacroix}, \citenamefont {Sargsyan}, \citenamefont {Adamian}, \citenamefont
  {Antonenko},\ and\ \citenamefont {Hovhannisyan}}]{Denis_PRA_2020}%
  \BibitemOpen
  \bibfield  {author} {\bibinfo {author} {\bibfnamefont {D.}~\bibnamefont
  {Lacroix}}, \bibinfo {author} {\bibfnamefont {V.~V.}\ \bibnamefont
  {Sargsyan}}, \bibinfo {author} {\bibfnamefont {G.~G.}\ \bibnamefont
  {Adamian}}, \bibinfo {author} {\bibfnamefont {N.~V.}\ \bibnamefont
  {Antonenko}},\ and\ \bibinfo {author} {\bibfnamefont {A.~A.}\ \bibnamefont
  {Hovhannisyan}},\ }\bibfield  {title} {\bibinfo {title} {Non-{M}arkovian
  modeling of {F}ermi-{B}ose systems coupled to one or several {F}ermi-{B}ose
  thermal baths},\ }\href {https://doi.org/10.1103/PhysRevA.102.022209}
  {\bibfield  {journal} {\bibinfo  {journal} {Phys. Rev. A}\ }\textbf {\bibinfo
  {volume} {102}},\ \bibinfo {pages} {022209} (\bibinfo {year}
  {2020})}\BibitemShut {NoStop}%
\bibitem [{\citenamefont {Zhang}\ \emph {et~al.}(2021)\citenamefont {Zhang},
  \citenamefont {Tao}, \citenamefont {He}, \citenamefont {Chen}, \citenamefont
  {Kong}, \citenamefont {Deng}, \citenamefont {Lambert},\ and\ \citenamefont
  {Ai}}]{Zhang_FP_2021}%
  \BibitemOpen
  \bibfield  {author} {\bibinfo {author} {\bibfnamefont {N.-N.}\ \bibnamefont
  {Zhang}}, \bibinfo {author} {\bibfnamefont {M.-J.}\ \bibnamefont {Tao}},
  \bibinfo {author} {\bibfnamefont {W.-T.}\ \bibnamefont {He}}, \bibinfo
  {author} {\bibfnamefont {X.-Y.}\ \bibnamefont {Chen}}, \bibinfo {author}
  {\bibfnamefont {X.-Y.}\ \bibnamefont {Kong}}, \bibinfo {author}
  {\bibfnamefont {F.-G.}\ \bibnamefont {Deng}}, \bibinfo {author}
  {\bibfnamefont {N.}~\bibnamefont {Lambert}},\ and\ \bibinfo {author}
  {\bibfnamefont {Q.}~\bibnamefont {Ai}},\ }\bibfield  {title} {\bibinfo
  {title} {Efficient quantum simulation of open quantum dynamics at various
  {H}amiltonians and spectral densities},\ }\href
  {https://doi.org/10.1007/s11467-021-1064-y} {\bibfield  {journal} {\bibinfo
  {journal} {Frontiers of Physics}\ }\textbf {\bibinfo {volume} {16}},\
  \bibinfo {pages} {51501} (\bibinfo {year} {2021})}\BibitemShut {NoStop}%
\bibitem [{\citenamefont {Wolf}\ \emph {et~al.}(2008)\citenamefont {Wolf},
  \citenamefont {Eisert}, \citenamefont {Cubitt},\ and\ \citenamefont
  {Cirac}}]{Wolf_PRL_2008}%
  \BibitemOpen
  \bibfield  {author} {\bibinfo {author} {\bibfnamefont {M.~M.}\ \bibnamefont
  {Wolf}}, \bibinfo {author} {\bibfnamefont {J.}~\bibnamefont {Eisert}},
  \bibinfo {author} {\bibfnamefont {T.~S.}\ \bibnamefont {Cubitt}},\ and\
  \bibinfo {author} {\bibfnamefont {J.~I.}\ \bibnamefont {Cirac}},\ }\bibfield
  {title} {\bibinfo {title} {Assessing non-{M}arkovian quantum dynamics},\
  }\href {https://doi.org/10.1103/PhysRevLett.101.150402} {\bibfield  {journal}
  {\bibinfo  {journal} {Phys. Rev. Lett.}\ }\textbf {\bibinfo {volume} {101}},\
  \bibinfo {pages} {150402} (\bibinfo {year} {2008})}\BibitemShut {NoStop}%
\bibitem [{\citenamefont {Breuer}\ \emph {et~al.}(2009)\citenamefont {Breuer},
  \citenamefont {Laine},\ and\ \citenamefont {Piilo}}]{Breuer_PRL_2009}%
  \BibitemOpen
  \bibfield  {author} {\bibinfo {author} {\bibfnamefont {H.-P.}\ \bibnamefont
  {Breuer}}, \bibinfo {author} {\bibfnamefont {E.-M.}\ \bibnamefont {Laine}},\
  and\ \bibinfo {author} {\bibfnamefont {J.}~\bibnamefont {Piilo}},\ }\bibfield
   {title} {\bibinfo {title} {Measure for the degree of non-{M}arkovian
  behavior of quantum processes in open systems},\ }\href
  {https://doi.org/10.1103/PhysRevLett.103.210401} {\bibfield  {journal}
  {\bibinfo  {journal} {Phys. Rev. Lett.}\ }\textbf {\bibinfo {volume} {103}},\
  \bibinfo {pages} {210401} (\bibinfo {year} {2009})}\BibitemShut {NoStop}%
\bibitem [{\citenamefont {Rivas}\ \emph {et~al.}(2010)\citenamefont {Rivas},
  \citenamefont {Huelga},\ and\ \citenamefont {Plenio}}]{Rivas_PRL_2010}%
  \BibitemOpen
  \bibfield  {author} {\bibinfo {author} {\bibfnamefont {{\'A}.}~\bibnamefont
  {Rivas}}, \bibinfo {author} {\bibfnamefont {S.~F.}\ \bibnamefont {Huelga}},\
  and\ \bibinfo {author} {\bibfnamefont {M.~B.}\ \bibnamefont {Plenio}},\
  }\bibfield  {title} {\bibinfo {title} {Entanglement and non-{M}arkovianity of
  quantum evolutions},\ }\href {https://doi.org/10.1103/PhysRevLett.105.050403}
  {\bibfield  {journal} {\bibinfo  {journal} {Phys. Rev. Lett.}\ }\textbf
  {\bibinfo {volume} {105}},\ \bibinfo {pages} {050403} (\bibinfo {year}
  {2010})}\BibitemShut {NoStop}%
\bibitem [{\citenamefont {Lu}\ \emph {et~al.}(2010)\citenamefont {Lu},
  \citenamefont {Wang},\ and\ \citenamefont {Sun}}]{Lu_PRA_2010}%
  \BibitemOpen
  \bibfield  {author} {\bibinfo {author} {\bibfnamefont {X.-M.}\ \bibnamefont
  {Lu}}, \bibinfo {author} {\bibfnamefont {X.}~\bibnamefont {Wang}},\ and\
  \bibinfo {author} {\bibfnamefont {C.~P.}\ \bibnamefont {Sun}},\ }\bibfield
  {title} {\bibinfo {title} {Quantum {F}isher information flow and
  non-{M}arkovian processes of open systems},\ }\href
  {https://doi.org/10.1103/PhysRevA.82.042103} {\bibfield  {journal} {\bibinfo
  {journal} {Phys. Rev. A}\ }\textbf {\bibinfo {volume} {82}},\ \bibinfo
  {pages} {042103} (\bibinfo {year} {2010})}\BibitemShut {NoStop}%
\bibitem [{\citenamefont {Lorenzo}\ \emph {et~al.}(2011)\citenamefont
  {Lorenzo}, \citenamefont {Plastina},\ and\ \citenamefont
  {Paternostro}}]{Lorenzo_PRA_2011}%
  \BibitemOpen
  \bibfield  {author} {\bibinfo {author} {\bibfnamefont {S.}~\bibnamefont
  {Lorenzo}}, \bibinfo {author} {\bibfnamefont {F.}~\bibnamefont {Plastina}},\
  and\ \bibinfo {author} {\bibfnamefont {M.}~\bibnamefont {Paternostro}},\
  }\bibfield  {title} {\bibinfo {title} {Role of environmental correlations in
  the non-{M}arkovian dynamics of a spin system},\ }\href
  {https://doi.org/10.1103/PhysRevA.84.032124} {\bibfield  {journal} {\bibinfo
  {journal} {Phys. Rev. A}\ }\textbf {\bibinfo {volume} {84}},\ \bibinfo
  {pages} {032124} (\bibinfo {year} {2011})}\BibitemShut {NoStop}%
\bibitem [{\citenamefont {Luo}\ \emph {et~al.}(2012)\citenamefont {Luo},
  \citenamefont {Fu},\ and\ \citenamefont {Song}}]{Luo_PRA_2012}%
  \BibitemOpen
  \bibfield  {author} {\bibinfo {author} {\bibfnamefont {S.}~\bibnamefont
  {Luo}}, \bibinfo {author} {\bibfnamefont {S.}~\bibnamefont {Fu}},\ and\
  \bibinfo {author} {\bibfnamefont {H.}~\bibnamefont {Song}},\ }\bibfield
  {title} {\bibinfo {title} {Quantifying non-{M}arkovianity via correlations},\
  }\href {https://doi.org/10.1103/PhysRevA.86.044101} {\bibfield  {journal}
  {\bibinfo  {journal} {Phys. Rev. A}\ }\textbf {\bibinfo {volume} {86}},\
  \bibinfo {pages} {044101} (\bibinfo {year} {2012})}\BibitemShut {NoStop}%
\bibitem [{\citenamefont {Zhong}\ \emph {et~al.}(2013)\citenamefont {Zhong},
  \citenamefont {Sun}, \citenamefont {Ma}, \citenamefont {Wang},\ and\
  \citenamefont {Nori}}]{Zhong_PRA_2013}%
  \BibitemOpen
  \bibfield  {author} {\bibinfo {author} {\bibfnamefont {W.}~\bibnamefont
  {Zhong}}, \bibinfo {author} {\bibfnamefont {Z.}~\bibnamefont {Sun}}, \bibinfo
  {author} {\bibfnamefont {J.}~\bibnamefont {Ma}}, \bibinfo {author}
  {\bibfnamefont {X.}~\bibnamefont {Wang}},\ and\ \bibinfo {author}
  {\bibfnamefont {F.}~\bibnamefont {Nori}},\ }\bibfield  {title} {\bibinfo
  {title} {Fisher information under decoherence in {B}loch representation},\
  }\href {https://doi.org/10.1103/PhysRevA.87.022337} {\bibfield  {journal}
  {\bibinfo  {journal} {Phys. Rev. A}\ }\textbf {\bibinfo {volume} {87}},\
  \bibinfo {pages} {022337} (\bibinfo {year} {2013})}\BibitemShut {NoStop}%
\bibitem [{\citenamefont {Liu}\ \emph {et~al.}(2013)\citenamefont {Liu},
  \citenamefont {Lu},\ and\ \citenamefont {Wang}}]{Liu_PRA_2013}%
  \BibitemOpen
  \bibfield  {author} {\bibinfo {author} {\bibfnamefont {J.}~\bibnamefont
  {Liu}}, \bibinfo {author} {\bibfnamefont {X.-M.}\ \bibnamefont {Lu}},\ and\
  \bibinfo {author} {\bibfnamefont {X.}~\bibnamefont {Wang}},\ }\bibfield
  {title} {\bibinfo {title} {Nonunital non-{M}arkovianity of quantum
  dynamics},\ }\href {https://doi.org/10.1103/PhysRevA.87.042103} {\bibfield
  {journal} {\bibinfo  {journal} {Phys. Rev. A}\ }\textbf {\bibinfo {volume}
  {87}},\ \bibinfo {pages} {042103} (\bibinfo {year} {2013})}\BibitemShut
  {NoStop}%
\bibitem [{\citenamefont {Lorenzo}\ \emph {et~al.}(2013)\citenamefont
  {Lorenzo}, \citenamefont {Plastina},\ and\ \citenamefont
  {Paternostro}}]{Lorenzo_PRA_2013}%
  \BibitemOpen
  \bibfield  {author} {\bibinfo {author} {\bibfnamefont {S.}~\bibnamefont
  {Lorenzo}}, \bibinfo {author} {\bibfnamefont {F.}~\bibnamefont {Plastina}},\
  and\ \bibinfo {author} {\bibfnamefont {M.}~\bibnamefont {Paternostro}},\
  }\bibfield  {title} {\bibinfo {title} {Geometrical characterization of
  non-{M}arkovianity},\ }\href {https://doi.org/10.1103/PhysRevA.88.020102}
  {\bibfield  {journal} {\bibinfo  {journal} {Phys. Rev. A}\ }\textbf {\bibinfo
  {volume} {88}},\ \bibinfo {pages} {020102} (\bibinfo {year}
  {2013})}\BibitemShut {NoStop}%
\bibitem [{\citenamefont {Rivas}\ \emph {et~al.}(2014)\citenamefont {Rivas},
  \citenamefont {Huelga},\ and\ \citenamefont {Plenio}}]{Rivas_RPP_2014}%
  \BibitemOpen
  \bibfield  {author} {\bibinfo {author} {\bibfnamefont {{\'A}.}~\bibnamefont
  {Rivas}}, \bibinfo {author} {\bibfnamefont {S.~F.}\ \bibnamefont {Huelga}},\
  and\ \bibinfo {author} {\bibfnamefont {M.~B.}\ \bibnamefont {Plenio}},\
  }\bibfield  {title} {\bibinfo {title} {Quantum non-{M}arkovianity:
  characterization, quantification and detection},\ }\href
  {https://doi.org/10.1088/0034-4885/77/9/094001} {\bibfield  {journal}
  {\bibinfo  {journal} {Reports on Progress in Physics}\ }\textbf {\bibinfo
  {volume} {77}},\ \bibinfo {pages} {094001} (\bibinfo {year}
  {2014})}\BibitemShut {NoStop}%
\bibitem [{\citenamefont {Debarba}\ and\ \citenamefont
  {Fanchini}(2017)}]{Debarba_PRA_2017}%
  \BibitemOpen
  \bibfield  {author} {\bibinfo {author} {\bibfnamefont {T.}~\bibnamefont
  {Debarba}}\ and\ \bibinfo {author} {\bibfnamefont {F.~F.}\ \bibnamefont
  {Fanchini}},\ }\bibfield  {title} {\bibinfo {title} {Non-{M}arkovianity
  quantifier of an arbitrary quantum process},\ }\href
  {https://doi.org/10.1103/PhysRevA.96.062118} {\bibfield  {journal} {\bibinfo
  {journal} {Phys. Rev. A}\ }\textbf {\bibinfo {volume} {96}},\ \bibinfo
  {pages} {062118} (\bibinfo {year} {2017})}\BibitemShut {NoStop}%
\bibitem [{\citenamefont {Strasberg}\ and\ \citenamefont
  {Esposito}(2018)}]{Strasberg_PRL_2018}%
  \BibitemOpen
  \bibfield  {author} {\bibinfo {author} {\bibfnamefont {P.}~\bibnamefont
  {Strasberg}}\ and\ \bibinfo {author} {\bibfnamefont {M.}~\bibnamefont
  {Esposito}},\ }\bibfield  {title} {\bibinfo {title} {Response functions as
  quantifiers of non-{M}arkovianity},\ }\href
  {https://doi.org/10.1103/PhysRevLett.121.040601} {\bibfield  {journal}
  {\bibinfo  {journal} {Phys. Rev. Lett.}\ }\textbf {\bibinfo {volume} {121}},\
  \bibinfo {pages} {040601} (\bibinfo {year} {2018})}\BibitemShut {NoStop}%
\bibitem [{\citenamefont {Huang}\ and\ \citenamefont
  {Guo}(2021)}]{Huang_PRA_2021}%
  \BibitemOpen
  \bibfield  {author} {\bibinfo {author} {\bibfnamefont {Z.}~\bibnamefont
  {Huang}}\ and\ \bibinfo {author} {\bibfnamefont {X.-K.}\ \bibnamefont
  {Guo}},\ }\bibfield  {title} {\bibinfo {title} {Quantifying
  non-{M}arkovianity via conditional mutual information},\ }\href
  {https://doi.org/10.1103/PhysRevA.104.032212} {\bibfield  {journal} {\bibinfo
   {journal} {Phys. Rev. A}\ }\textbf {\bibinfo {volume} {104}},\ \bibinfo
  {pages} {032212} (\bibinfo {year} {2021})}\BibitemShut {NoStop}%
\bibitem [{\citenamefont {Ghoshal}\ and\ \citenamefont
  {Sen}()}]{Ahana_arxiv_2022}%
  \BibitemOpen
  \bibfield  {author} {\bibinfo {author} {\bibfnamefont {A.}~\bibnamefont
  {Ghoshal}}\ and\ \bibinfo {author} {\bibfnamefont {U.}~\bibnamefont {Sen}},\
  }\href@noop {} {\bibinfo {title} {Multiparty {S}pohn's theorem for mixed
  local {M}arkovian and non-{M}arkovian quantum dynamics}},\ \Eprint
  {https://arxiv.org/abs/2208.13026} {arXiv:2208.13026 [quant-ph]} \BibitemShut
  {NoStop}%
\bibitem [{\citenamefont {Breuer}\ \emph {et~al.}(1999)\citenamefont {Breuer},
  \citenamefont {Kappler},\ and\ \citenamefont {Petruccione}}]{Breuer1999}%
  \BibitemOpen
  \bibfield  {author} {\bibinfo {author} {\bibfnamefont {H.-P.}\ \bibnamefont
  {Breuer}}, \bibinfo {author} {\bibfnamefont {B.}~\bibnamefont {Kappler}},\
  and\ \bibinfo {author} {\bibfnamefont {F.}~\bibnamefont {Petruccione}},\
  }\bibfield  {title} {\bibinfo {title} {Stochastic wave-function method for
  non-markovian quantum master equations},\ }\href
  {https://doi.org/10.1103/PhysRevA.59.1633} {\bibfield  {journal} {\bibinfo
  {journal} {Phys. Rev. A}\ }\textbf {\bibinfo {volume} {59}},\ \bibinfo
  {pages} {1633} (\bibinfo {year} {1999})}\BibitemShut {NoStop}%
\bibitem [{\citenamefont {Maniscalco}\ \emph {et~al.}(2004)\citenamefont
  {Maniscalco}, \citenamefont {Intravaia}, \citenamefont {Piilo},\ and\
  \citenamefont {Messina}}]{Maniscalco_2004}%
  \BibitemOpen
  \bibfield  {author} {\bibinfo {author} {\bibfnamefont {S.}~\bibnamefont
  {Maniscalco}}, \bibinfo {author} {\bibfnamefont {F.}~\bibnamefont
  {Intravaia}}, \bibinfo {author} {\bibfnamefont {J.}~\bibnamefont {Piilo}},\
  and\ \bibinfo {author} {\bibfnamefont {A.}~\bibnamefont {Messina}},\
  }\bibfield  {title} {\bibinfo {title} {Misbeliefs and misunderstandings about
  the non-markovian dynamics of a damped harmonic oscillator},\ }\href
  {https://doi.org/10.1088/1464-4266/6/3/016} {\bibfield  {journal} {\bibinfo
  {journal} {Journal of Optics B: Quantum and Semiclassical Optics}\ }\textbf
  {\bibinfo {volume} {6}},\ \bibinfo {pages} {S98} (\bibinfo {year}
  {2004})}\BibitemShut {NoStop}%
\bibitem [{\citenamefont {Erika~Andersson}\ and\ \citenamefont
  {Hall}(2007)}]{Andersson2007}%
  \BibitemOpen
  \bibfield  {author} {\bibinfo {author} {\bibfnamefont {J.~D.~C.}\
  \bibnamefont {Erika~Andersson}}\ and\ \bibinfo {author} {\bibfnamefont
  {M.~J.~W.}\ \bibnamefont {Hall}},\ }\bibfield  {title} {\bibinfo {title}
  {Finding the kraus decomposition from a master equation and vice versa},\
  }\href {https://doi.org/10.1080/09500340701352581} {\bibfield  {journal}
  {\bibinfo  {journal} {Journal of Modern Optics}\ }\textbf {\bibinfo {volume}
  {54}},\ \bibinfo {pages} {1695} (\bibinfo {year} {2007})}\BibitemShut
  {NoStop}%
\bibitem [{\citenamefont {Piilo}\ \emph {et~al.}(2008)\citenamefont {Piilo},
  \citenamefont {Maniscalco}, \citenamefont {H\"ark\"onen},\ and\ \citenamefont
  {Suominen}}]{Piilo2008}%
  \BibitemOpen
  \bibfield  {author} {\bibinfo {author} {\bibfnamefont {J.}~\bibnamefont
  {Piilo}}, \bibinfo {author} {\bibfnamefont {S.}~\bibnamefont {Maniscalco}},
  \bibinfo {author} {\bibfnamefont {K.}~\bibnamefont {H\"ark\"onen}},\ and\
  \bibinfo {author} {\bibfnamefont {K.-A.}\ \bibnamefont {Suominen}},\
  }\bibfield  {title} {\bibinfo {title} {Non-markovian quantum jumps},\ }\href
  {https://doi.org/10.1103/PhysRevLett.100.180402} {\bibfield  {journal}
  {\bibinfo  {journal} {Phys. Rev. Lett.}\ }\textbf {\bibinfo {volume} {100}},\
  \bibinfo {pages} {180402} (\bibinfo {year} {2008})}\BibitemShut {NoStop}%
\bibitem [{\citenamefont {Hall}\ \emph {et~al.}(2014)\citenamefont {Hall},
  \citenamefont {Cresser}, \citenamefont {Li},\ and\ \citenamefont
  {Andersson}}]{Anderson_PRA_2014}%
  \BibitemOpen
  \bibfield  {author} {\bibinfo {author} {\bibfnamefont {M.~J.~W.}\
  \bibnamefont {Hall}}, \bibinfo {author} {\bibfnamefont {J.~D.}\ \bibnamefont
  {Cresser}}, \bibinfo {author} {\bibfnamefont {L.}~\bibnamefont {Li}},\ and\
  \bibinfo {author} {\bibfnamefont {E.}~\bibnamefont {Andersson}},\ }\bibfield
  {title} {\bibinfo {title} {Canonical form of master equations and
  characterization of non-{M}arkovianity},\ }\href
  {https://doi.org/10.1103/PhysRevA.89.042120} {\bibfield  {journal} {\bibinfo
  {journal} {Phys. Rev. A}\ }\textbf {\bibinfo {volume} {89}},\ \bibinfo
  {pages} {042120} (\bibinfo {year} {2014})}\BibitemShut {NoStop}%
\bibitem [{\citenamefont {Siudzińska}\ and\ \citenamefont
  {Chruściński}(2020)}]{Siudzinska_2020}%
  \BibitemOpen
  \bibfield  {author} {\bibinfo {author} {\bibfnamefont {K.}~\bibnamefont
  {Siudzińska}}\ and\ \bibinfo {author} {\bibfnamefont {D.}~\bibnamefont
  {Chruściński}},\ }\bibfield  {title} {\bibinfo {title} {Quantum evolution
  with a large number of negative decoherence rates},\ }\href
  {https://doi.org/10.1088/1751-8121/aba7f2} {\bibfield  {journal} {\bibinfo
  {journal} {Journal of Physics A: Mathematical and Theoretical}\ }\textbf
  {\bibinfo {volume} {53}},\ \bibinfo {pages} {375305} (\bibinfo {year}
  {2020})}\BibitemShut {NoStop}%
\bibitem [{\citenamefont {Sanderson}\ and\ \citenamefont
  {Curtin}(2016)}]{Sanderson}%
  \BibitemOpen
  \bibfield  {author} {\bibinfo {author} {\bibfnamefont {C.}~\bibnamefont
  {Sanderson}}\ and\ \bibinfo {author} {\bibfnamefont {R.}~\bibnamefont
  {Curtin}},\ }\bibfield  {title} {\bibinfo {title} {Armadillo: a
  template-based {C}++ library for linear algebra},\ }\href
  {https://doi.org/10.21105/joss.00026} {\bibfield  {journal} {\bibinfo
  {journal} {J. Open Source Softw}\ }\textbf {\bibinfo {volume} {1}},\ \bibinfo
  {pages} {26} (\bibinfo {year} {2016})}\BibitemShut {NoStop}%
\bibitem [{\citenamefont {Sanderson}\ and\ \citenamefont
  {Curtin}(2018)}]{Sanderson1}%
  \BibitemOpen
  \bibfield  {author} {\bibinfo {author} {\bibfnamefont {C.}~\bibnamefont
  {Sanderson}}\ and\ \bibinfo {author} {\bibfnamefont {R.}~\bibnamefont
  {Curtin}},\ }\bibfield  {title} {\bibinfo {title} {A user-friendly hybrid
  sparse matrix class in {C}++},\ }\href
  {https://doi.org/10.1007/978-3-319-96418-8_50} {\bibfield  {journal}
  {\bibinfo  {journal} {LNCS}\ }\textbf {\bibinfo {volume} {10931}},\ \bibinfo
  {pages} {430} (\bibinfo {year} {2018})}\BibitemShut {NoStop}%
\end{thebibliography}%


\end{document}